\documentclass[12pt,preprint]{aastex}
\begin{document}
\title{The Black Hole-Bulge Relationship in Luminous Broad-Line Active Galactic Nuclei and Host Galaxies}
\author{Jiajian Shen}
\author{Daniel E. Vanden Berk}
\author{Donald P. Schneider}
\affil{\rm Department of Astronomy and Astrophysics, The Pennsylvania State University, University Park, PA 16802}
\and
\author{Patrick B. Hall}
\affil{\rm Department of Physics and Astronomy, York University, 4700 Keele St., \\
  Toronto, ON, M3J 1P3, Canada}
\begin{abstract}
We have measured the stellar velocity dispersions ($\sigma_*$) and
estimated the central black hole (BH) masses for over 900 broad-line
active galactic nuclei (AGNs) observed with the Sloan Digital Sky
Survey. The sample includes objects which have redshifts up to
$z=0.452$, high quality spectra, and host galaxy spectra dominated
by an early-type (bulge) component.  The AGN and host galaxy
spectral components were decomposed using an eigenspectrum
technique. The BH masses ($M_{\rm BH}$) were estimated from the AGN
broad-line widths, and the velocity dispersions were measured from
the stellar absorption spectra of the host galaxies.  The range of
black hole masses covered by the sample is approximately $10^{6} <
M_{\rm BH} < 10^{9} M_{\Sun}$.  The host galaxy luminosity-velocity
dispersion relationship follows the well-known Faber-Jackson
relation for early-type galaxies, with a power-law slope
$4.33\pm0.21$.  The estimated BH masses are correlated with both the
host luminosities ($L_{H}$) and the stellar velocity dispersions
($\sigma_{*}$), similar to the relationships found for low-redshift,
bulge-dominated galaxies. The intrinsic scatter in the correlations
are large ($\sim$~0.4 dex), but the very large sample size allows
tight constraints to be placed on the mean relationships: $M_{\rm
BH} \propto L_H^{0.73\pm0.05}$ and $M_{\rm BH} \propto
\sigma_*^{3.34\pm0.24}$.  The amplitude of the $M_{\rm BH}-\sigma_*$
relation depends on the estimated Eddington ratio, such that objects
with larger Eddington ratios have smaller black hole masses than
expected at a given velocity dispersion.  While this dependence is
probably caused at least in part by sample selection effects, it can
account for the intrinsic scatter in the $M_{\rm BH}-\sigma_*$
relation, and may tie together the accretion rate with physical
properties of the host bulge component.  We find no significant
evolution in the $M_{\rm BH}-\sigma_*$ relation with redshift, up to
$z\approx0.4$, after controlling for possible dependencies on other
variables.  Interested readers can contact the authors to obtain the
eigenspectrum decomposition coefficients of our objects.
\end{abstract}


\section{Introduction}
It is now widely accepted that all galaxies with a massive bulge component
contain a central massive black hole (BH).  Application of stellar
dynamical and gas dynamical techniques for measuring masses of central
BHs in normal galaxies has led to the identification of correlations
between the BHs and host galaxies \citep{KR95, M98}.  The tightest of
these relationships are the correlations of BH mass ($M_{\rm BH}$) with
the galaxy bulge luminosity and the bulge stellar velocity dispersion
($\sigma_{*}$) \citep{FM00,G00a,T02}.

Massive BHs have also been postulated in quasars and active galaxies
\citep{L69,R84}.  The question of whether AGNs follow a similar BH-bulge
relation as normal galaxies is a very interesting one, since it may
elucidate the connection between the host galaxy and the active nucleus.
Comparisons between the bulge and BH properties in quasars and bright
Seyfert galaxies is observationally very difficult, however, since
the stellar component near the BH is easily lost in the glare of the
active nucleus.

Reverberation mapping (e.g. Blandford $\&$ McKee 1982; Peterson 1993;
Netzer $\&$ Peterson 1997; Peterson et al. 2004 and references therein)
of the broad emission lines in AGNs has been used to determine the size
of the broad line region (BLR), $R_{\rm BLR}$=c$\tau$, where $\tau$ is the
time lag between continuum and emission line variations.  Assuming that
the widths of permitted emission lines (e.g., H\,$\beta$) are due to
virialized gas motions in the BH potential, the BH mass can be estimated
from the velocity dispersion $\Delta V$ and the radius at the location
of the gas $R_{\rm BLR}$.  The central BH mass is then given by
\begin{equation}
M_{\rm BH} = f\frac{R_{BLR}\Delta V^2}{G},
\label{masseq}
\end{equation}
where $f$ is a factor of order unity that depends on the structure and
geometry of the BLR.

Multiple line observations show an anticorrelation between
emission-line lags and line widths, i.e., $\tau \propto \Delta V^{-2}$
\citep{Peterson99,Peterson00,Onken02,Kollatschny03}, which strongly
supports the interpretation that the BLR dynamics are dominated by the
BH gravitational potential.  Thus, reverberation mapping provides an
important technique for estimating the central BH masses of AGNs.
The most complete reverberation-based BH masses can be found in
\citet{Peterson04}.  Another important result of reverberation
mapping is the discovery of a simple power law relationship between
the continuum luminosity and the size of BLR, $R_{BLR} \propto
L^{\alpha}$ \citep{K00,K05,Bentz06}.  The value of $\alpha$ using
optical wavelengths and the broad Balmer emission lines, and after
correcting for the contribution of host galaxy light,  is found to
be $\approx 0.5$, which is consistent with simple photoionization
expectations \,\citep{Bentz06}.  The radius-luminosity relationship
provides a secondary method of estimating the BH masses, suitable for
use with single-epoch spectra.  The single-epoch spectra technique has
been applied to large samples of broad line AGNs to estimate BH masses
\citep{Vestergaard02,Mclure02,Netzer03,Warner03,Wu04,Vestergaard04,
Vestergaard06}.

The difficulty of measuring the host bulge properties remains,
however. Image decomposition techniques can separate the host from the
central AGNs in some cases (e.g. Mclure et al. 2000; Peng et al. 2002,
Kulbroadt et al. 2004, Bentz et al. 2006), but they do not provide the
necessary spectroscopic information, such as the stellar absorption lines
that are necessary for measuring host galaxy stellar velocity dispersions
($\sigma_*$).  Host galaxy stellar velocity dispersions have been directly
measured in some cases by fitting the stellar absorption lines of the
Ca\,{\sc ii} triplet at a wavelength of $\lambda \approx 8550${\AA}
\citep[e.g.][]{Ferrarese01, Barth02, Nelson04, Onken04, Greene06b}.
However, the Ca\,{\sc ii} triplet is difficult to observe for large
samples of AGNs, for the following three reasons.  First, the lines are
often redshifted out of the observable spectral range. Second, the lines
are usually contaminated by night sky emission and absorption features.
Finally, the host galaxy is usually very faint as compared to the active
nucleus, causing the absorption lines to be very weak in contrast to
the nuclear continuum.  When the Ca\,{\sc ii} triplet is not available,
the [O\,{\sc iii}]$\lambda 5007$ narrow emission line is sometimes
used as a surrogate for estimating the stellar velocity dispersion
\citep{Nelson00, Shields03, grupe04, Salviander06}. However, while there
is a correlation between the width of [O\,{\sc iii}] and $\sigma_*$
measured from absorption lines, the scatter in the relationship is
substantial \citep{Nelson96, Onken04, Botte05, Woo06}. So while the
width of [O\,{\sc iii}] may be used as a surrogate for $\sigma_*$
in a statistical sense, the BH mass estimated from [O\,{\sc iii}] can
be uncertain by a factor of 5  \citep{Nelson00, Boroson03, Bonning05,
Boroson05}.  Thus, the published measurements of host galaxy velocity
dispersions for Seyfert 1 galaxies and quasars are few in number ($<100$),
which prevent statistically significant investigations of the $M_{\rm
BH}$-$\sigma_*$ relationship.

Principal component analysis (PCA) has been performed on both galaxy
\citep{YiP04a} and quasar \citep{YiP04b} samples from the Sloan
Digital Sky Survey (SDSS; \citealt{York00}).  These studies have
shown that galaxies and quasars can be classified based on only a few
eigencoefficients in each case. The galaxy spectroscopic sample variance
is strongly concentrated in just the first few eigenspectra, with over
$98\%$ of the information contained in the first three eigenspectra
\citep{YiP04a}. The quasar information is not as strongly concentrated
as in the case of galaxies, but the first 10 quasar eigenspectra
account for $\approx 92\%$ of the sample variance \citep{YiP04b}.
At low luminosities, the quasar second eigenspectrum has a strong
galactic component that resembles the first galaxy eigenspectrum.
\citet[][hereafter VB06]{V06} used separate sets of galaxy and quasar
eigenspectra to efficiently and reliably separate the AGNs and host
spectroscopic components.  Their tests showed that the technique
accurately reproduces the host galaxy spectrum, its contributing flux
fraction, and its classification.  Inspired by the success of the
spectral decomposition, we have separated the host galaxies from the
broad line AGNs in a large sample of spectra using the technique of
VB06. Among the merits of this method is that a large sample of host
galaxy spectra is available from which the stellar velocity dispersions
can be measured by the stellar absorption lines.  Another advantage of
this method is that the BH masses are more accurately measured by using
the host-subtracted AGNs than by using composite spectra, especially
when host galaxy contributions are not negligible \citep{Bentz06}. The
purpose of this work is to study the BH-bulge relationships for active
galaxies in a more statistically meaningful way, given that we can obtain
both host and AGN properties for a large number of objects.

The structure of this paper is as follows.  In \S\,2 we describe our
sample selection, and in \S\,3 we describe the data analysis.  The main
results of the AGN and host parameter correlations are included in \S\,4.
A discussion and summary follow in \S\,5.  Throughout the paper we assume
a cosmology consistent with recent results from the WMAP experiment
\citep{SD06}: $\Omega_m$=0.3, $\Omega_{\Lambda}$ = 0.7 and H${}_0$ =
70~km~s${}^{-1}$ Mpc${}^{-1}$.

\section{Observations and Sample definition \label{observations}}
The broad-line AGNs used for this study were selected from the SDSS.
The SDSS is a project to image approximately 10${}^4$ deg${}^2$ of sky,
in five broad photometric bands ($u, g, r, i, z$) to a depth of $r \sim
23$, and to obtain spectra of $10^6$ galaxies and $10^5$ quasars selected
from the imaging survey \citep{F96,Hogg01,I02,Smith02,pier03}. Imaging
observations are made with a dedicated 2.5 m telescope \citep{gunn06},
using a large mosaic CCD camera \citep{gunn98} in a drift-scanning
mode.  The SDSS spectra are obtained using a pair of multi-object
spectrographs that simultaneously accept 320 optical fibers each
\citep[see][]{stoughton02,blanton03}.  Each fiber subtends a $3''$ diameter
on the sky (see \S\,\ref{sigma} for a discussion of fiber size effects).
The wavelength range of each spectrum is approximately 3800-9200{\AA},
with a resolution of $\lambda/\Delta\lambda\approx1850$. The total
spectral integration time is at least 45 minutes.

We define broad-line AGNs here to mean any extragalactic object with
at least one spectroscopic emission line with a FWHM of at least
$1000\,{\rm km\,s^{-1}}$, regardless of the object luminosity or
morphology.  Broad-line AGN candidates are selected from SDSS color
space, or as unresolved matches to sources in the FIRST radio catalog
\citep{becker95}, as described by \citet{richards02}. Quasars are also
often identified because the objects were targeted for spectroscopy
by non-quasar selection algorithms, such as optical matches to ROSAT
sources \citep{stoughton02,anderson03}, various classes of stars
\citep{stoughton02}, so-called serendipity objects \citep{stoughton02},
and galaxies \citep{strauss02}.  The completeness of the SDSS quasar
selection algorithm is close to $95\%$ up to the $i$ band limiting survey
magnitude of $19.1$ \citep{vandenberk05}.

The set of AGNs studied here were selected from the list given by VB06. In that study, the AGNs were drawn from
the catalog of SDSS Data Release Three quasars described by \citet{Schneider05}, and from an extension to that
catalog that was constructed by including objects with absolute magnitudes fainter than $M_{i}=-22$, which is the
limit imposed by \citet{Schneider05}.  From that sample, VB06 selected $4666$ AGNs for which a host galaxy
spectral component could be reliably decomposed, using their eigenspectrum technique.      Those AGNs have
redshifts of $z<0.752$,     and host galaxy fractional flux contributions, $F_{H}$, of greater than $10\%$ between
$4160$ and $4210${\AA}.  The selected spectra were corrected for foreground Galactic extinction using the Milky
Way extinction curve described by \citet{fitzpatrick99}, and the reddening maps provided by \citet{schlegel98}.
The decomposition technique uses separate sets of five galaxy eigenspectra and ten quasar eigenspectra to
efficiently and reliably separate the AGN and host spectroscopic components.  The technique accurately reproduces
the host galaxy spectrum, its contributing fraction, and its classification. The details of the AGN and host
galaxy decomposition are given by VB06, along with the decomposition parameters and various derived parameters for
each object.  For clarity, throughout this paper we define the original spectra before decomposition as the active
galaxy spectra, the host galaxy components reconstructed with galaxy eigenspectra as the {\em reconstructed} host
galaxy spectra, and the AGN spectra reconstructed with quasar eigenspectra as the {\em reconstructed} AGN spectra.

\begin{figure}
  \plotone{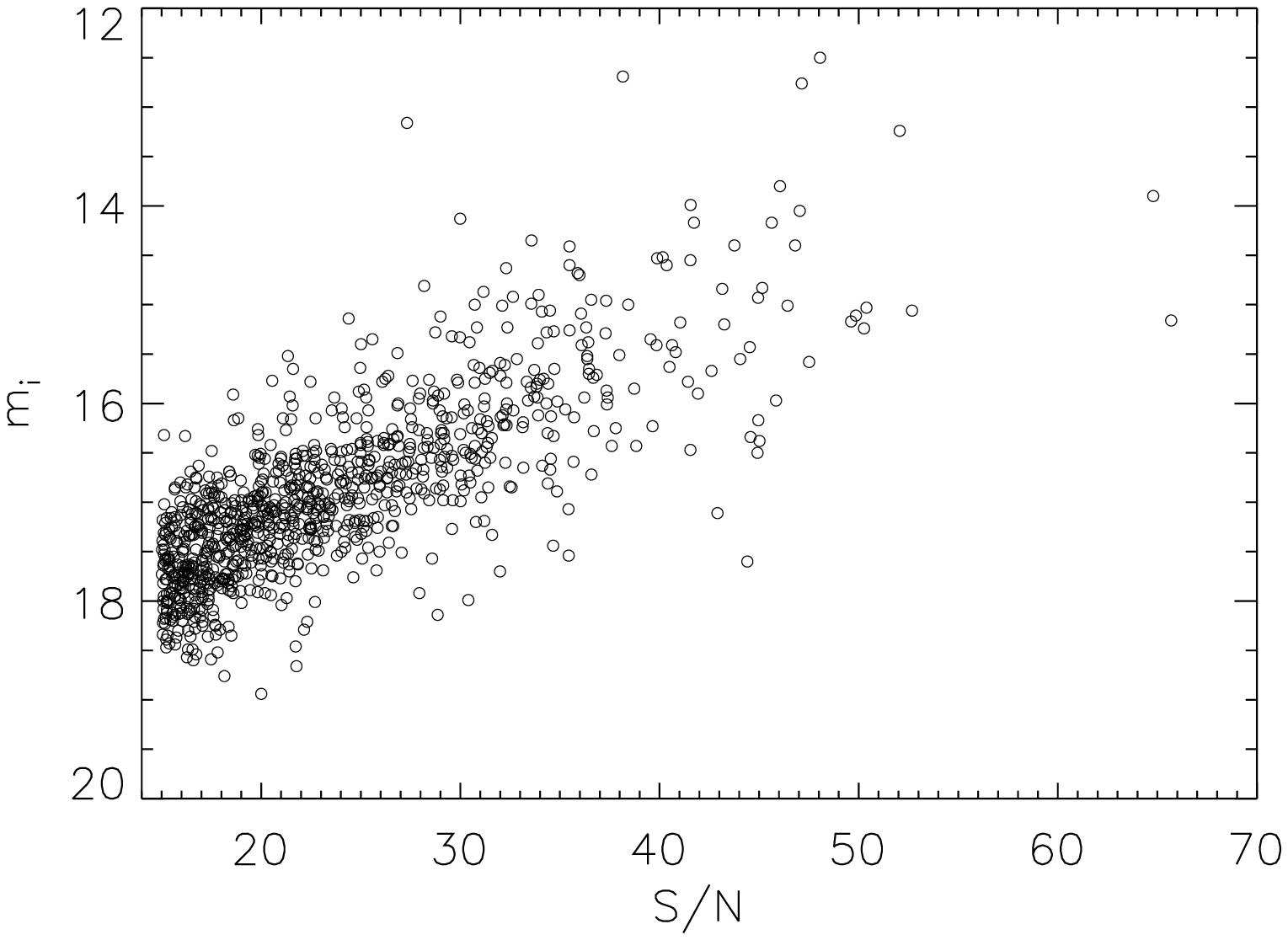}
\caption{Sample object apparent $i$ band magnitudes $m_i$ vs. the
  spectroscopic $S/N$ averaged over the $i$ band.  The median $S/N$ is
  21.6. \label{lsn}}
\end{figure}

\begin{figure}
  \plotone{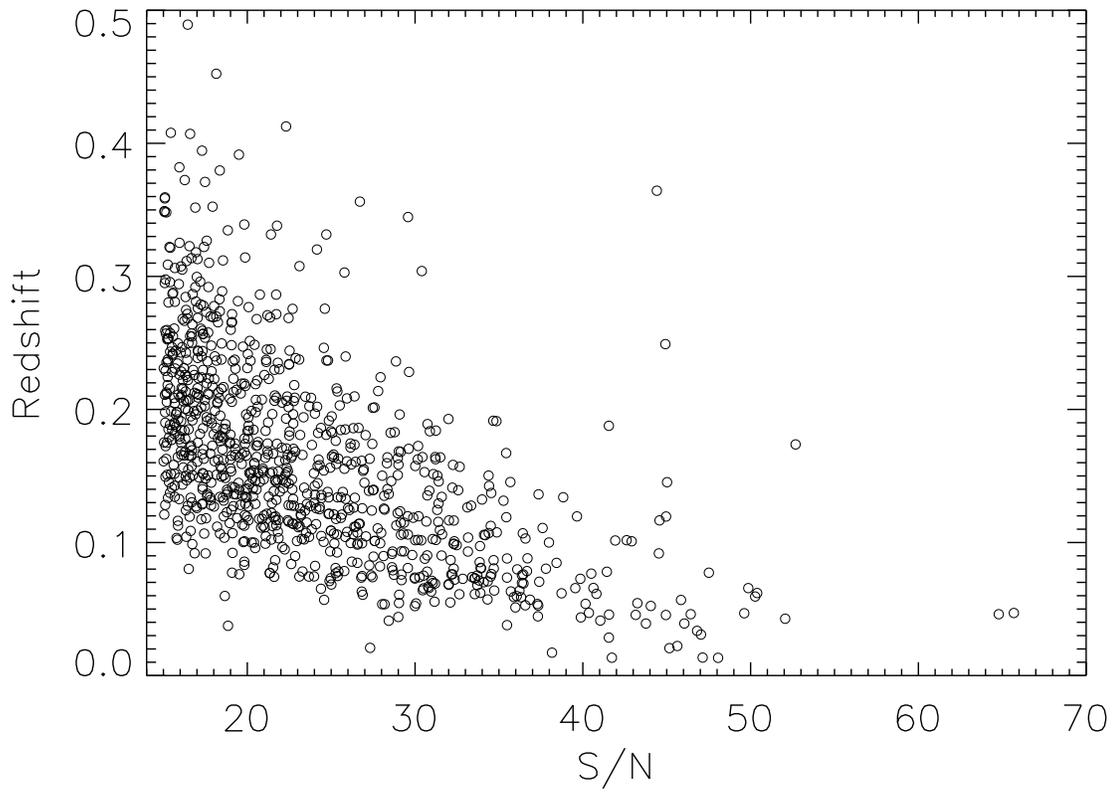}
\caption{Sample object redshift vs. the spectroscopic $S/N$ averaged
  over the $i$ band. \label{zsn}}
\end{figure}

\begin{figure}
  \plotone{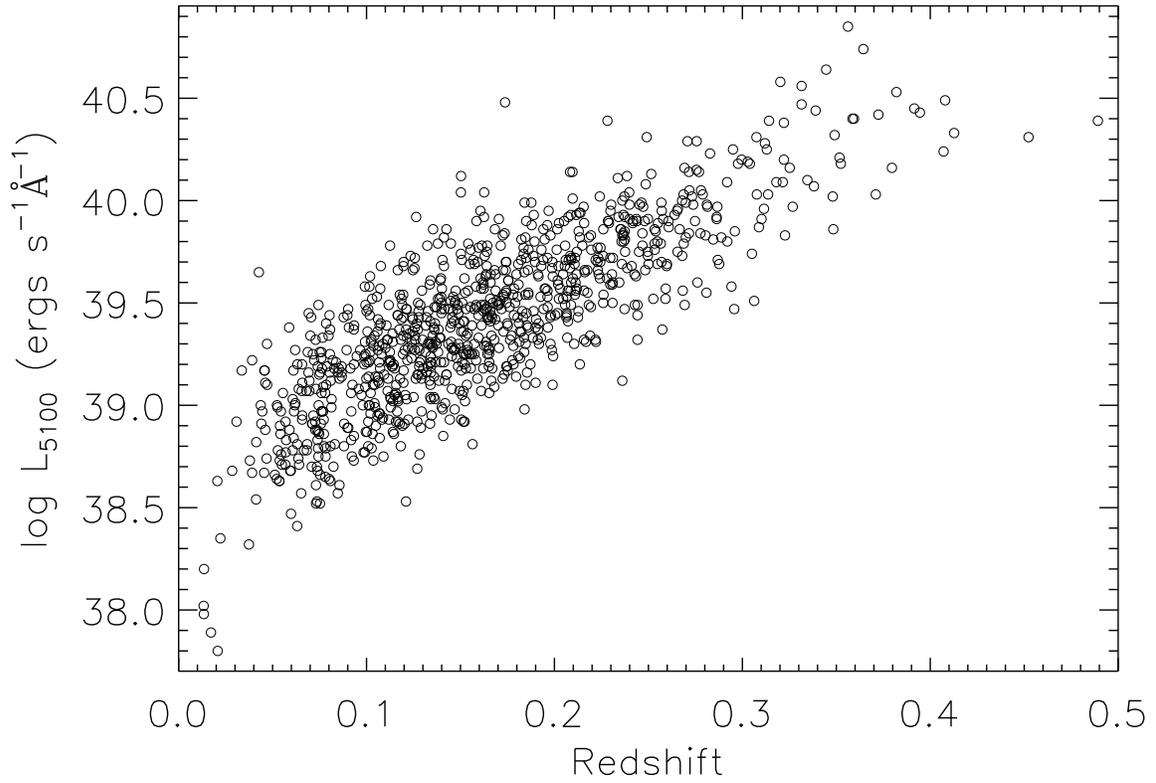}
  \caption{AGN monochromatic continuum luminosity at a rest wavelength of
  5100\AA, $L_{5100}$, as a function of redshift for the 960 objects
  in our sample.  The AGN luminosity was measured after host galaxy
  subtraction; the measurement of this quantity is discussed in
  \S\,\ref{mbhmeasure}. \label{zl}}
\end{figure}

For the current study we made additional restrictions on the initial
set of $4666$ AGNs.  Each of the decomposed components (AGN and host
galaxy) must have a sufficiently high signal-to-noise ratio ($S/N$) to
reliably estimate the black hole masses and stellar velocity dispersions,
respectively.  Because the $M_{\rm BH}-\sigma_*$ relation applies only
to the bulge component of the host galaxies, we also required that
the {\em reconstructed} host galaxy components be dominated by bulge
or ``early-type'' spectra.  Quantitatively, the following criteria
were applied: 1) The host flux fraction $F_{H}$, must be between
0.2 and 0.8, which guarantees that the AGN and host galaxy are both
significant contributors to the active galaxy spectrum.  2) The host
galaxy classification angle $\phi_{H}$, defined using the values of the
first two eigencoefficients (see VB06 and \citet{YiP04a}), must be in the
range $0\arcdeg<\phi_{H}<15\arcdeg$, which selects bulge-dominated galaxy
spectra.  The spectrum classification  is based  on the  flux within SDSS
fiber apertures ($3''$).  For a galaxy dominated  by a bulge component in
the SDSS spectrum, the morphology of the entire galaxy image, extending
beyond  the  $3''$ aperture, may or may not be classified as early
type.  Thus, this criterion does not guarantee that the host galaxy
would be classified {\em morphologically} as a bulge-dominated galaxy.
However, this spectroscopic study focuses on the bulge  component of  the
galaxies,  regardless  of  the  extended morphology.   Within the  $3''$
fiber aperture,  the selected  sample galaxies are all bulge dominated.
3) The active galaxy spectrum must have a mean $S/N$ per pixel $> 15$,
averaged over the SDSS $i$\,band sensitivity function.  We also required
the redshifted position of the H\,$\beta$ line to be covered by the
spectra, but that constraint was already satisfied by the $z<0.752$
limit in the VB06 sample.  In practice the $S/N$ requirement limited
the maximum redshift to be far below 0.75.  These criteria resulted in
a sample of 960 spectra, with redshifts from 0.013 to 0.489 (the median
redshift is 0.153) and $S/N$ from 15.0 to 65.7 (the median $S/N$ is 21.6).
Figure\,\ref{lsn} shows the relation between the $i$ band $S/N$ and the
apparent magnitude $m_i$.  Figure\,\ref{zsn} shows the $S/N$ level as a
function of redshift.  Figure\,\ref{zl} shows the AGN component luminosity
$L_{5100}$ (the monochromatic continuum luminosity at a rest wavelength
of 5100{\AA} (see \S\,\ref{emWidth})), as a function of redshift for the
sample. The correlation of luminosity with redshift is due mainly to the
apparent magnitude limit imposed by the SDSS quasar selection algorithm.

It is possible that some low redshift, extraordinarily luminous objects
are missed because of the SDSS upper flux limits ($i=15$).  However,
these should be relatively rare objects since the SDSS completeness to
broad-line AGN is extremely high, $\sim 95\%$ (Vanden Berk et al. 2005).
Moreover, since the relative brightness of the AGN and host is one of
our selection criteria ($0.2<F_H<0.8$), these most luminous objects
would not be selected for our sample since the flux is dominated
by nuclear light.  At the other extreme, the sample certainly omits
the least luminous nuclear sources, both because of the SDSS faint
flux limits, and because host galaxy light would dominate the flux.
These objects would presumably, though not certainly, have properties
that lie somewhere between those of the relatively brighter objects in
our sample, and the galaxies with quiescent supermassive black holes
examined in other studies \citep[e.g.][]{T02}.

\section{Data Analysis\label{analysis}}
%
\subsection{Estimation of Black Hole Mass\label{mbhmeasure}}
The BH masses were estimated by equation (\ref{masseq}), which requires
the widths of the broad emission lines $\Delta V$ or FWHM, the distance
of the broad emission line gas from the central gravitational source
$R_{\rm BLR}$, and the scale factor $f$.  The broad emission line widths
can be measured directly from the spectra.  The gas radius cannot
be measured directly from single-epoch spectra, so we use the result
based on reverberation mapping studies that the continuum luminosity
and radius are strongly correlated.  The radius-luminosity (R-L)
relationship has been calibrated using the gas distances from samples
of objects for which reverberation mapping techniques could be used.
The power-law index for the correlation between gas radius and the optical
luminosity at $5100${\AA}, $L_{5100}$, was reported to be about 0.67-0.70
by \citet{K00,K05}.  However, the R-L relationship can be affected by
host galaxy contamination; using new $Hubble \ Space \ Telescope \ (HST)$
imaging of reverberation-mapped AGNs, \citet{Bentz06} corrected the host
galaxy contamination of the optical continuum luminosity and reported a
flatter power-law index of 0.52.  Since our study deals with the nuclear
luminosities, separated from the host stellar contribution, we adopted the
R-L relationship from \citet{Bentz06}.  The value of the scale factor $f$
is unknown.  However, using the reanalyzed reverberation mapping BH masses
of \citet{Peterson04} and {\em assuming} that the active galaxies obey the
same $M_{\rm BH}-\sigma_*$ relationship as quiescent galaxies \citep{T02},
\citet{Onken04} derived a value for $f$ by requiring the two techniques
yield statistically consistent BH masses. In this paper we use the most
recently determined mass scaling relationship from \citet{Vestergaard06},
which was calibrated by combining the R-L relationship by \citet{Bentz06}
and the scale factor by \citet{Onken04}.  The mass scaling relationship
for single epoch spectra, using the FWHM of the broad H\,$\beta$ emission
line, and the monochromatic continuum luminosity at 5100{\AA}, $L_{5100}$,
is given as \citep{Vestergaard06}
\begin{equation}
  \log M_{\rm BH}=\log  \biggl\{
    {\Bigl(\frac{{\rm FWHM}_{{\rm H}\beta}}{10^3 {\rm ~km~s}^{-1}}\Bigr)}^2
    {\Bigl( \frac{\lambda L_{5100}}
    {10^{44} {\rm ergs}}\Bigr)}^{0.5} \biggr\}+(6.91 \pm 0.02) \,,
  \label{Mbh}
\end{equation}
where $M_{\rm BH}$ is the BH mass in solar units.  The zero point
error in equation (\ref{Mbh}) is a formal statistical error.  A more
representative uncertainty in the zero point for an individual object
is the intrinsic scatter about the relationship of $\pm 0.43$ dex.
The measurements of the H$\beta$ FWHM and $L_{5100}$ for our sample of
AGNs are described in the following subsections.

%
\subsection{Emission Line Width Measurement\label{emWidth}}
\begin{figure}
  \plotone{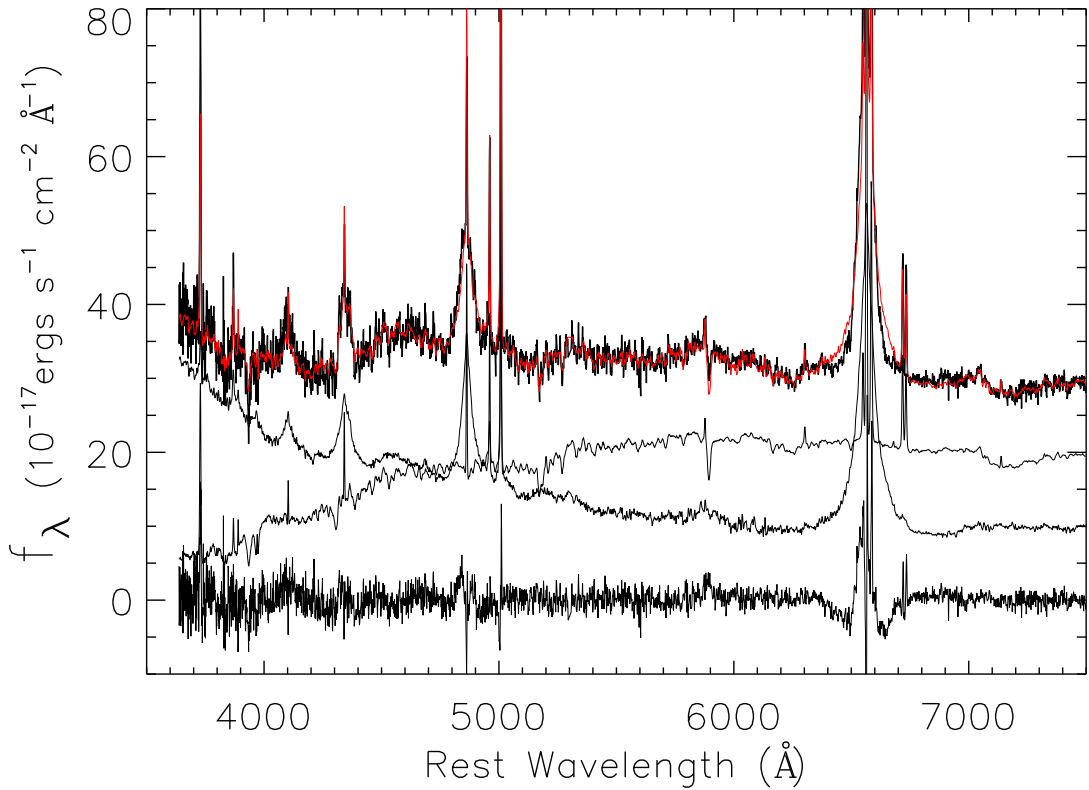}
  \caption{The SDSS spectrum of SDSS\,J104252.94+041441.1, with $z=0.053$,
  a typical example of a high-$S/N$ object in our sample, with $S/N=37.3$
  (the spectra have a resolution of $\Delta \lambda/\lambda \sim$1850).
  The middle components are the {\em reconstructed} AGN and host galaxy
  spectra, the top histogram is the original active galaxy spectrum,
  and the bottom histogram is the fit residual. The gray curve is the
  modeled active galaxy spectrum, which is red in the electronic version.
  \label{decom}}
\end{figure}

\begin{figure}
  \plotone{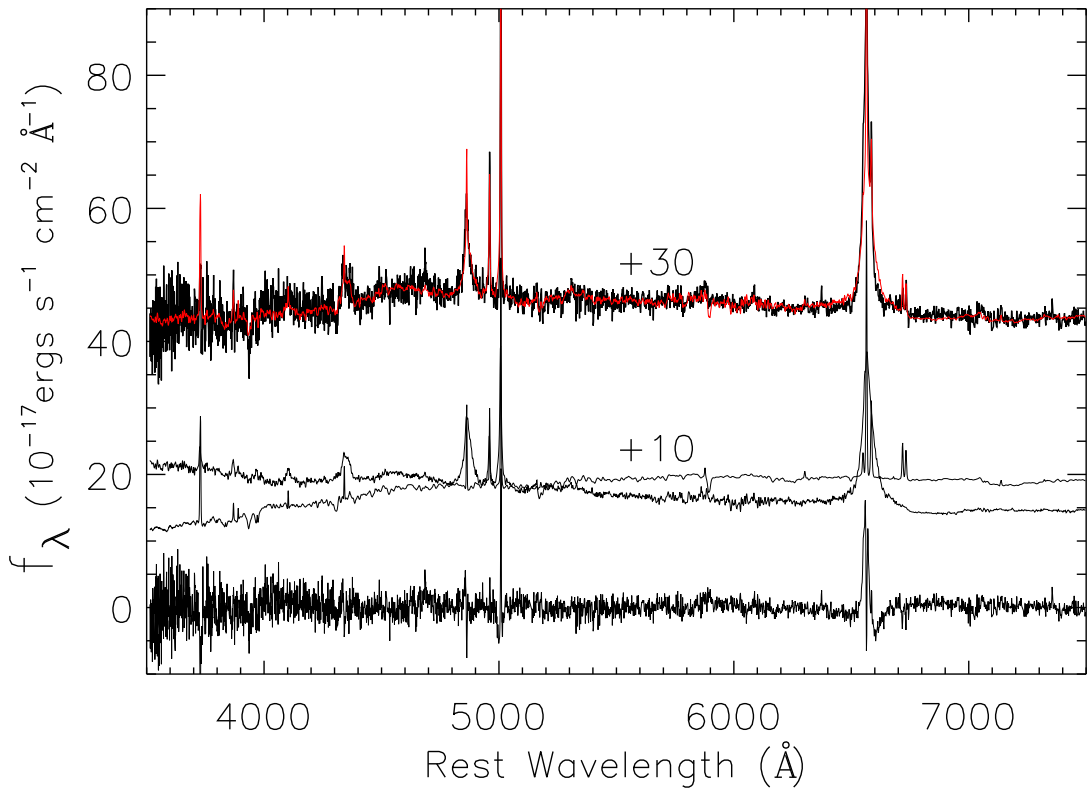}
  \caption{The SDSS spectrum of SDSS\,J212348.61+105348.2, with $z=0.087$,
  a typical example of an object with average $S/N$ in our sample, with
  $S/N=21.7$. The middle components are the {\em reconstructed} AGN and
  host galaxy spectra with an offset for clarity, the top histogram
  is the original active galaxy spectrum with an offset for clarity,
  and the bottom histogram is the fit residual.  The gray curve is
  the modeled active galaxy spectrum, which is red in the electronic
  version.\label{decom_845}}
\end{figure}

\begin{figure}
  \plotone{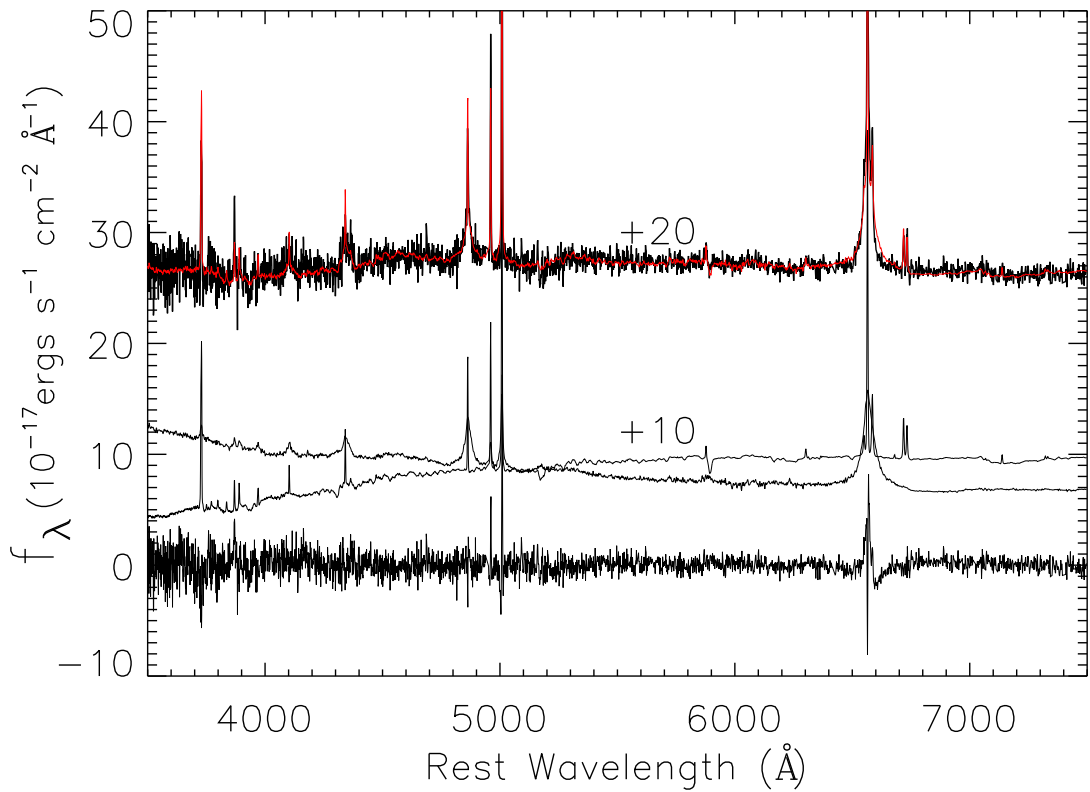}
  \caption{The SDSS spectrum of SDSS J094813.80+401325.9, with $z=0.128$,
  a typical example of a low-$S/N$ object in our sample, with
  $S/N=15.1$. The middle components are the {\em reconstructed} AGN and
  host galaxy spectra with an offset for clarity, the top histogram
  is the original active galaxy spectrum with an offset for clarity,
  and the bottom histogram is the fit residual.  The gray curve is
  the modeled active galaxy spectrum, which is red in the electronic
  version.\label{decom_303}}
\end{figure}

The 960 spectra were first decomposed into host galaxy and AGN spectra
using the eigenspectrum method described by VB06. Examples of the
spectroscopic decomposition of the objects with different $S/N$ are
shown in Figs.~\ref{decom}, \ref{decom_845}, and \ref{decom_303}. One
can also refer to Fig.\,9 in VB06 for more decomposition examples. In
each of the examples, the original spectrum is shown along with the
{\em reconstructed} AGN and host galaxy spectra and the fit residuals.
The {\em reconstructed} spectrum is a close match to the original active
galaxy spectrum, except for some larger residuals in the regions of narrow
and broad emission lines.  Accurately reconstructing the emission lines
can require a large number of eigenspectra, which may introduce spurious
features into the reconstructed continuum (VB06).

\begin{figure}
  \plotone{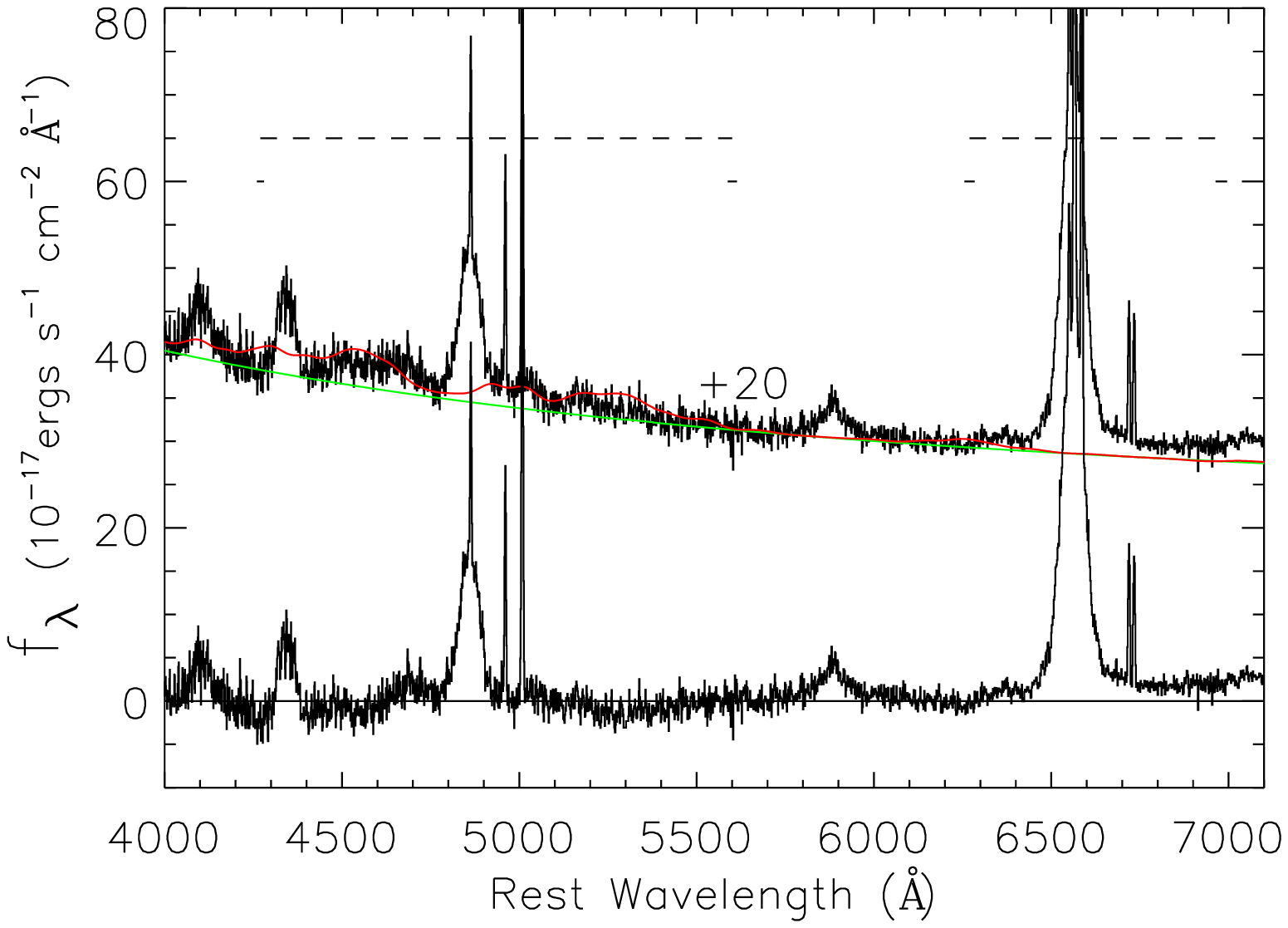}
  \caption{Example continuum and FeII template fit, for the object described
  in Figure \ref{decom}.  The top spectrum is the host-subtracted
  AGN spectrum with an offset for clarity, the middle lines are
  the featureless continuum and the iron template superposed on the
  continuum with an offset for clarity, and the bottom histogram is the
  AGN spectrum after subtracting the continuum and iron template. The
  four short horizontal solid lines denote the end points used for continuum
  fitting.  The two horizontal dashed lines show the two continuum
  regions used to fit the broken power law.  In the electronic version,
  the green curve is the modeled continuum and the red curve is the iron
  template superposed on the continuum.
  \label{coniron}}
\end{figure}

\begin{figure}
  \plotone{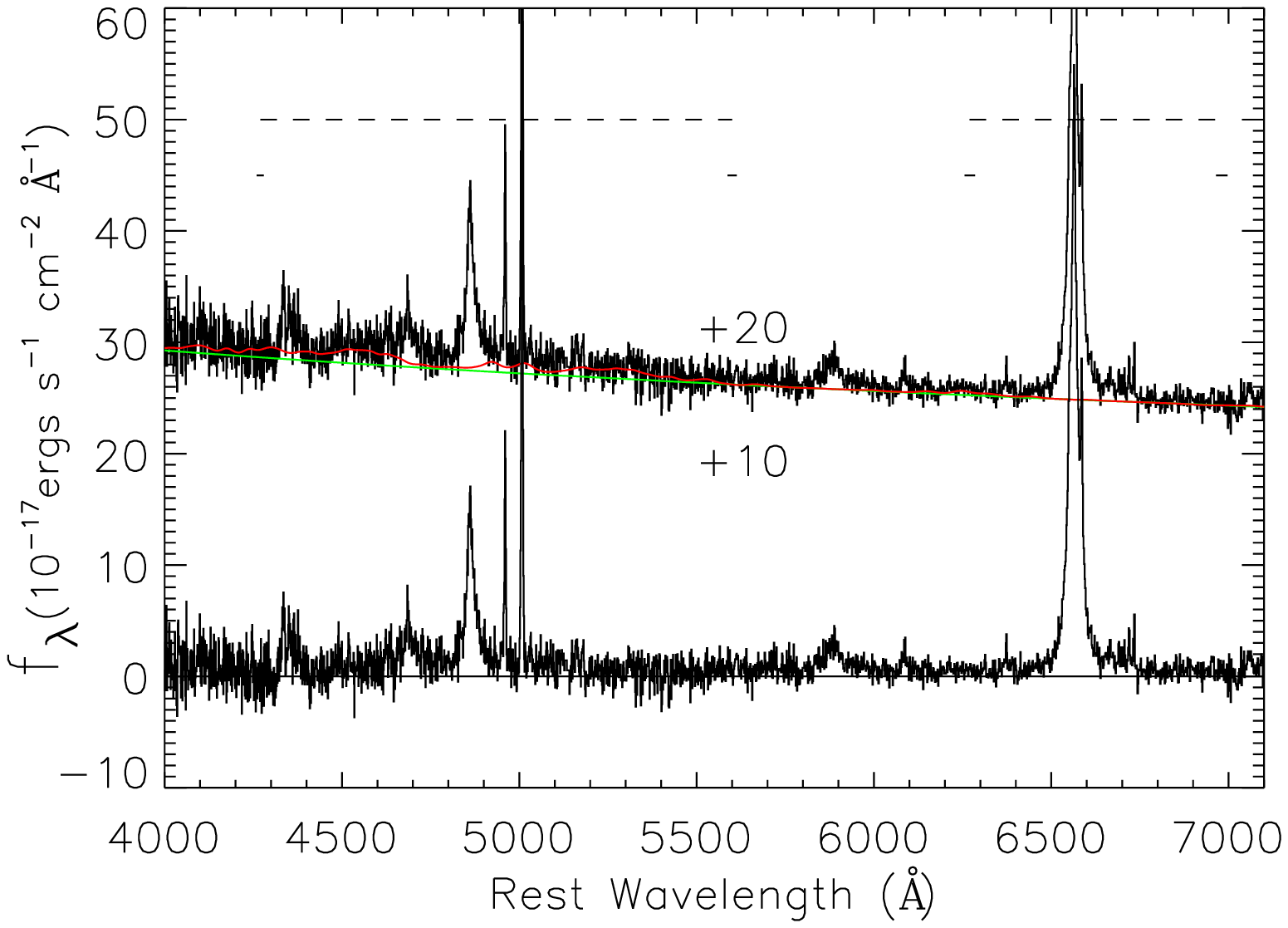}
  \caption{Example continuum and FeII template fit, for the object described
  in Figure \ref{decom_845}.  The top spectrum is the host-subtracted
  AGN spectrum with an offset for clarity, the middle lines are
  the featureless continuum and the iron template superposed on the
  continuum with an offset for clarity, and the bottom histogram is
  the AGN spectrum after subtracting the continuum and iron template.
  The four short horizontal solid lines denote the end points used for
  continuum fitting.  The two horizontal dashed lines show the
  continuum regions used to fit the broken power law.  In the electronic
  version, the green curve is the modeled continuum and the red curve
  is the iron template superposed on the continuum.
  \label{coniron_845}}
\end{figure}

\begin{figure}
  \plotone{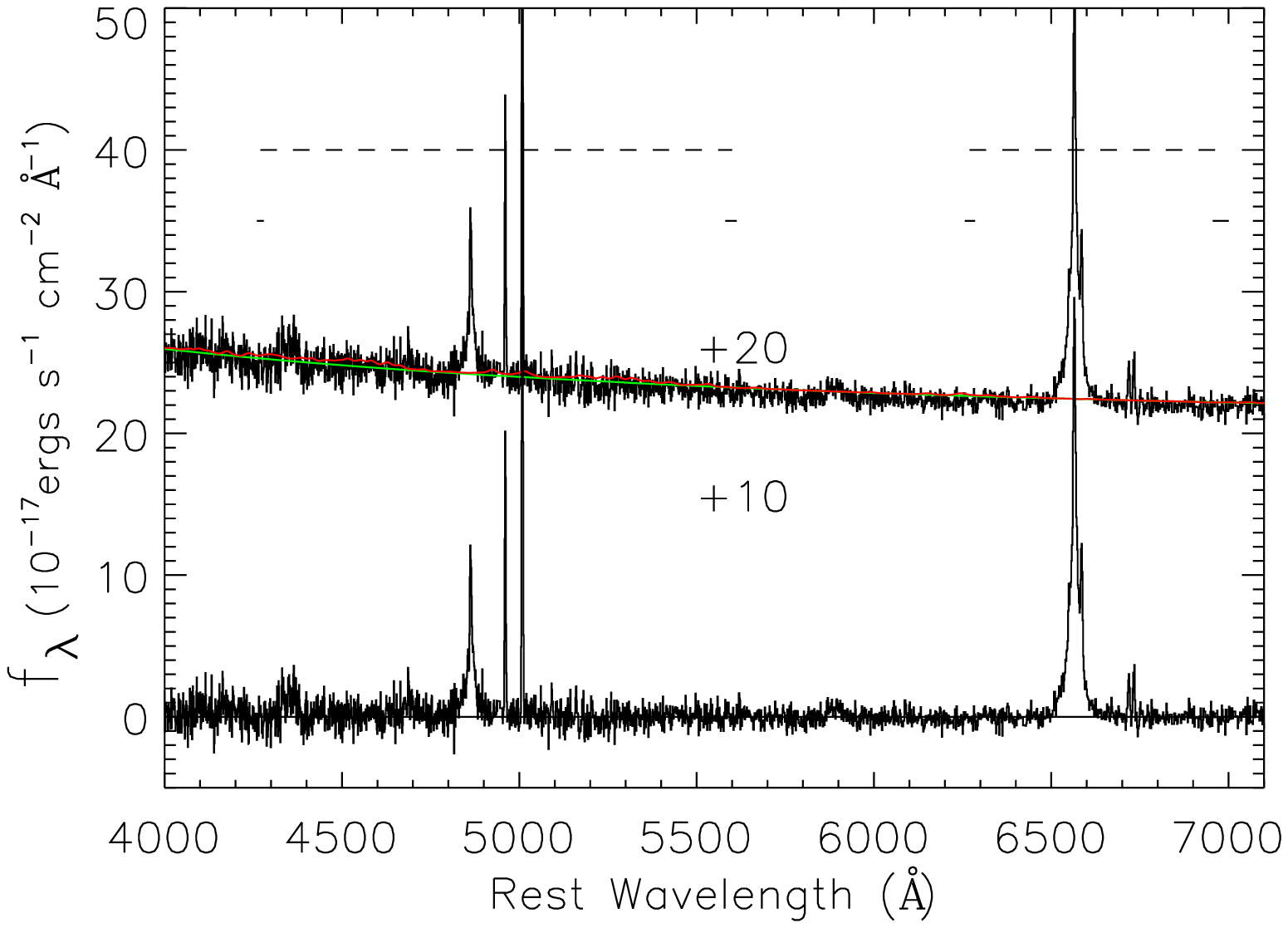}
  \caption{Example continuum and FeII template fit, for the object described
  in Figure \ref{decom_303}.  The top spectrum is the host-subtracted
  AGN spectrum with an offset for clarity, the middle lines are
  the featureless continuum and the iron template superposed on the
  continuum with an offset for clarity, and the bottom histogram is
  the AGN spectrum after subtracting the continuum and iron template.
  The four short horizontal solid lines denote the end points used for
  continuum fitting.  The two horizontal dashed lines show the
  continuum regions used to fit the broken power law.  In the electronic
  version, the green curve is the modeled continuum and the red curve
  is the iron template superposed on the continuum.
  \label{coniron_303}}
\end{figure}

To measure the AGN emission lines, we do not use the {\em reconstructed}
AGN spectrum, but instead use the original active galaxy spectrum with
the {\em reconstructed} host spectrum subtracted.  Because we do not use
the {\em reconstructed} AGN spectrum for our analysis of the emission
lines, the imperfect modeling of the AGN broad lines as seen in the
residuals in Fig.\,\ref{decom} does not affect the FWHM measurements of
the broad lines.  In subtracting the host spectrum, we are very careful
in dealing with the narrow lines.  After decomposing an active galaxy
spectrum, the {\em reconstructed} host galaxy spectrum is interpolated
across the narrow line regions, effectively removing the narrow lines.
The interpolated {\em reconstructed} host spectrum is then subtracted
from the original active galaxy spectrum.  The resulting AGN spectrum
thus contains both the host and AGN components of the narrow line regions.
This method is better at preserving the details of the AGN broad emission
line profile, and it retains the noise of the original spectrum, which is
necessary to assess the quality of the spectral measurements.  We define
the host-subtracted active galaxy spectrum as the {\em decomposed} AGN
spectrum.  The {\em decomposed} AGN spectrum should be differentiated
from the {\em reconstructed} AGN spectrum.  As mentioned above, the
{\em decomposed} AGN spectrum has had the host spectrum subtracted,
and it retains all of the necessary original spectrum information,
such as emission lines and noise.  However, the {\em reconstructed} AGN
spectrum is reconstructed by the eigenspectra and the eigencoefficients
--- this spectrum is not used for any spectroscopic measurements.
Figs.\,\ref{coniron}, \ref{coniron_845}, and \ref{coniron_303} show
examples of {\em decomposed} AGN spectra, which clearly show that
the emission lines are preserved, even in the narrow line regions.
The monochromatic luminosities at 5100{\AA} of the sample objects were
measured from the rest frame {\em decomposed} AGN spectra, by averaging
over a 10.57{\AA} (10-pixel) wide region centered on 5100{\AA}.

The H\,$\beta$ profile is not always sufficiently strong in our spectra to
accurately determine its FWHM.  We also measured the FWHM of H\,$\alpha$
for all objects possible because, as shown by \citet{Greene05}, one
can use the H\,$\alpha$ line width to approximate the width of the
H\,$\beta$ line when the latter is unavailable.  This issue and the
results are discussed further in \S\,\ref{hahb}.  The widths of the
emission lines were determined by fits to the line profiles in the
regions near each of the lines.  To fit the H\,$\alpha$ and H\,$\beta$
lines, a featureless power-law continuum and Fe\,{\sc ii} template
were first fit to the {\em decomposed} AGN spectrum.  The H\,$\alpha$
and H\,$\beta$ lines were fitted separately.  We adopted a local power
law to fit the underlying continuum; the ostensibly line-free continuum
end points were selected as 4270{\AA} and 5600{\AA} for H\,$\beta$ lines,
and as 6270{\AA} and 6980{\AA} for H\,$\alpha$ lines.  The average fluxes
of the end points, averaged over the 21 pixel wide regions centered on
the end points, were used to fit the power law continuum.  Following
\citet{BG92}, an effective Fe\,{\sc ii} template can be generated by
simply broadening and scaling the Fe\,{\sc ii} spectrum derived from
observations of the narrow line Seyfert~1 galaxy I~Zw~1 (kindly provided
by T.~Boroson).  Although a newer template is available \citep{veron04},
we still use the template of \citet{BG92} because one of our aims is to
compare our results to those of previous studies, all of which used the
\citet{BG92} template.  The Fe\,{\sc ii} template strength is free to
vary, but it is broadened to be consistent with the FWHM of H\,$\beta$
for each object \citep{BG92,Shang05}. In order to avoid contamination
from strong spectral lines, we include only the following regions in the
Fe\,{\sc ii} fit: 4450-4750{\AA} and 5100-5350{\AA} \citep{Greene05}. The
best Fe\,{\sc ii} template is found by the minimization of the $\chi^2$
values in the fit region. The Fe\,{\sc ii} template is very weak and may
be negligible in the local region of the H\,$\alpha$ line.  Examples of
continua and Fe\,{\sc ii} template fits to {\em decomposed} AGN spectra
in our sample are shown in Figs.\,\ref{coniron}, \ref{coniron_845},
and \ref{coniron_303}.

\begin{figure}
  \plotone{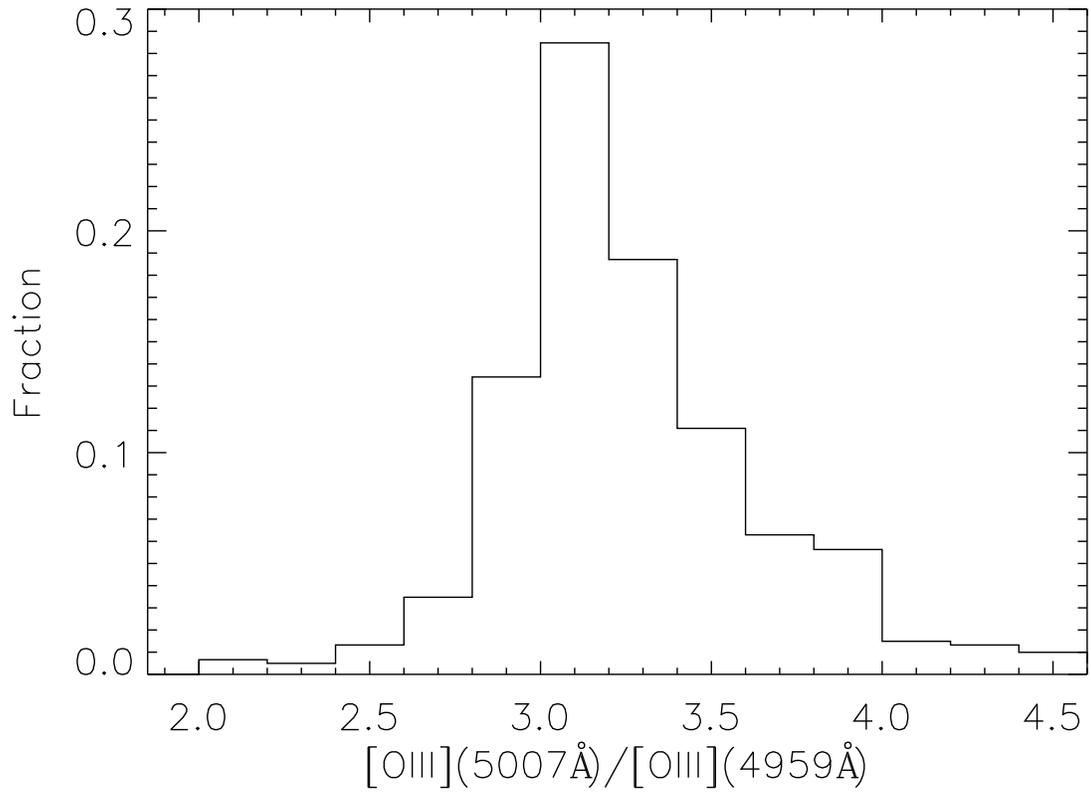}
  \caption{Distribution of the \ion{O}{3} doublet flux ratio
  (F(5007\AA)/F(4959\AA)).  The distribution shows that most
  \ion{O}{3} doublet ratios are close to the theoretical value of
  3:1. \label{oratio}}
\end{figure}

\begin{figure}
  \begin{center}
  \plottwo{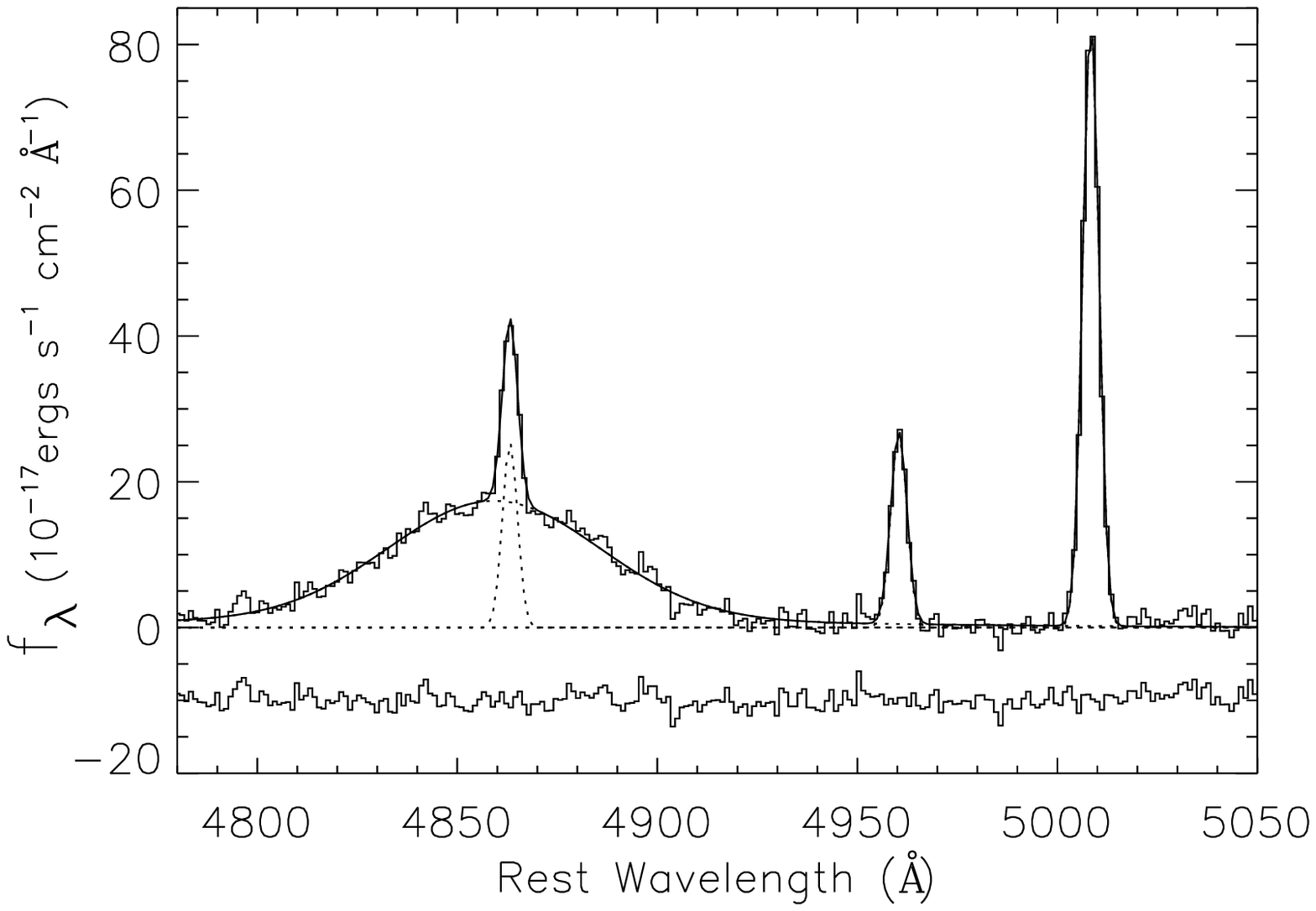}{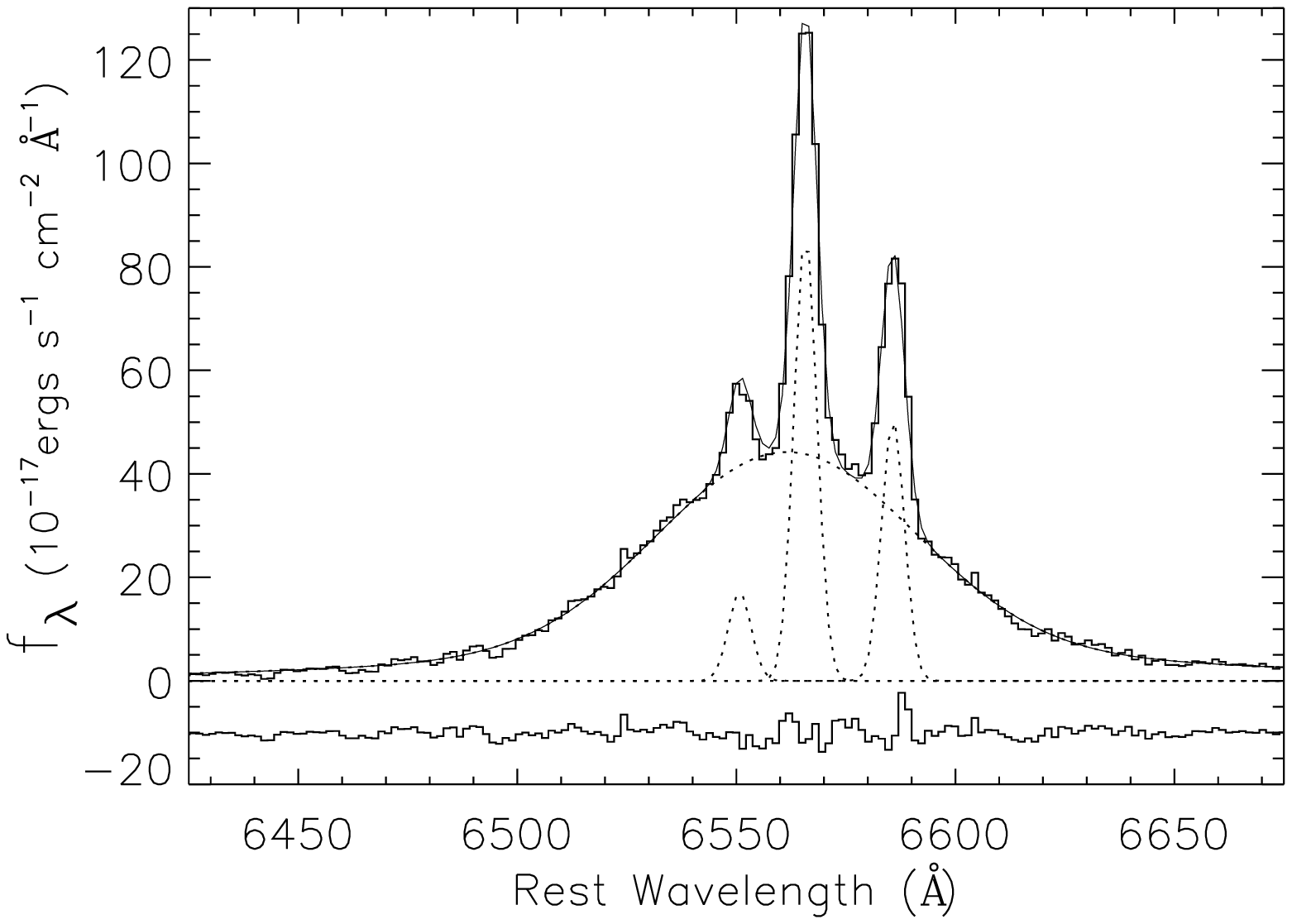}
  \end{center}
  \caption{Example line fit for the object described in Figs.~\ref{decom}
  and \ref{coniron}.  Left panel: Fits to the
  H$\beta$+\ion{O}{3}$\lambda\lambda$ 4959, 5007{\AA} region.  The data
  are shown by the histogram, the fitted broad and narrow components
  are shown as dotted lines, and the sum of all components is shown as a
  solid line.  The bottom histogram is the fit residual with an offset
  of -10 for clarity.  Right panel:  Same as the figure to the left,
  but for the H$\alpha$+[NII]$\lambda\lambda$ 6548,6583{\AA} \ complex.
  \label{fwhmfit}}
\end{figure}

After subtracting the continuum and iron template, the remaining broad and
narrow emission features near H\,$\beta$ and H\,$\alpha$ were fit with a
series of Gaussian profiles.  The broad component of the H\,$\beta$ line
was fit with two Gaussians, which are usually sufficient to account for
the width and possible asymmetry of the line \citep{Shang05}.  Each of
the [O\,{\sc iii}]$\lambda\lambda 4959, 5007${\AA} lines and the narrow
component of the H\,$\beta$ line were fit by single Gaussians.   The broad
component of the H\,$\alpha$ line was also fit by two Gaussians, and a
single Gaussian was used for each of the [N\,{\sc ii}]$\lambda\lambda
6548, 6583${\AA} lines and the narrow component of the H${\alpha}$ line.
The relative positions of the narrow H$\alpha$ and [N\,{\sc ii}] lines
were constrained by their laboratory values, and the relative ratio of the
two [N\,{\sc ii}] components was fixed to 2.96.  For the H\,$\beta$ line
fits, however, we did not set as many strict constraints as we did for
H\,$\alpha$. Instead, we allow a few {\AA} shift for each line position,
and let the peaks of the [\ion{O}{3}] doublet lines vary independently.
The H\,$\beta$ constraints are looser than those for H\,$\alpha$ because
H\,$\beta$ is weaker than H\,$\alpha$.  For example, if many restrictions
are placed on the fit to the [O\,{\sc iii}] doublets,  the fit to the
broad H\,$\beta$ line can become unreliable in some cases. Although the
peaks and fluxes of the [\ion{O}{3}] doublet lines were  allowed to vary
independently, the doublet flux ratios in our sample (Fig.\,\ref{oratio})
are close to the theoretical value of 3:1.  All of the Gaussian components
within either the H\,$\beta$ or H\,$\alpha$ set were fit simultaneously,
with the narrow-line dispersions all constrained to be smaller than
5\AA \,(308.6 km s${}^{-1}$).  The best-fit parameters were determined
through a minimization of the $\chi^2$ statistic, using the non-linear
Levenberg-Marquarrdt minimization algorithm as implemented by $mpfit$
in IDL\footnote{IDL is a trademark of Research Systems, Inc.}. Using
the fitted model profiles and excluding the narrow components, we
calculated the final FWHM of the H\,$\beta$ and H$\,\alpha$ lines
based on an analytic solution from the sum of the two fitted broad
Gaussian components.  The SDSS spectrograph wavelength dispersion (160
km s${}^{-1}$) does not increase the broad emission line widths by any
more than 1\% even in the worst  case.  Example fits to the H\,$\beta$
and H$\,\alpha$ line regions are shown in Fig.\,\ref{fwhmfit}, which
shows the {\em decomposed} AGN spectrum, the fit, the narrow and broad
components of the fit, and the fit residuals.

To estimate the errors in the line width measurements, we used a
method similar to that of \citet{Greene05}.  An artificial spectrum was
generated for each object by combining the best-fit model emission lines
with Gaussian random errors according to the object's error array. Then
the artificial spectra were fit using our fitting procedure, as outlined
above.  Each spectrum was simulated by 100 realizations (a larger number
of realizations did not significantly improve the error estimation),
and the estimated error was simply taken to be the dispersion in the
100 measurements.  The error measured in this way is the formal error,
which accounts only for the random noise error in the FWHM measurement.
Additional uncertainties come from the uncertainty of the fitted local
continuum, and imperfect removal of the Fe\,{\sc ii} components. Assuming
a single Gaussian profile, if the continuum is lowered by 10\% of the
line peak, the estimated FWHM will increase by 7\%.  If we do not remove
the Fe\,{\sc ii} components at all, the typical measured FWHM changes
by about 2\%.

%
\subsection{The Relationship Between the FWHM of H$\alpha$ and
  H$\beta$ \label{hahb}}
Inspection of the H\,$\beta$ region of the spectra showed that some of
the broad H\,$\beta$ lines are either missing (e.g. due to problems
with that part of spectrum) or are not detectable in the spectra.
In contrast, when H\,$\alpha$ is in the observable spectral range
($z<0.36$), the broad component is almost always easily detectable.
The H\,$\alpha$ line can be used to estimate the width of the H\,$\beta$
line in cases where the H\,$\beta$ line cannot be measured.  We used the
H\,$\beta$ line to estimate $M_{\rm BH}$ when the $S/N$ of the peak of
its broad component in the {\em decomposed} AGN spectrum is at least 3.
That criterion is satisfied in 640 of the 960 objects in the full sample.
Among the 960 objects, the H\,$\alpha$ line is not measurable due to bad
pixels in 11 cases and because it is redshifted outside of the observed
wavelength range in 17 cases. In 38 cases, the FWHM of the H\,$\alpha$
line is smaller than 25{\AA} ($1143\,{\rm km\,s^{-1}}$), so that the
narrow and broad components could not be unambiguously separated; these
spectra were excluded from further H\,$\alpha$ analysis.  In eight cases,
the profiles of the H\,$\alpha$ and H\,$\beta$ lines possess double
emission profiles and extremely broad wings, making the meaning of FWHM
ambiguous in these cases; these spectra were also excluded from further
analysis.  In total, there are 886 spectra for which the H\,$\alpha$
line width was reliably measured.

\begin{figure}
  \plotone{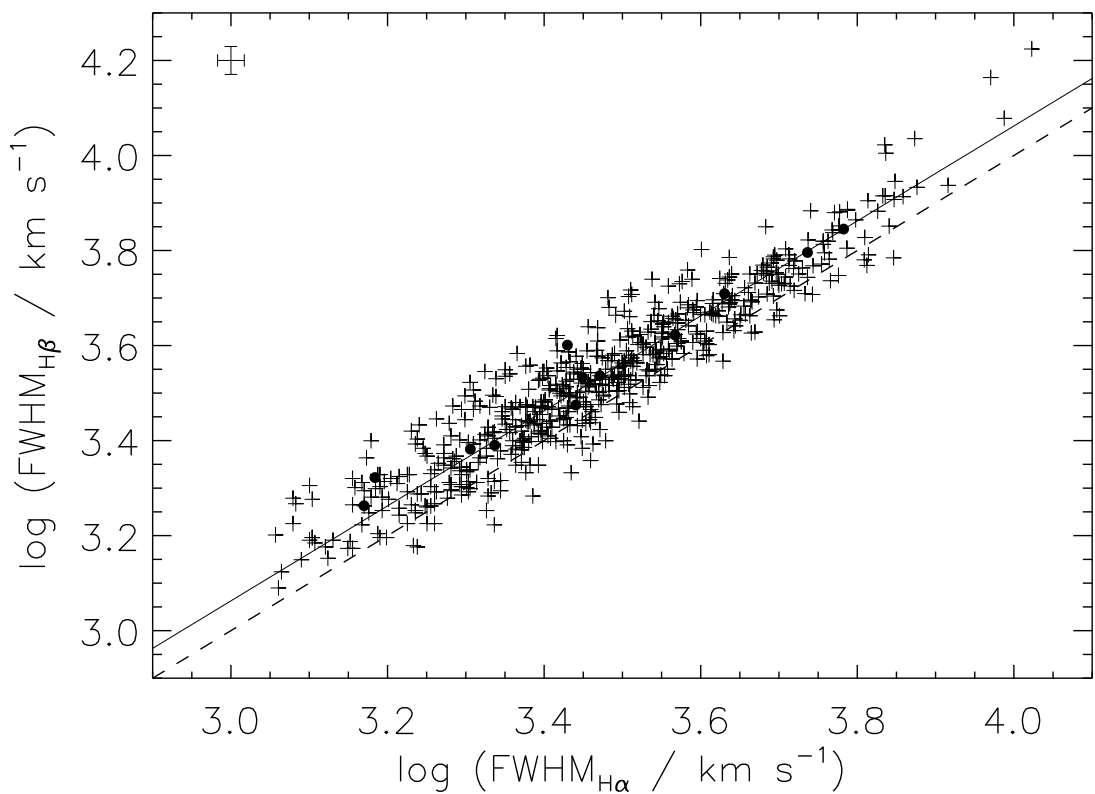}
  \caption{Correlation between H$\alpha$ and H$\beta$ line widths.  The
  solid line gives the BCES bisector fit to 597 objects.  The dashed line
  denotes FWHM${}_{H\alpha}$=FWHM${}_{H\beta}$.  The filled dots are the
  data from \citet{K00}, which are the mean FWHM values.  A typical FWHM
  error bar for an individual measurement, determined by the mean of all
  the formal errors of FWHM measurements, is shown in the top left part
  of the panel. \label{fwhmab}}
\end{figure}

Among the 640 good H\,$\beta$ and 886 good H\,$\alpha$ spectra, there
are 597 spectra for which the widths of both lines can be measured. The
FWHM of H\,$\alpha$ and H\,$\beta$ for these spectra are plotted
in Fig.\,\ref{fwhmab}; it is clear that there is a relatively tight
relationship between the two line widths, FWHM${}_{{\rm H}\,\alpha}$
and FWHM${}_{{\rm H}\,\beta}$.  When fitting the relationship, neither
FWHM${}_{{\rm H}\,\alpha}$ nor FWHM${}_{{\rm H}\,\beta}$ can be regarded
as the independent variable, so we calculate the best fit line to the
logarithmic quantities using the ``BCES'' (Bivariate Correlated Error
and intrinsic Scatter) bisector estimator described by \citet{AB96},
which takes uncertainties in both parameters into account. The best fit
relationship using the BCES bisector estimator is
\begin{equation}
 {{\frac{{\rm FWHM}_{{\rm H}\,\beta}}{10^3 {\rm \,km\,s^{-1}}}}}
  =(1.16 \pm 0.02)\times
 {\left({\frac{{\rm FWHM}_{{\rm H}\,\alpha}}
 {10^3 {\rm \,km\,s^{-1}}}}\right)}^{0.99 \pm 0.02}{\rm \,km\,s^{-1}}
  \label{eqfwhmab}
\end{equation}
This result is similar to that found by \citet{Greene05}, and the scatter
of equation\,(\ref{eqfwhmab}) is about 0.06dex ($\sim$ 470 km s${}^{-1}$).
In Fig.\,\ref{fwhmab}, the solid line shows the best fit, and the dashed
line denotes FWHM${}_{{\rm H}\,\alpha}$= FWHM${}_{{\rm H}\,\beta}$. The
filled dots are the mean spectrum FWHM from \citet{K00} for comparison;
those points are also described well by the best fit relation.

%
\subsection{Uncertainty in BH Mass Estimates \label{BHuncertain}}
The uncertainties of the FWHM measurements for H\,$\alpha$ and H\,$\beta$
and of the continuum luminosity include not only the errors from the
fitting routine, but also the errors that arise in the decomposition of
the AGN spectrum from the original active galaxy spectrum.  We estimated
the total uncertainty by using simulations in the following way.  For a
typical object with the median $S/N$ of the sample, we simulated 100
spectra by adding random noise to the original active galaxy spectrum
according to its error array.  Each of the 100 simulated spectra was
decomposed, and the continuum luminosity and FWHM of the emission lines
were fit as described above.

The uncertainty in each parameter was calculated by measuring the rms
dispersion of the 100 measurements.  The typical rms dispersions for
the FWHM measurement of H\,$\alpha$ and H\,$\beta$ are 4$\%$ and 7$\%$
respectively; the luminosity at 5100{\AA} has a typical dispersion of
4$\%$.  The final uncertainty for the BH mass by equation (\ref{Mbh}) is
7$\%$ and 14$\%$ respectively, using the FWHM of H$\alpha$ and H$\beta$.
A representative value of the uncertainty in the BH mass for an individual
object is the zero point uncertainty of the mass scaling relationship,
e.g. equation\,(\ref{Mbh}).  The standard deviation of the zero point
(intrinsic scatter) is $\pm 0.43$dex \citep{Vestergaard06}.

\begin{figure}
  \plotone{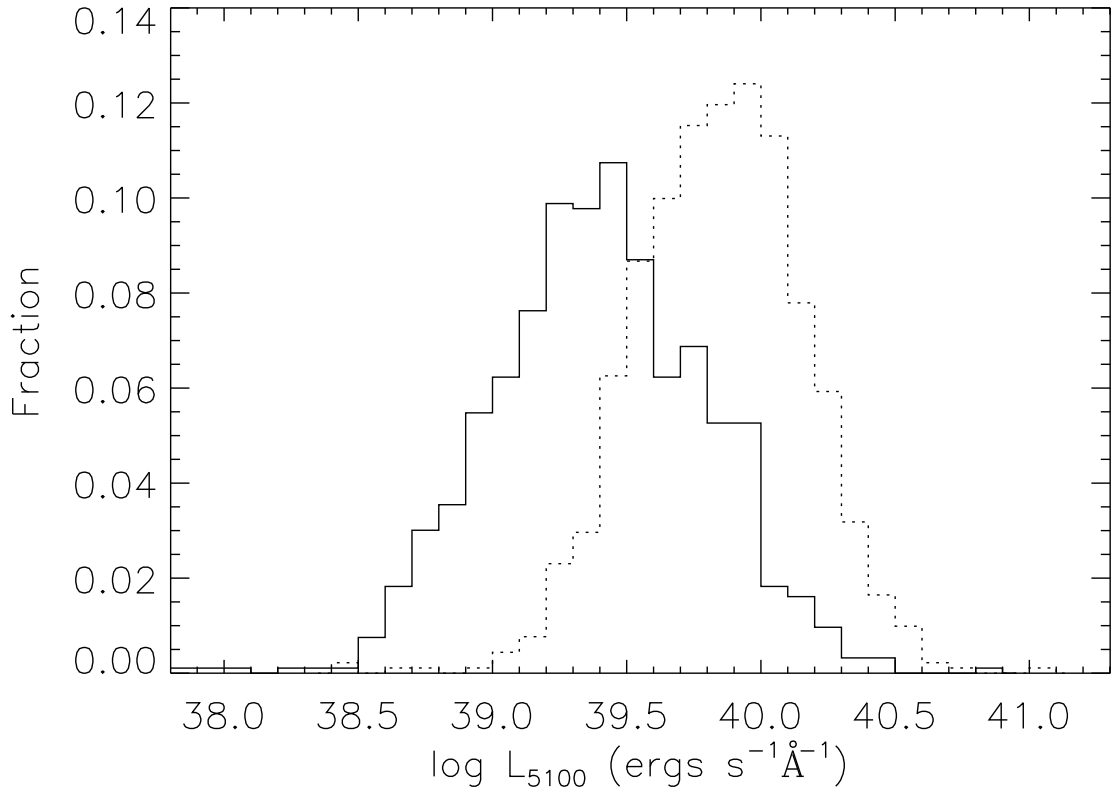}
  \caption{Distribution of AGN monochromatic luminosity at 5100\AA,
  $L_{5100}$.  The solid line shows the luminosity distribution after
  subtraction of the host galaxy component, and the dotted line shows
  the distribution without subtraction of the host galaxy. \label{L5100}}
\end{figure}

Figure\,\ref{L5100} shows the distribution of the AGN monochromatic
continuum luminosity, $L_{5100}$.  The solid line is the distribution of
$L_{5100}$ after subtracting the host galaxy component.  The dotted line
shows the estimated $L_{5100}$ values when the host galaxy component is
not subtracted; these values were estimated directly from the extinction
corrected active galaxy spectrum at rest-frame 5100{\AA}, averaged over
a 10-pixel wide region.  It is clear that the host galaxy component is
non-negligible in our sample; the difference between the luminosity with
and without host galaxy subtraction is close to 0.5 dex.  This implies
that the host galaxy components can significantly affect the accuracy
of the BH mass estimates at the luminosities present in our sample.
The host galaxy contribution will also affect the luminosity-radius
relationship, as discussed by \citet{Bentz06}.

%
\subsection{BH Mass Distribution \label{bhdist}}
\begin{figure}
  \plotone{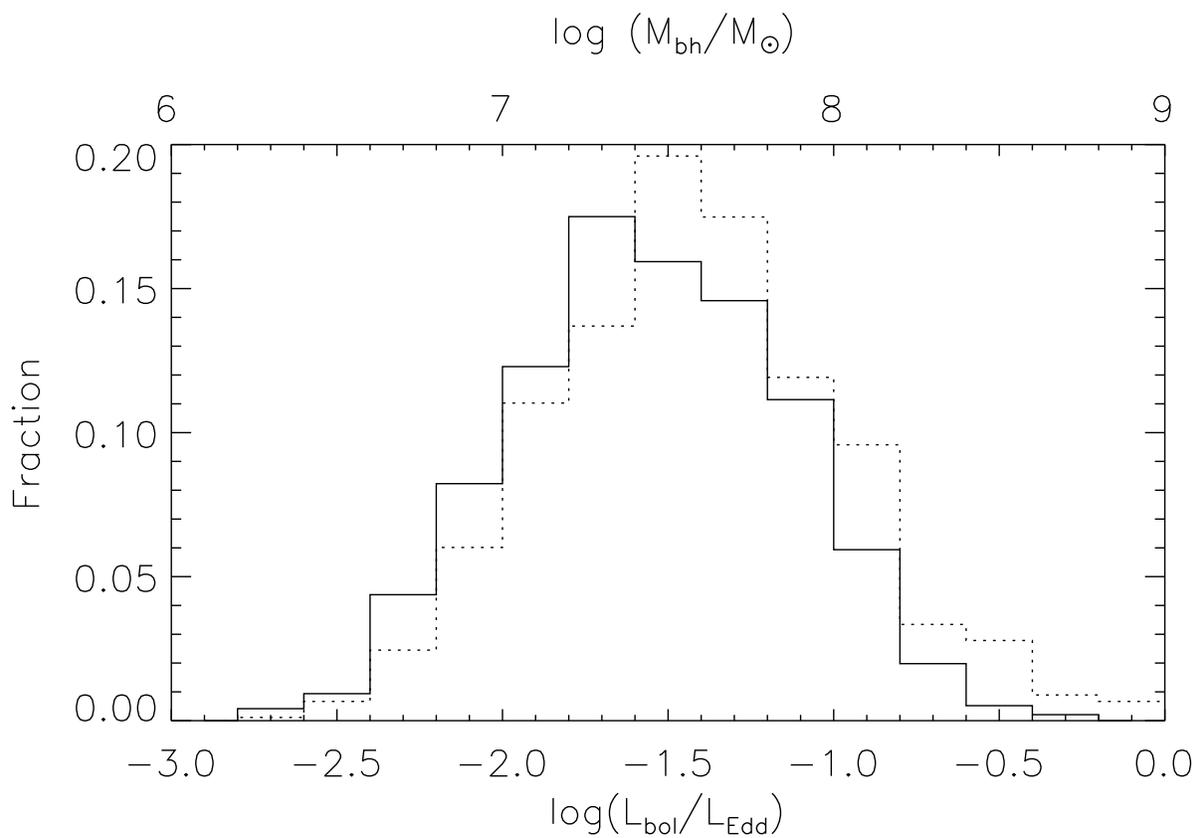}
  \caption{The distributions of the Eddington ratio,
  $\log (L_{\rm bol}/L_{\rm Edd})$ (solid histogram, units labeled
  at bottom of plot) and the BH mass, $M_{\rm BH}/M_{\odot}$ (dotted
  histogram, units labeled at top of plot). The AGN bolometric luminosity,
  $L_{\rm bol}$, was estimated according to the approximate relation
  $L_{\rm bol}\approx10\lambda L_{\lambda}$(5100\AA). \label{Ledd}}
\end{figure}

The black hole masses were calculated according to
equation\,(\ref{Mbh}). The FWHM of the H\,$\beta$ line was used directly
in the 640 cases for which the line could be measured.  For the 286 cases
in which the H\,$\alpha$ line was measurable when H\,$\beta$ was not,
the H\,$\alpha$ FWHM was used to estimate the FWHM of the H\,$\beta$
line according to equation\,(\ref{eqfwhmab}).  The distribution of
black hole masses is shown by the dotted histogram in Fig.\,\ref{Ledd}.
Most of the black hole masses are in the range $10^{6} < M_{\rm BH}
< 10^{9} M_{\Sun}$. From the black hole masses and the continuum
luminosities, we can estimate the Eddington ratio, $L_{\rm bol}/L_{\rm
Edd}$ --- the ratio of the AGN bolometric luminosity to the Eddington
luminosity for the black hole --- using the approximate relation $L_{\rm
bol}\approx10\lambda L_{\lambda}$(5100\AA) \citep[e.g.,][]{Wandel99}.
The estimated Eddington ratio has a standard deviation of 25$\%$ due to
variations in the bolometric correction \citep{richards06}.  Since $L_{\rm
Edd}$ is proportional to $M_{\rm BH}$, the Eddington ratio would have
a standard deviation of $\approx28\%$, assuming the typical error in
our BH mass estimates (\S\,\ref{BHuncertain}).  The solid histogram in
Fig.\,\ref{Ledd} shows the distribution of $L_{\rm bol}/L_{\rm Edd}$; the
Eddington ratios also fall within a limited range, mostly at small values.
The small Eddington ratios and limited range of BH masses are due to
the sample selection criteria, which require that the stellar absorption
features are detectable at a level that allows the host galaxy velocity
dispersions to be measured.

\begin{figure}
  \plotone{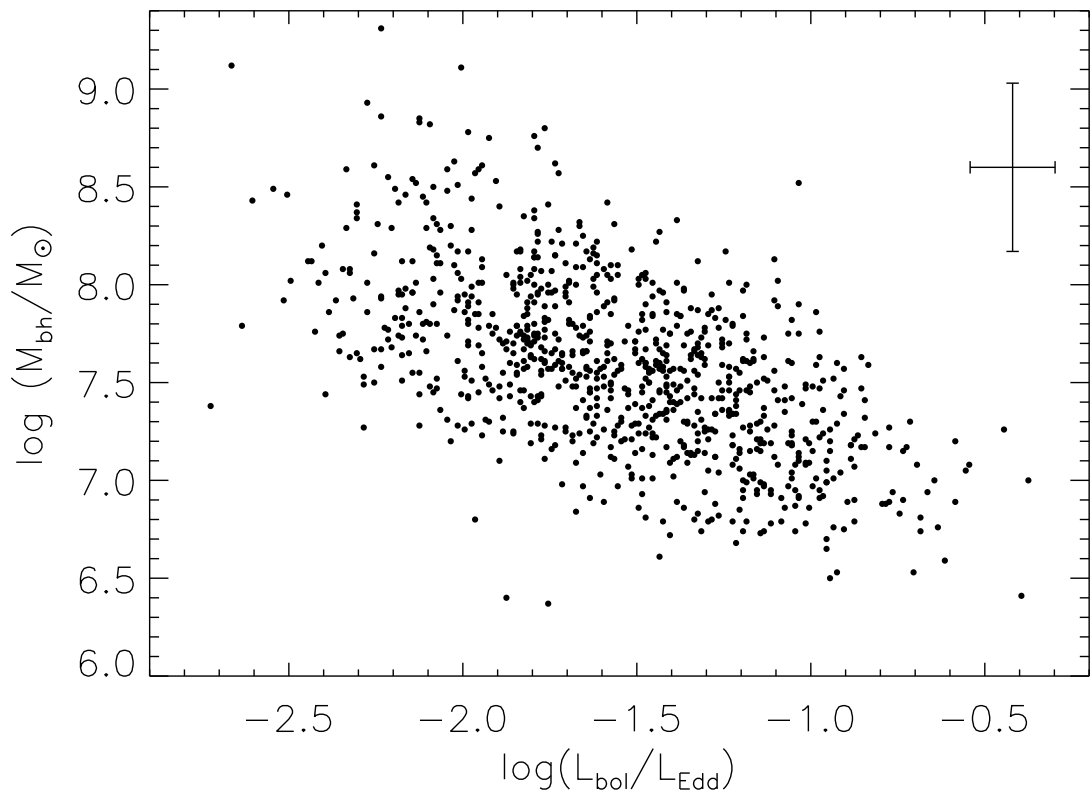}
  \caption{The distribution of the BH masses with the estimated Eddington
  ratio $L_{\rm bol}/L_{\rm Edd}$.  The apparent anti-correlation is
  due to parameter interdependencies, and sample selection effects
  (see \S\,\ref{bhdist}).  The right top horizontal line shows
  the typical uncertainty of the estimated Eddington ratio $L_{\rm
  bol}/L_{\rm Edd}$, due to the $25\%$ standard deviation in the
  bolometric correction \citep{richards06} and the BH mass estimation
  error.  The right top vertical line shows the statistical BH masses
  uncertainties. \label{MbhLedd}}
\end{figure}

Fig.\,\ref{MbhLedd} shows the relationship between BH mass and the
estimated Eddington ratio $L_{\rm bol}/L_{\rm Edd}$.  In our data
sample the BH mass is inversely correlated with the Eddington ratio.
This is not unexpected if the range of bolometric luminosities does
not change proportionally with $M_{\rm BH}$.  Both $M_{\rm BH}$
and $L_{\rm bol}/L_{\rm Edd}$ are parameters calculated from AGN
continuum luminosity $L_{AGN} \approx \lambda L_{\lambda}$(5100 \AA),
and  emission line FWHM.  Therefore, Fig.\,\ref{MbhLedd} is simply a
remapping of the sample $L_{AGN}$ and FWHM values.  The correlation
is due in part to sample selection effects in the SDSS and our sample.
Since our sample is selected only from SDSS AGN with detectable stellar
features ($0.2<F_H<0.8$), the luminosity cannot be above the SDSS bright
limit for spectroscopy ($i=15$).  Our sample is neither so bright that
the flux is dominated by nuclear light, nor so faint that the flux is
dominated by host galaxy light.  Thus, objects with higher BH  masses
tend to have a lower range of Eddington ratios, and vice versa. The
dispersion in the relationship is largely due to the range of measured
emission line widths that are used to calculate $M_{\rm BH}$.

%
\subsection{Host Galaxy Velocity Dispersion Measurement \label{sigma}}
Two accurate and objective methods, the {\it Fourier-fitting}
method and the {\it direct-fitting} method, have been
developed for measuring the stellar velocity dispersion
\citep{S77,TD79,F89,B90,RW92,Barth02,Greene06b}.  They are all based on
a comparison between broadened template spectra and the spectrum of the
galaxy whose velocity dispersion is to be determined.  Fourier space
is the natural choice to estimate the velocity dispersions because
a galaxy's spectrum is a mix of stellar spectra convolved with the
distribution of velocities within the galaxy.  However, there are several
advantages to treating the problem entirely in pixel space, rather
than Fourier space.  Specifically, the effects of noise are much more
easily incorporated in the pixel-space-based direct-fitting method. We
have used the ``direct-fitting'' method \citep[e.g.][]{RW92,B03a}, in
which the spectrum is directly fit in pixel space.  We use the public
IDL program $vdispfit$, written by David Schlegel, to find the velocity
dispersions for the host galaxies.  The templates consist of the first
four eigenspectra from a principal component analysis of the echelle
stellar spectra in the Elodie database \citep{MJ04}.  The best-fitting
dispersion value was determined by minimizing $\chi^2$ for the fit.

The {\em reconstructed} host galaxy spectrum was not used to measure
the velocity dispersion, because features such as narrow absorption
line profiles may not be well reconstructed without invoking a much
larger number of galaxy eigenspectra than we have used here.  Instead,
we used the observed active galaxy spectrum after subtracting the {\em
reconstructed} AGN component, which we define as the {\em decomposed}
host galaxy spectrum. This method allows us to use the error array of
the original spectrum to determine measurement uncertainties.  In the
fitting routine, all of the narrow {\em emission} lines were blocked,
because there is very little information about the absorption spectrum
in those regions, and the AGN fit to the narrow emission lines may be
poor.  We also blocked the broad emission line regions of H\,$\alpha$
and H\,$\beta$ to avoid any artifacts introduced by imperfectly {\em
reconstructed} AGN components.

\begin{figure}
  \plotone{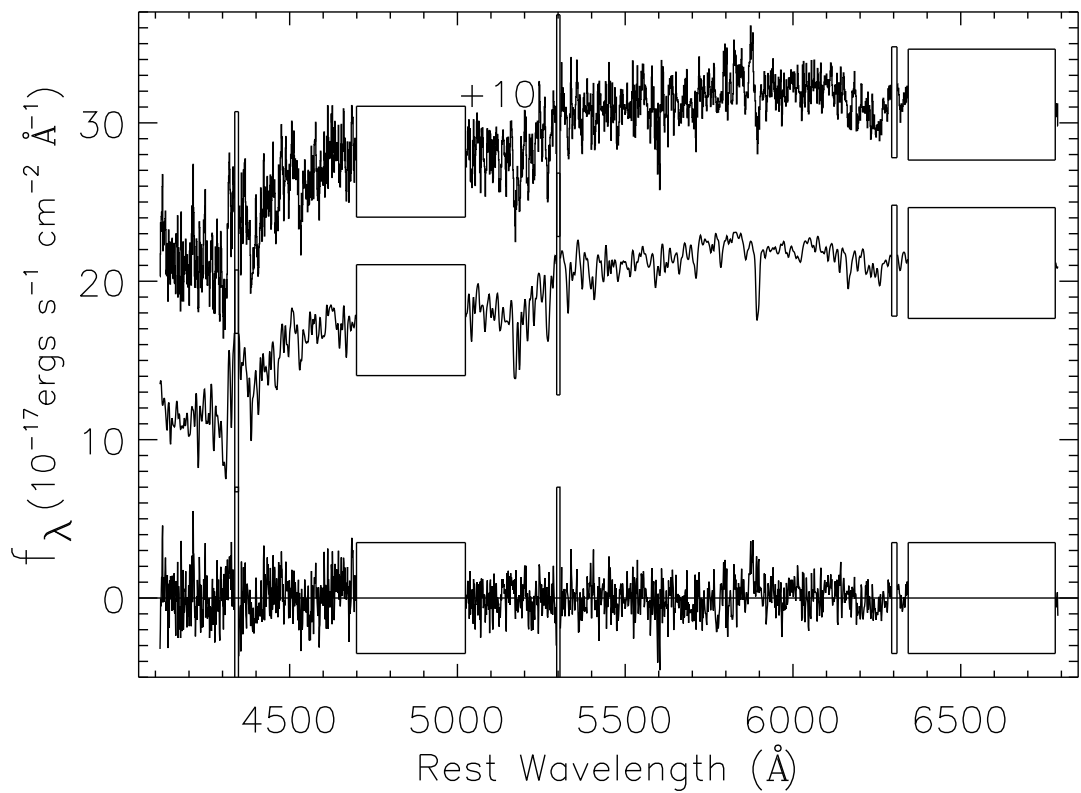}
  \caption{An example host galaxy fit to measure the velocity dispersion.
  The top line is the {\em decomposed} host galaxy spectrum with an offset
  10 for clarity, the middle solid line is the fit, and the bottom solid
  line is the fit residual.  The boxes show the masked regions. The two
  large boxes are masks for H$\,\beta$ and H$\,\alpha$, the left and
  right narrow boxes are masks for H$\,\gamma$ and [O{\sc i}], and the
  middle box is a mask for bad pixels.  \label{vmask}}
\end{figure}

An example fit of a host galaxy (from among the same objects used
as examples in Fig.\,\ref{decom}), resulting in a measurement of the
velocity dispersion, is shown in Fig.\,\ref{vmask}.  In the figure,
the two large boxes show the masked broad H\,$\alpha$ and H\,$\beta$
line regions; the left and the right small boxes show the masked narrow
emission line regions (H\,$\gamma$ and [O\,{\sc i}]), and the middle
small box shows the masked bad pixel region.

The SDSS spectra were obtained by using a fixed $3\arcsec$ fiber,
which covers different fractions of the projected areas of different
objects, due both to different intrinsic galaxy sizes and different
redshifts. Thus, the velocity dispersions should not be compared without
accounting for the finite fiber diameter, which is discussed here. The
dispersion is scaled to a standard relative circular aperture, defined
to be one-eighth of the effective galaxy radius.  The correction formula
was applied following \citet{J95} and \citet{Wegner99}
\begin{equation}
  \frac{\sigma_{\rm cor}}{\sigma_{\rm est}}
   ={\Bigl(\frac{r_{\rm fiber}}{{r_0}/8}\Bigr)}^{0.04}\,,
  \label{corr}
\end{equation}
where $r_{\rm fiber} = 1\farcs5$ and $r_0$ is the effective radius
of the galaxy measured in arcseconds. \citet{B03a} studied early type
galaxies up to a redshift of $z=0.3$ and showed that most galaxies have
$r_0 \ge 1\farcs5$.  Our sample has almost the same redshift limit as
that of \citet{B03a} and the correction depends only weakly on $r_0$,
so the correction is not large.  The effective radii of the galaxies
in our sample have not yet been measured, so we assume that all of the
early type galaxies (or the galaxy bulges) in our sample have typical
effective radii, which corresponds to an angular size of $r_0=1\farcs5$
at a redshift of $z=0.3$.  At redshifts other than $z=0.3$, the angular
sizes were scaled with distance.  The correction has a maximum value of
10.8$\%$ and a median value of 5.4$\%$.  For a more accurate correction,
the effective radii of the hosts must be measured.  In the case of
inactive galaxies, estimations of effective radii are publicly available
parameters measured from SDSS images.  For active galaxies, one should
first remove the central nucleus to get the host galaxy effective radius,
which requires image decomposition to separate the nucleus and host.
We have not pursued this additional correction here, owing to the limited
improvements possible with SDSS imaging data in relation to the required
workload.

%
\subsection{Uncertainty in the Velocity Dispersion Measurement}
It was shown by \citet{B03a} that the direct-fitting method does not
produce large systematic errors, but the same conclusion may not apply
to our measurements given the presence of AGN components.  Even though
a measured velocity dispersion may be accurate for a {\em decomposed}
host galaxy spectrum, it may be different from the true value if the
{\em decomposed} host galaxy is not an accurate representation of the
true host galaxy. A test for systematic errors in the velocity dispersion
measurement was made by generating simulated data.  We first constructed
an early type galaxy template using the stellar echelle spectra in
the Elodie database. The template early type galaxy has a high $S/N$
and a very small velocity dispersion.  The template was broadened
with different values of the velocity dispersion, ranging from 60 to
$400\,{\rm km\,s^{-1}}$, and rebinned to the same wavelength scale as the
SDSS spectra ($69\,{\rm km\,s^{-1}}$/pixel).  To each of the broadened
galaxy templates, a template AGN spectrum was added to form simulated
active galaxy spectra. In all cases, the fractional contribution of the
host galaxy was set to $F_H=50\%$.

\begin{figure}
  \plotone{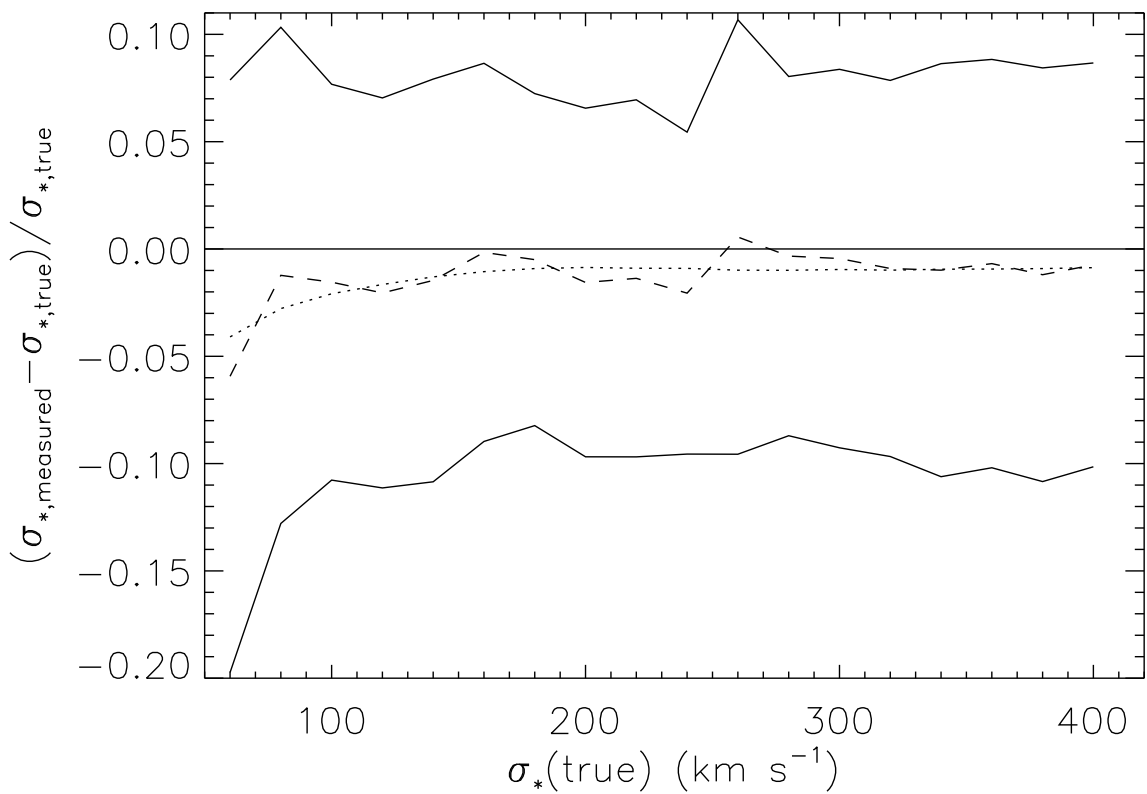}
  \caption{The relative systematic error and random error in the measurement
  of velocity dispersion, found from simulations.  The dotted line
  shows the relative error between the true velocity and the measured
  noiseless velocity.  The dashed line shows the relative error between
  the true velocity and the mean velocity in 100 simulations with noise
  added, for the case of spectroscopic $S/N=15$, and host galaxy flux
  fraction $F_{H}=50\%$.  The solid curves show the standard deviations
  of the relative velocity measurements in 100 simulations.  \label{sys2}}
\end{figure}

The simulated active galaxy spectra were decomposed into host galaxy
and AGN components, and the direct-fitting method was used to make
measurements of the velocity dispersions.  No noise was added to the
spectra at this point, so any deviations of the measured velocity
dispersions from the true values represent systematic errors in the
measurements.  Figure\,\ref{sys2} (dotted line) shows the relative
difference between the true velocity dispersions and the measured values,
$\delta \sigma_*=(\sigma_{*,measured}-\sigma_{*,true})/\sigma_{*,true}$.
The measured velocity dispersion is systematically approximately $1\%$
smaller than the true velocity dispersion.  This systematic error
steadily increases to about $4\%$ below velocity dispersions of about
$120\,{\rm km\,s^{-1}}$.

To test the effects of noise, we added random noise to the template
active galaxy spectra.  In our sample, the lowest spectroscopic $S/N$
is 15, which sets the worst case limit in the simulations (of course,
objects with a host galaxy fraction of less than the simulation value of
$F_H=50\%$ will usually return poorer results; see below).  Random noise
corresponding to a $S/N$ of 15 was added to each template 100 times.
The noisy spectra were decomposed and the velocity dispersions were
measured as usual. The dashed line in Fig.\,\ref{sys2} shows the relative
error between the true velocity dispersion and the mean measurement in
each set of 100 simulated spectra.  The solid curves in Fig.\,\ref{sys2}
show the rms scatter of the velocity measurements from the mean, due to
the random error in 100 simulations.  Fig.\,\ref{sys2} shows that the
random error of the velocity dispersion measurement is around 10$\%$
for the spectra with the lowest $S/N$ in our sample.  The mean relative
errors track the noiseless relative errors very closely, showing again
an approximately $1\%$ systematic offset.  The systematic relative error
is larger at smaller velocity dispersions, reaching a maximum absolute
value of about $4\%$ at $60\,{\rm km\,s^{-1}}$, which is close to the
dispersion per pixel of the SDSS spectrographs (69 km s${}^{-1}$). Values
of the velocity dispersions smaller than this limit are very uncertain.

\begin{figure}
  \plotone{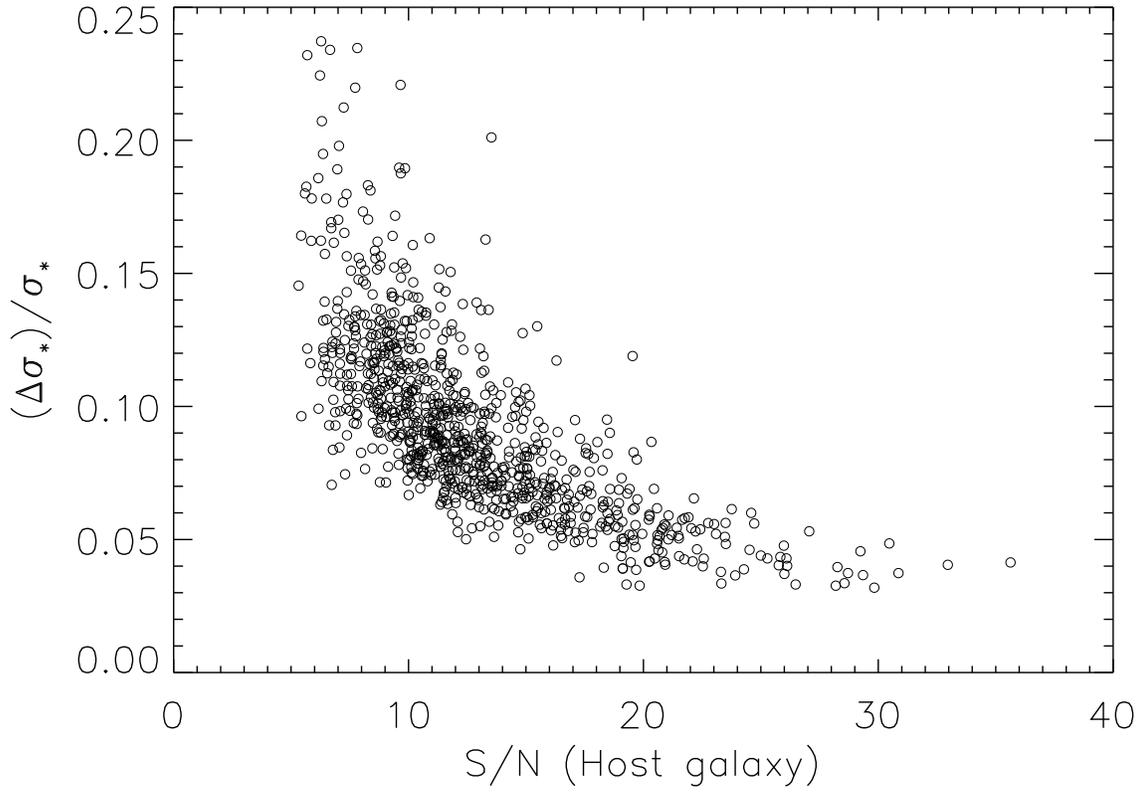}
  \caption{The measurement error of the velocity dispersion in the host
  galaxy sample, returned from the $vdispfit$ routine, as a function of
  host galaxy spectroscopic $S/N$.} \label{verr}
\end{figure}

\begin{figure}
  \plotone{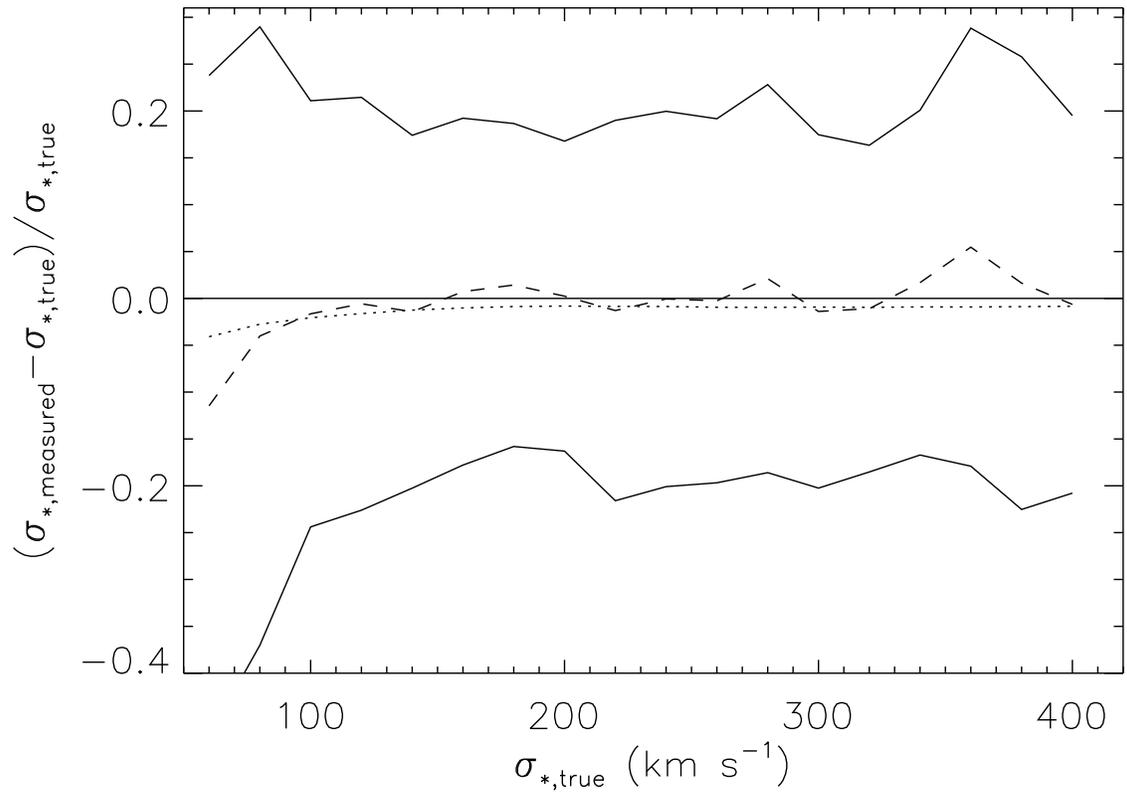}
  \caption{Same as Fig.\,\ref{sys2}, but for the worst case
  of spectroscopic $S/N=15$, and host galaxy flux fraction
  $F_{H}=20\%$. \label{sys1}}
\end{figure}

The $vdispfit$ routine returns an estimate of the uncertainty in the
velocity dispersion, based on the spectral noise, which we find to be
consistent with the results of the simulations. Figure\,\ref{verr}
shows the estimated uncertainty in the measurement of the velocity
dispersion, returned from the $vdispfit$ routine, as a function of
the {\em decomposed} host galaxy $S/N$.  As expected, the errors
are generally smaller with increasing $S/N$.  Some of the errors are
greater than the worst case results from the simulations, because their
host galaxy fractions are less than $F_H=50\%$. Figure\,\ref{sys1}
is the simulation of the the worst case ($F_H=20\%$ and $S/N=15$),
which shows that the maximum error can be 30$\%$ for small velocity
dispersions.  Error may arise if the templates are not good matches
to the host galaxy spectra. For example, the templates may not be
able to account for a strong post-starburst component.  However, all
of our host galaxies are chosen so as to be dominated by old stars.
In addition, the simulations did not account for bad pixels, night sky
line residuals, and other artifacts.  The reduced $\chi ^2$ values in
the velocity dispersion fittings are close to 1, which proves that any
template mismatch is negligible in our sample. The vast majority of the
objects have velocity dispersion errors of less than $15\%$.

%
\subsection{The AGN and Host Galaxy Data Set}
Spectra whose velocity dispersions are less than $69\,{\rm km\,s^{-1}}$
were excluded from further analysis.  After all of the selection criteria
were applied, there were 617 objects with reliable H\,$\beta$ and 840
with reliable H\,$\alpha$ line measurements, all of which also have
reliable velocity dispersion measurements.

Black hole masses were calculated using the H$\beta$ FWHM in the 617
cases for which it was measurable, and in another 286 cases, using an
estimate of the H$\beta$ FWHM based on the H\,$\alpha$ measurements
and equation\,(\ref{eqfwhmab}).  The total number of objects (out of
the initial 960) with reliable parameter measurements is 903. The
electronic version of Table~\ref{table1} lists all of the active
galaxies and their measured or derived parameters, including redshift,
$L_{5100}$, H\,$\alpha$ and H$\beta$ FWHM, $M_{\rm BH}$, $\sigma_*$,
and host luminosity (as calculated by VB06).  A partial table is shown
in the printed version of Table~\ref{table1}.

\section{Results\label{results}}
\subsection{The fitting algorithm}
In the following sections, we will study how the BH masses, host galaxy
luminosities and the host velocity dispersions are correlated.  We assume
that there is an underlying linear relation in the logarithmic quantities
for each pair in the form $y=\alpha +\beta x$.  The best fit parameters
$\alpha$ and $\beta$ were found by the minimization of $\chi ^2$, defined,
following \citet{T02}, as
\begin{equation}
\chi ^2 \equiv \sum _1^N \frac{(y_i-\alpha-\beta x_i)^2}
{\epsilon_{yi}^2+\beta^2\epsilon_{xi}^2}
\label{chi2}
\end{equation}
where $x_i$, $y_i$ correspond to the measurements for each object,
and $\epsilon_i$ is the formal uncertainty in the measurement.  The 1
$\sigma$ uncertainties in $\alpha$ and $\beta$ are given by the maximum
range of $\alpha$ and $\beta$ for which $\chi^2-\chi_{\rm min}^2 \le
1$.  To minimize $\chi ^2$, we use the Levenberg-Marquardt algorithm
as implemented by $mpfit$ in IDL, which recovers the results of the
Numerical Recipes routine $fitexy$ \citep{press92} as implemented in
IDL for the case of symmetric errors.

%
\subsection{Host Galaxy Bulge Luminosity and Velocity Dispersion
  Relation \label{faberjackson}}
The luminosities and velocity dispersions of early type galaxies have long
been known to be correlated --- a phenomenon known as the Faber-Jackson
relation \citep{FJ76}.  The host galaxy absolute magnitude in the
SDSS $g$ band, $M_{g}$ (calculated by VB06) is shown as a function
of velocity dispersion in Fig.\,\ref{vl}. The host galaxy $M_g$
range of our sample is very similar to that of the normal elliptical
galaxies studied by \citet{B03a}, except ours are a bit more luminous.
There is a wide intrinsic scatter in the Faber-Jackson relation, which
we also find in our sample, so for clarity we have binned the values
of the host galaxy luminosity in the figure (similar to the procedure
used by \citealt{B03b}). Each point represents the average value for
at least 50 objects.  The bars in the figure show the rms scatter in
the velocity dispersion in each bin (the errors on the mean values
would be approximately $\sqrt{50}$ times smaller).  There is clearly
a strong correlation between $M_{g}$ and $\sigma_*$, although we find
a slightly larger dispersion about the relation relative to studies
of inactive early type galaxies \citep[e.g.][]{B03b}, most likely due
to the typically smaller $S/N$ in our {\em decomposed} host spectra.
The solid line shows the best power-law fit to the relation
\begin{equation}
  L\propto \sigma_*^{4.33\pm0.21}\,,
\end{equation}
which is consistent with the Faber-Jackson relation for inactive early
type galaxies \citep{B03b}.  Table~\ref{table2} lists the mean values
of the host luminosity, stellar velocity dispersion, black hole mass,
and rms dispersions for the data sets in each bin.  The binned values
will be used in the analysis of the following subsections.

%
\subsection{Black Hole Mass and Bulge Luminosity Relation}
Figure \ref{ml} shows the relationship between BH mass and absolute $g$
band magnitude $M_{g}$ of the host galaxies.  Crosses show the mean
value in each bin, and error bars show the rms scatter around the mean
value. Each point represents the average value for about 50 objects.
A linear fit using the logarithmic $M_{\rm BH}$ values gives the relation
\begin{equation}
\log M_{\rm BH} = - 0.29 (\pm 0.02) M_g + 1.46 (\pm 0.33)\,.
\label{Eq_MbhMg}
\end{equation}

From the relation in Eq.\,\ref{Eq_MbhMg}, we can also derive the expected
relationship between $M_{\rm BH}$ and host bulge mass $M_{bulge}$ by
assuming that only the host bulge contributes to $M_{g}$.  We adopt a
$g$-band mass-luminosity relation $M_{bulge}\propto L_{bulge}^{1.18\pm
0.03}$, consistent with the $V$-band results of Magorrian et
al. (1998). Substituting in, we find a relation $M_{\rm BH}\propto
M_{bulge}^{0.61\pm 0.05}$.  This implies that the $M_{\rm BH} -M_{\rm
bulge}$ relationship is non-linear in our sample, and it is consistent
with the result found by \citet{wandel02}.  However, it is in contrast
to the linear relationship found by others \citep[e.g.][]{MD01,MD02}.
Verification of our result will require more precise measurements of the
bulge luminosity, since our measurements rely on spectral decomposition,
rather than image decomposition, and we have assumed that the host
component of the SDSS $g$ band ``cmodel'' magnitude contains only the
bulge flux.  The ``cmodel'' magnitude uses a best fit linear combination
of an exponential and de Vauc. profile, which gives a robust estimate
of the total flux from an extended object, regardless of morphology.
The bulge flux was estimated by subtracting an appropriately scaled
nuclear component flux, derived from the decomposed AGN spectrum, from the
``cmodel'' magnitude flux, as described by VB06.  A more precise bulge
luminosity could be obtained from image decomposition, which directly
separates out the host galaxy bulge and nuclear AGN components.  As a
complement to the spectral decomposition, image decomposition is planed
for a future paper.

%
\subsection{Black Hole Mass and Velocity Dispersion Relationship}
\begin{figure}
  \plotone{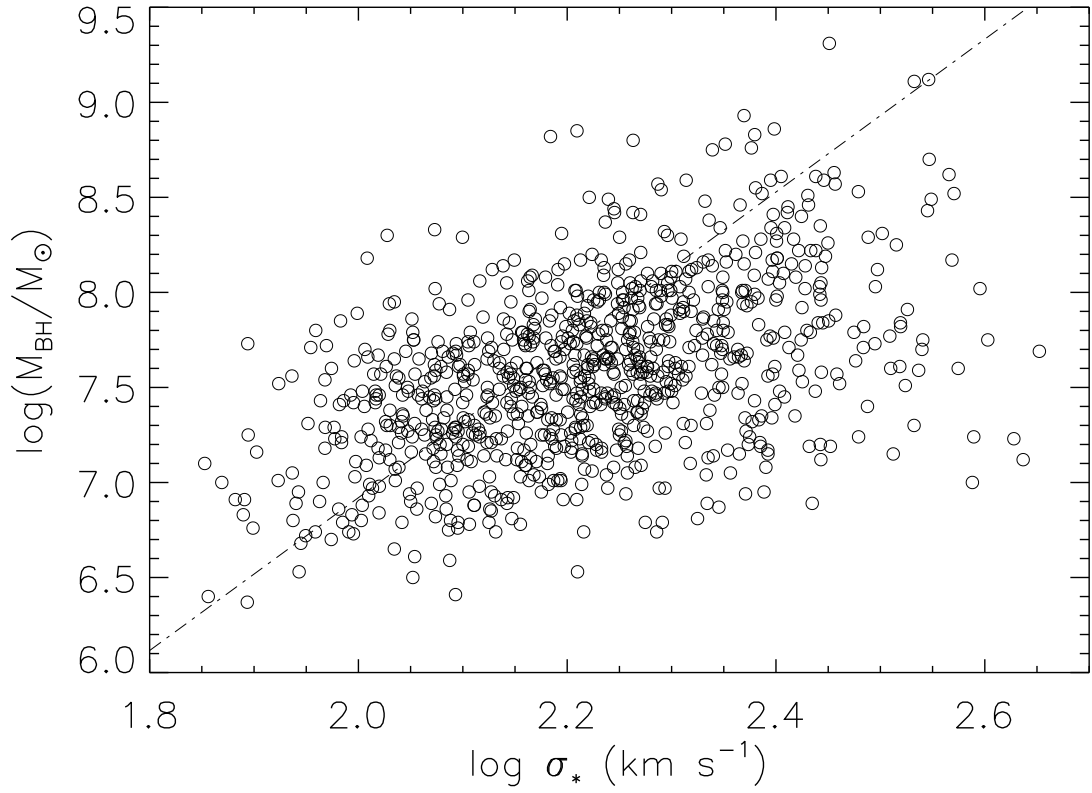}
  \caption{Black hole mass vs.\ stellar velocity dispersion for individual
  points.  The dot-dashed line shows the best fit for inactive galaxies
  \citep{T02}\label{vmall}}
\end{figure}

Figure \ref{vmall} shows the BH mass and $\sigma_*$ relation for all
the individual points, and the dot dashed line is the best fit to
the $M_{\rm BH}-\sigma_*$ relation for inactive galaxies \citep{T02}.
For $\log \sigma_* > 2.2$, objects in our sample have significantly
smaller average BH masses than do inactive galaxies at the same value
of $\sigma_*$.  The measured values of slope $\beta$ in the $M_{\rm BH}
\propto \sigma_*^{\beta}$ relation cover a wide range from 3.5 to 5.0
\citep{FM00,G00a}.  As studied by \citet{T02}, the wide uncertainties
in the slope are due to the use of different statistical algorithms,
the unknown intrinsic scatter in the BH mass estimation, and the
uncertainties in the velocity dispersions. Our large sample helps reduce
some of these uncertainties.

\begin{figure}
  \plotone{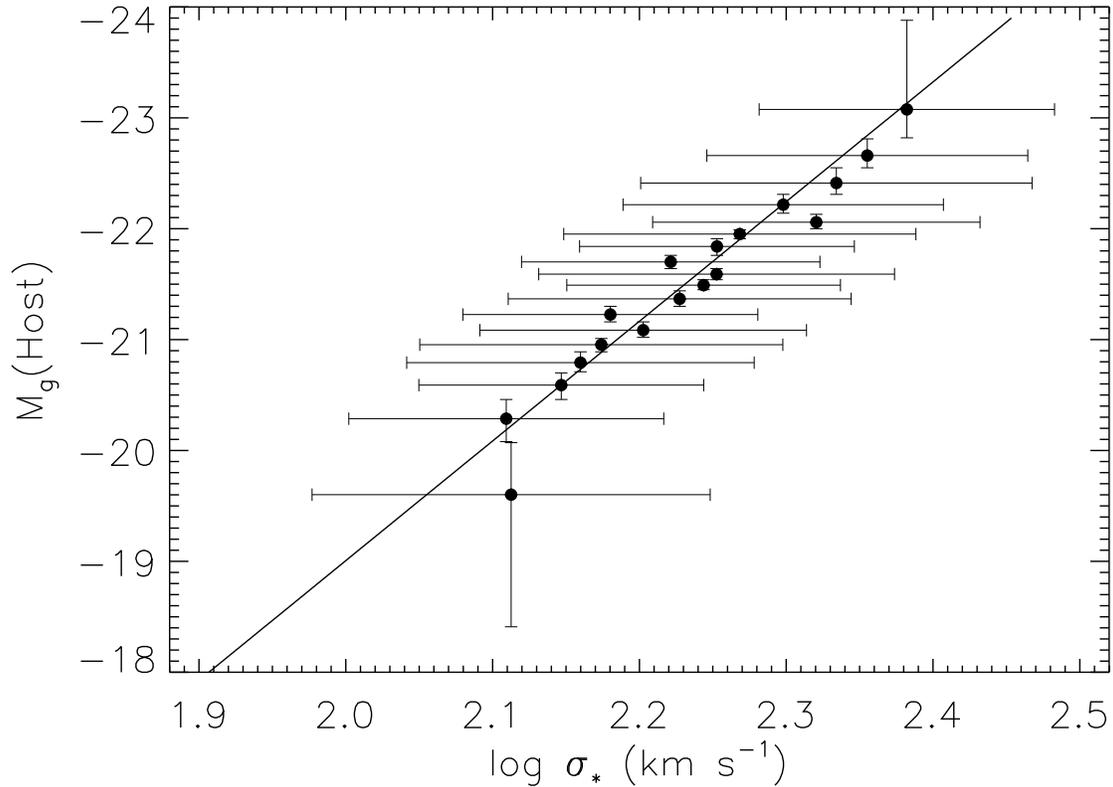}
  \caption{Relation between host galaxy $g$ band absolute magnitudes
  $M_g$, and stellar velocity dispersion $\sigma_*$.  Filled circles
  show the mean values of $\log \sigma_*$ in sets of objects that span
  small ranges in absolute magnitude; each set contains 50 host galaxies.
  Vertical error bars show the $M_g$ bin size, and horizontal error bars
  show the rms scatter around the mean values of $\sigma_*$.  The solid
  line shows the best fit linear relation (in logarithmic space).
  \label{vl}}
\end{figure}

\begin{figure}
  \plotone{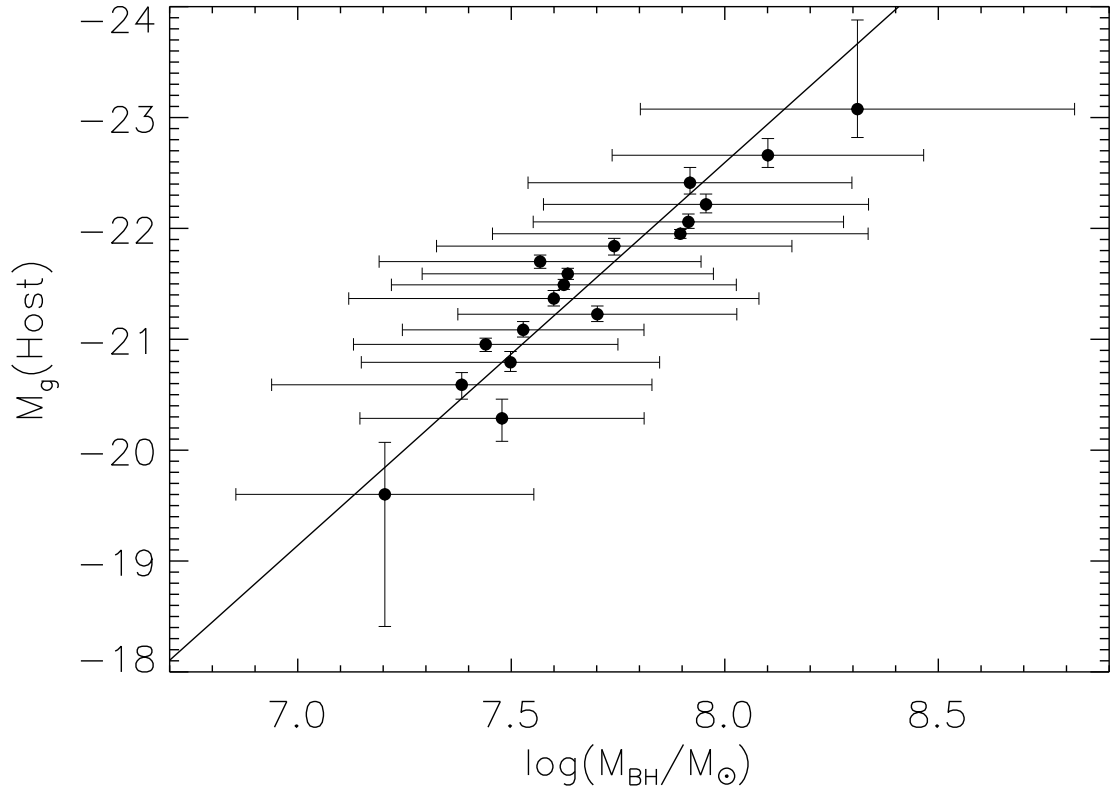}
  \caption{Host galaxy absolute g band magnitude vs.\ BH mass.  The data
  are binned by $M_{g}$ in the same way as in Fig.\,\ref{vl}.  Filled
  circles show the mean value of $M_{\rm BH}/M_{\odot}$ in each bin.
  The horizontal error bars show the rms scatter around the mean
  BH values, and the vertical error bars show the $M_g$ bin size.
  \label{ml}}
\end{figure}

\begin{figure}
  \plotone{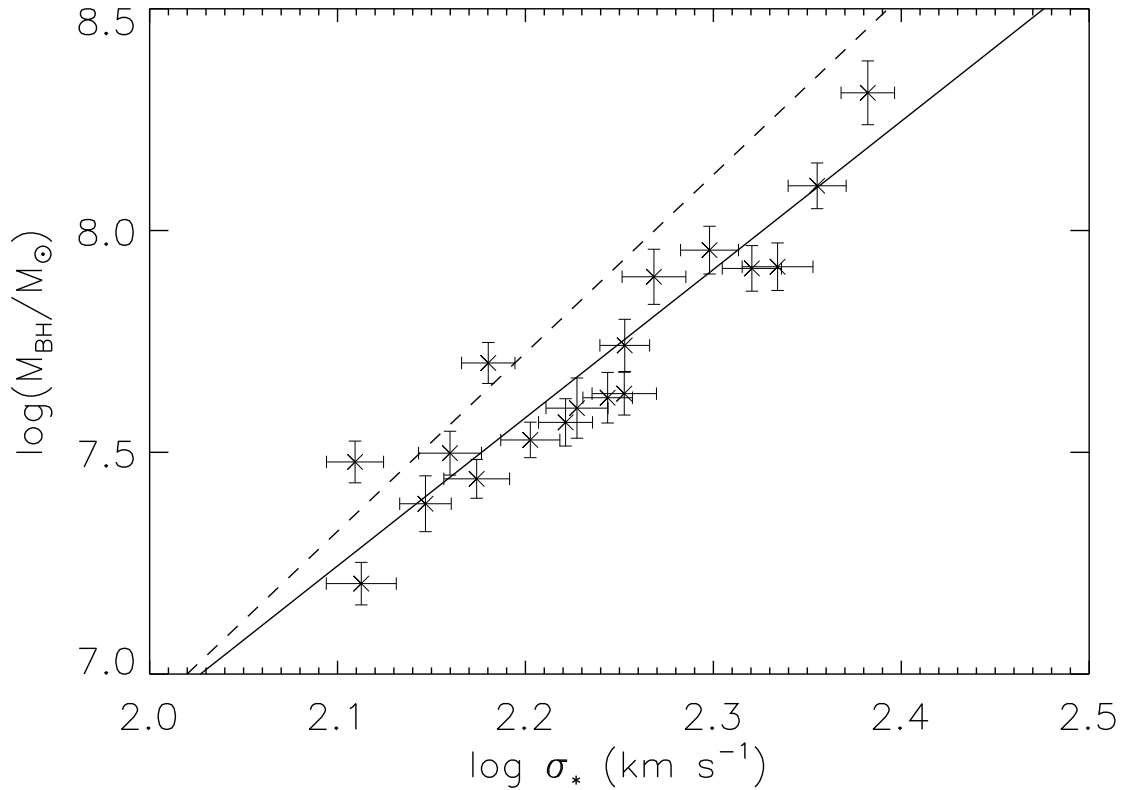}
  \caption{Black hole mass vs.\ stellar velocity dispersion in binned
  data.  The solid line is the best-fitting linear relation (in
  logarithmic space), which gives the slope 3.34$\pm$0.24 and intercept
  7.92 $\pm$0.02.  The dashed line is the \citet{T02} relation for
  inactive galaxies.  Each point represents 50 objects which have been
  binned by host galaxy absolute magnitude, as in Figs.\,\ref{vl} and
  \ref{ml}. The crosses show the mean value in each bin, and error bars
  are the standard deviation of the mean in each bin.  \label{vm}}
\end{figure}

The relationship between the black hole mass and host velocity dispersion
was found by averaging the values of the two quantities in bins of host
absolute magnitude.  That is, the values represented by the points in
each $M_{g}$ bin in Figures~\ref{vl} and \ref{ml} were compared to find
the $M_{\rm BH}-\sigma_*$ relation, which is shown in Fig.\,\ref{vm}. The
crosses show the mean value in each bin and error bars are the standard
deviation of the mean in each bin.  Each point represents the mean value
for about 50 objects, and the error bars are those in Figures\,\ref{vl}
and \ref{ml} divided by the square root of the number of objects in
each bin.  The solid line is the best-fitting linear relation (to the
logarithmic quantities)
\begin{equation}
  \log M_{\rm BH} = 3.34 (\pm 0.24) \log (\sigma_*/200{\rm km\,s^{-1}})
  + 7.92 (\pm 0.02)\,.
\label{mveq1}
\end{equation}

The slope found here is $1.7\sigma$ flatter than the one found for
inactive galaxies by \citet{T02}, who found a value of $4.02 \pm 0.32$.
Fixing the slope to that value, the intercept of the best fit to our
data becomes 7.96$\pm$0.02, which is smaller than the value of $8.13 \pm
0.06$ found by \citet{T02} by 0.17 dex, a significance of $2.7 \sigma$.
(In both cases $\sigma$ is the quadrature sum of the uncertainties of
both measurements.) The same $M_{BH}-\sigma_*$ relation as equation
(\ref{mveq1}) can also be obtained by fitting the individual data
points in Figure \ref{vmall}, when an intrinsic scatter of 0.36 dex
on BH masses and 0.12 dex on the velocity dispersions are included.
This scatter is consistent with the dispersions in each binned data set
(see Table~\ref{table2}).

\begin{figure}
  \plotone{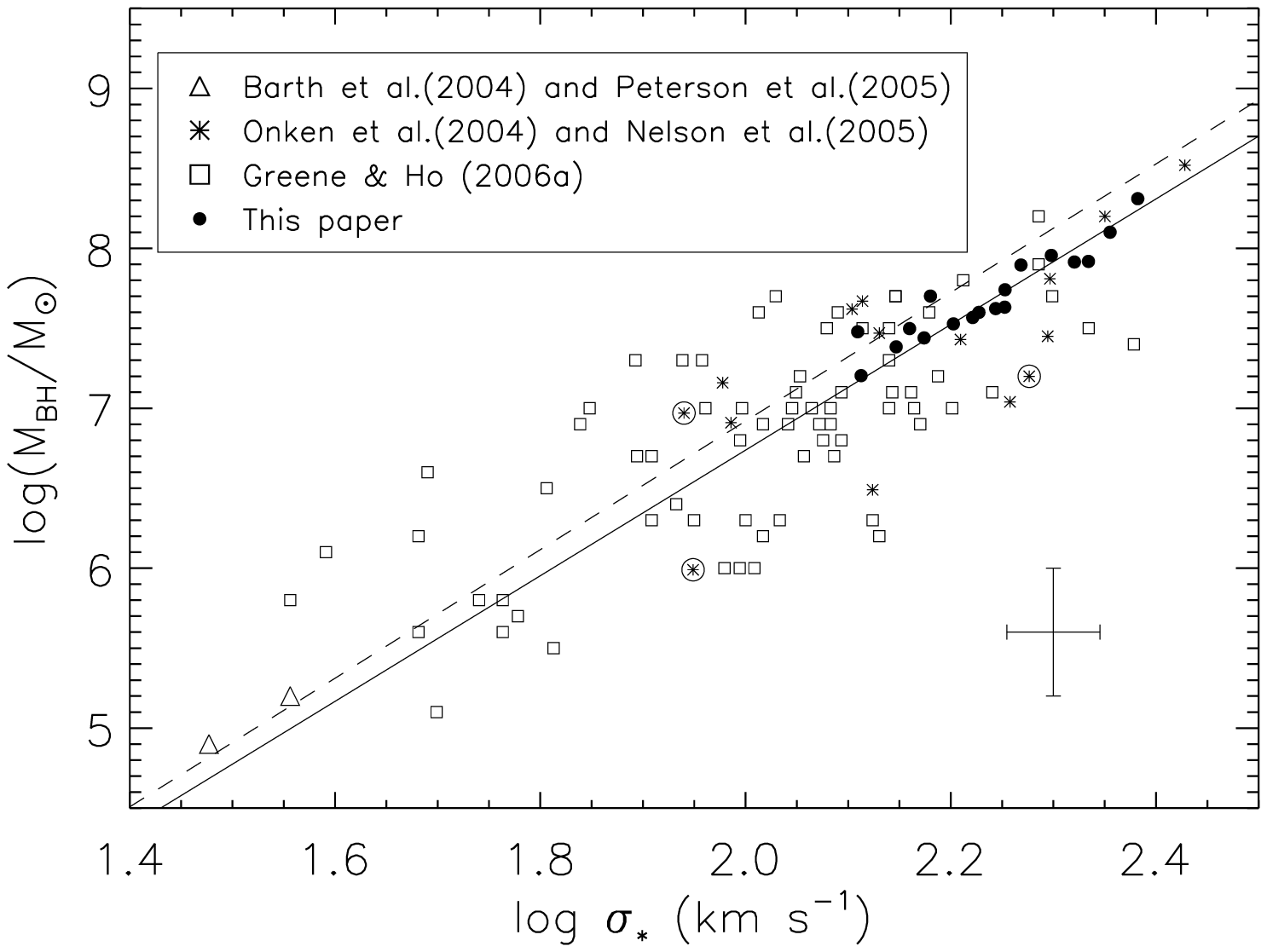}
  \caption{Black hole mass vs.\ stellar velocity dispersion for several
  different data sets, covering a larger dynamic range than
  Fig.\,\ref{vm}.  The filled circles are the data from this paper. The
  asterisks are the data from \citet{Onken04} and \citet{Nelson04},
  who used reverberation mapping measurements.  The open squares are
  the data from \citet{Greene06a}, and the  open triangles are the data
  from \citet{Peterson05} and \citet{Barth04} for NGC 4395 and POX 52
  respectively.  The encircled objects are NLS1s. All the literature
  data values and their uncertainties are available in the table from
  \citet{Greene06a}. The solid line is the best-fit relation for all of
  the data (this paper and the literature data combined).  For comparison,
  the dashed line shows the best-fit relation for inactive galaxies,
  as found by \citet{T02}.  Typical uncertainties in the measurements
  of BH mass and velocity dispersion for individual objects are shown
  in the bottom right corner. \label{vm2}}
\end{figure}

Figure \ref{vm2} shows the same relationship as Figure \ref{vm} but
with values from the literature added in the plot.   The additional
data extends well beyond the dynamic range of our values, particularly
toward smaller values of $M_{\rm BH}$ and $\sigma_*$.  The literature
data and uncertainties are all obtained from the table given by
\citet{Greene06a}. The open squares are the data estimated by
\citet{Greene06a}, the asterisks are the data from \citet{Onken04}
and \citet{Nelson04}, and the open triangles are the data from
\citet{Peterson05} and \citet{Barth04} for NGC\,4395 and POX\,52
respectively.  The filled circles are the binned data from this paper
as shown in Fig.\,\ref{vm}.  In Fig.\,\ref{vm2} the solid line is the
best fit using all of the data.  The dashed line is the fit for inactive
galaxies from \citet{T02}.

The best fit to the $M_{\rm BH}-\sigma_*$ relation using all of the data is
\begin{equation}
  \log M_{\rm BH} = 3.93 (\pm 0.10) \log (\sigma_*/200{\rm km\,s^{-1}})
  + 7.92 (\pm 0.02)\,.
\label{mveq2}
\end{equation}
This slope is entirely consistent with that measured for inactive galaxies
\citep{T02}.  Fits were also made with a fixed slope of $4.02$ to all of
the data. The fixed-slope intercept is smaller than the inactive galaxy
value of $8.13$ by 0.19 dex, a significance of $3 \sigma$.

Based on our analysis, we find that the $M_{\rm BH}-\sigma_*$ relationship
for active galaxies in our sample is marginally flatter than that
of inactive galaxies, but that it is consistent with that of inactive
galaxies if the literature data are added.  The flattening is consistent
with the study of \citet{Greene06a}.  The intercept of the active galaxy
relationship is smaller than that for inactive galaxies.  This holds
whether or not the literature data are included in the analysis, and
is consistent with \citet{Nelson04} and \citet{Greene06a}, though with
uncertainties $\sim$5 times smaller than in the former reference and
$\sim$2 times smaller than the latter.

We caution that the results presented in this section depended in part on
the selection of the mass-scaling relationship used to estimate $M_{\rm
BH}$, and that not all of the literature values were derived using the
same relationship that we use here.  In particular the intercept of the
$M_{\rm BH}-\sigma_*$ relationship would be even smaller had we used the
commonly adopted scale factor value $f=3$ (which is inconsistent with the
value derived empirically by \citet{Onken04}).  The power-law index of
the radius-luminosity relationship also affects the slope of the $M_{\rm
BH}-\sigma_*$ relationship.  A higher radius-luminosity power-law index
(which would be inappropriate for our study since we account for host
galaxy flux contamination) would increase the slope.

%
\subsection{The Dependence of the $M_{\rm BH}-\sigma_*$ Relation on the
  Eddington Ratio and Redshift \label{dependence}}
The distribution of data points in the $M_{\rm BH}-\sigma_*$ plane may
be affected by many factors, such as the details of the BLR physics
and galaxy evolution.  We can begin to examine the dependence of the
distribution of points in the plane on some of these factors, at least
in a statistical sense, thanks to the large size of our data set.

\begin{figure}
  \plotone{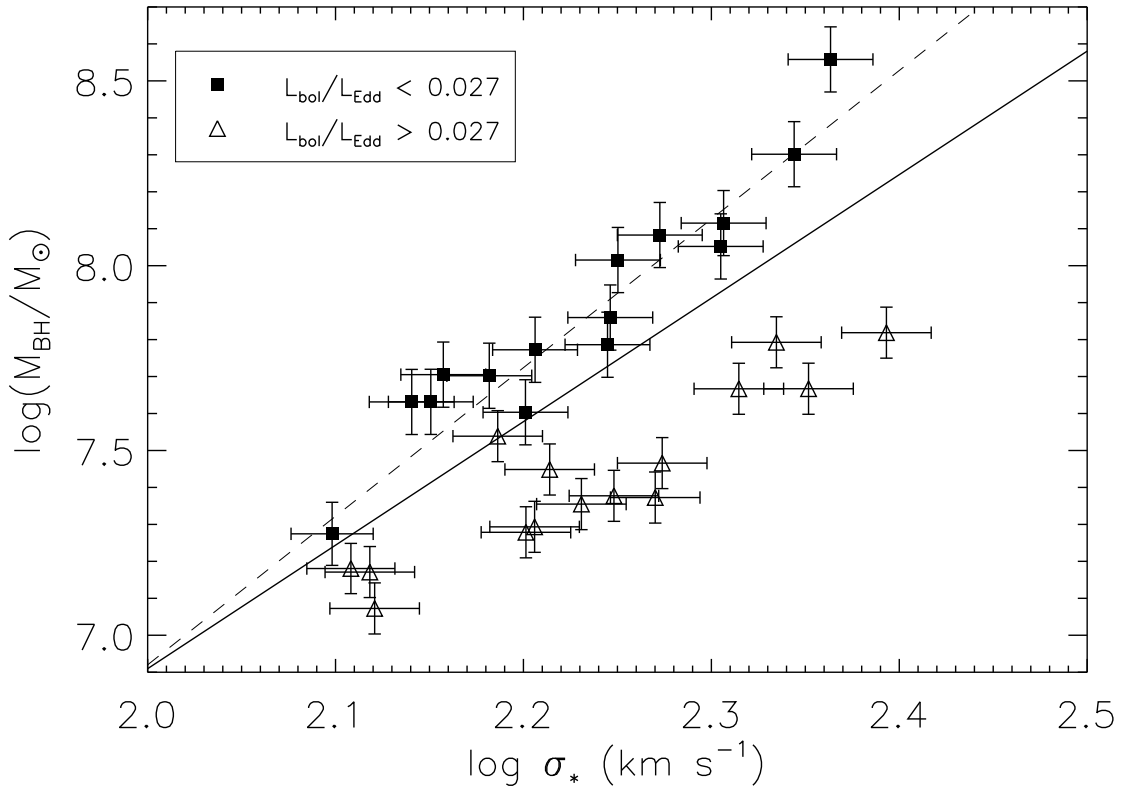}
  \caption{The $M_{\rm BH}-\sigma_*$ relationship for two samples
  divided by the mean value of the Eddington ratio, $L_{\rm bol}/L_{\rm
  Edd}$. Mean values of $M_{\rm BH}$ and $\sigma_*$ for objects with
  $L_{\rm bol}/L_{\rm Edd}$ below the median value of $0.027$, are shown
  with filled squares; open triangles show the mean values for objects
  with $L_{\rm bol}/L_{\rm Edd}$ above the median value.  Each sample
  has contributions from 451 objects, and each point represents the
  mean values for at least 30 objects.  The error bars are the standard
  deviation of the mean in each bin.  The solid line shows the best
  fit relation from equation\,(\ref{mveq1}).  The dashed line is the
  \citet{T02} relation for inactive galaxies.  \label{vmledd}}
\end{figure}

Figure\,\ref{vmledd} shows the $M_{\rm BH}-\sigma_*$ relationship
for two different ranges of $L_{\rm bol}/L_{\rm Edd}$.  The AGN
sample was divided at the median value of the Eddington ratio,
$L_{\rm bol}/L_{\rm Edd}=0.027$.  Objects with $L_{\rm bol}/L_{\rm
Edd}$ values below the median are shown as filled squares, and those
with values above the median are shown as open triangles.  The
median values for the low and high $L_{\rm bol}/L_{\rm Edd}$ groups
are 0.015 and 0.072 respectively.  Each point represents the mean
value for at least 25 objects. There is a clear separation in the
$M_{\rm BH}-\sigma_*$ relationship between the high and low $L_{\rm
bol}/L_{\rm Edd}$ samples. For a given velocity dispersion, those
objects with lower values of $L_{\rm bol}/L_{\rm Edd}$ have larger
black hole masses. The slopes of the relationship in the two samples
are not significantly different, but the intercepts differ greatly.
With the slope fixed at 4.02 as for inactive galaxies, the
intercepts are 8.15$\pm 0.02$ and 7.68$\pm 0.02$ for the samples
with low and high $L_{\rm bol}/L_{\rm Edd}$ respectively. The
intercepts differ by about 0.47\,dex in $\log (M_{\rm BH})$ in the
two groups. The intercept of the low $L_{\rm bol}/L_{\rm Edd}$
sample is a bit higher than that of inactive galaxies \citep{T02},
which of course have extremely low $L_{\rm bol}/L_{\rm Edd}$ values.
Generally then, in the sample studied here, BH masses are smaller
relative to the mean relation for objects with larger Eddington
ratios.

The 0.47~dex difference of the intercepts in the two groups is
statistically significant ($>10 \sigma$), given the precision with
which they can be measured with the large samples.  At face value,
this result would tie together the accretion rate with physical
properties of the host bulge component.  However, there may be
selection effects and parameter interdependencies that could cause
the apparent difference.  For example, Fig.\,\ref{MbhLedd} shows an
anticorrelation between $M_{\rm BH}$ and $L_{\rm bol}/L_{\rm Edd}$,
but as discussed in \S\,\ref{bhdist}, it is caused by selection
effects and the functional interdependence of the two parameters.
Therefore, to determine the reality of the apparent dependence of
the $M_{\rm BH}-\sigma_*$ relationship on $L_{\rm bol}/L_{\rm Edd}$,
correlations among the other parameters that are involved --- AGN
monochromatic luminosity $L_{5100}$, emission line FWHM, host
luminosity $L_{H}$, stellar velocity dispersion $\sigma_{*}$, and
redshift --- must be taken into account.  To test the reality of the
results, we have used partial correlation analysis to account for
the interdependencies.

\begin{figure}
  \plotone{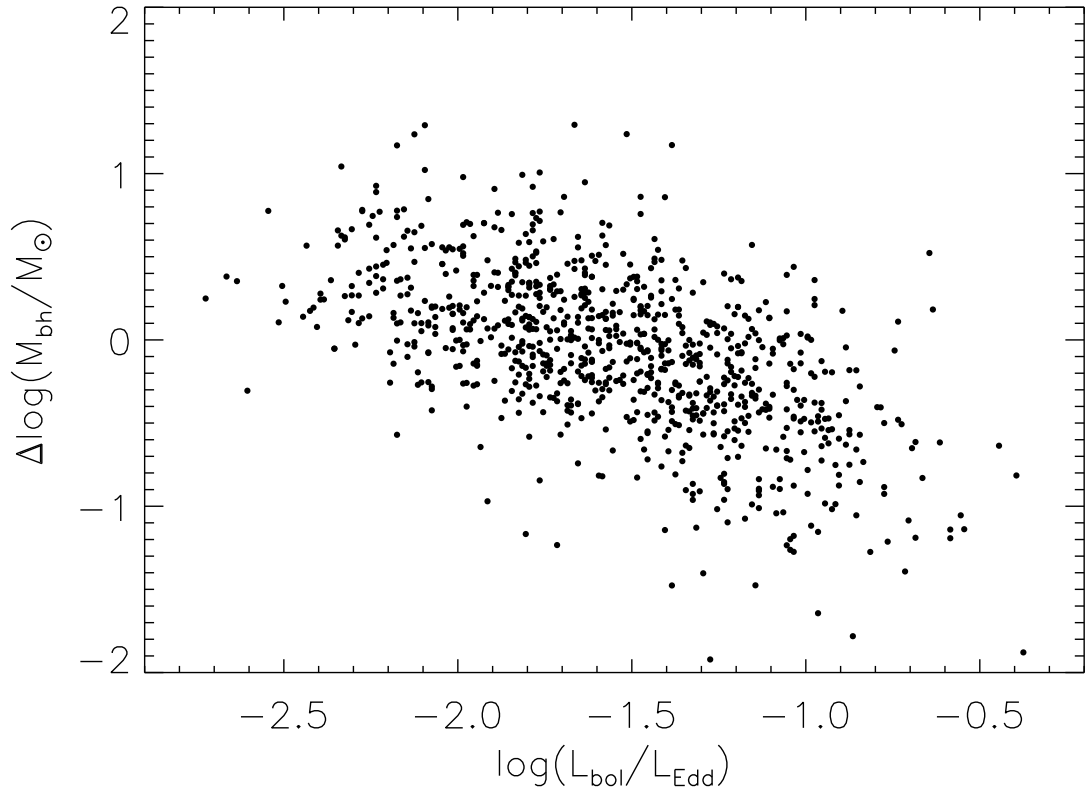}
  \caption{The BH mass difference between the measured BH mass and BH mass
  predicted by equation (\ref{mveq1}) versus the Eddington ratio,
  $L_{\rm bol}/L_{\rm Edd}$.  \label{dMbh_ledd}}
\end{figure}

For the partial correlation tests, we defined a new variable $\Delta
M_{\rm BH}$,
\begin{eqnarray}
\Delta M_{\rm BH} = \log(M_{\rm BH})-3.34
  \log (\sigma_*/200{\rm km\,s^{-1}}) - 7.92 \,,
  \label{eq:dMbh}
\end{eqnarray}
which is the difference, in logarithmic space, between the measured
BH mass and the BH mass predicted by equation (\ref{mveq1}).
Fig.\,\ref{dMbh_ledd} shows the relationship between $\Delta M_{\rm BH}$
and Eddington ratio; there is apparently a strong anti-correlation,
with a Pearson product-moment correlation coefficient of $-0.576$. A
partial correlation analysis \citep[e.g.][]{wall} was performed on $\Delta
M_{\rm BH}$ and $L_{\rm bol}/L_{\rm Edd}$, controlling for $L_{5100}$,
FWHM, $\sigma_{*}$, $L_{H}$, and redshift; the analysis was performed
in logarithmic space, because many of the relationships are then more
closely linear.  The results are shown in Table~\ref{table3}.  Each row of
Table~\ref{table3} shows the $\Delta M_{\rm BH}$ vs.\ $L_{\rm bol}/L_{\rm
Edd}$ partial correlation coefficient, $r_{{\Delta}M,Eratio;x}$, where
$x$ is one of the control variables, and the probability that such
a coefficient value or less would occur by chance.  Controlling for
each of the other variables, there is less than a $0.01\%$ probability
that the ${\Delta}M_{\rm BH}$ anti-correlation with Eddington ratio is
due to chance.  Table~\ref{table3} lists the correlation coefficients,
$r_{Eratio,x}$ and $r_{{\Delta}M,x}$, between Eddington ratio and $\Delta
M_{\rm BH}$ respectively, and the control variables.  It is easy to
show from the partial correlation results that the anti-correlation of
$\Delta M_{\rm BH}$ with $L_{\rm bol}/L_{\rm Edd}$ is not due to the
interdependencies of $M_{\rm BH}$, $L_{5100}$, and $L_{\rm bol}/L_{\rm
Edd}$.  For a fixed $FWHM$, both $M_{BH}$ and $L_{\rm bol}/L_{\rm Edd}$
are positively correlated with $L_{5100}$, by their definitions.  Thus,
by equation (\ref{eq:dMbh}), one would expect a positive correlation
between $\Delta M_{\rm BH}$ and $L_{\rm bol}/L_{\rm Edd}$ for a given
$FWHM$, assuming no other dependencies. However, the partial correlation
coefficient, controlling for $FWHM$, is negative (Table~\ref{table3})
and still highly significant. Therefore, the anti-correlation of
$\Delta M_{\rm BH}$ with $L_{\rm bol}/L_{\rm Edd}$ is not due to the
interdependencies of $M_{\rm BH}$,$L_{5100}$, and $L_{\rm bol}/Ledd$.

\begin{figure}
  \plotone{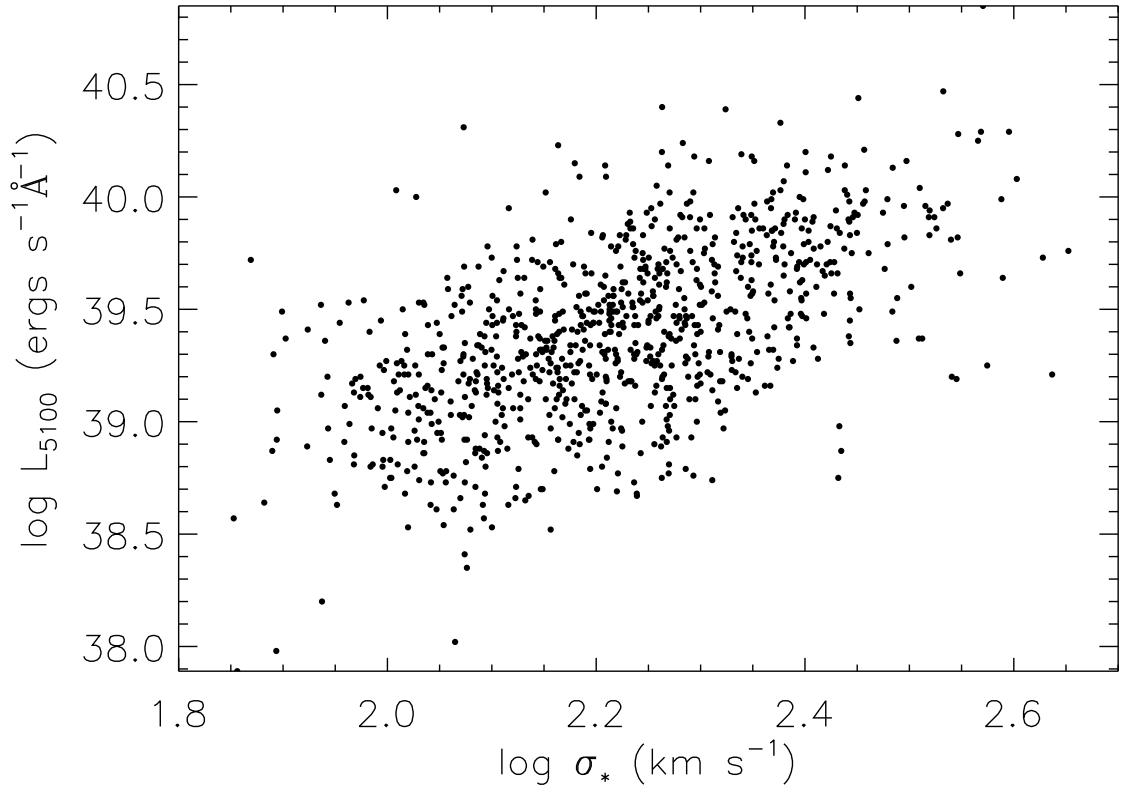}
  \caption{The distribution of the host galaxy velocity dispersion with
  the AGN monochromatic luminosity at 5100{\AA}. \label{vlagn}}
\end{figure}

\begin{figure}
  \plotone{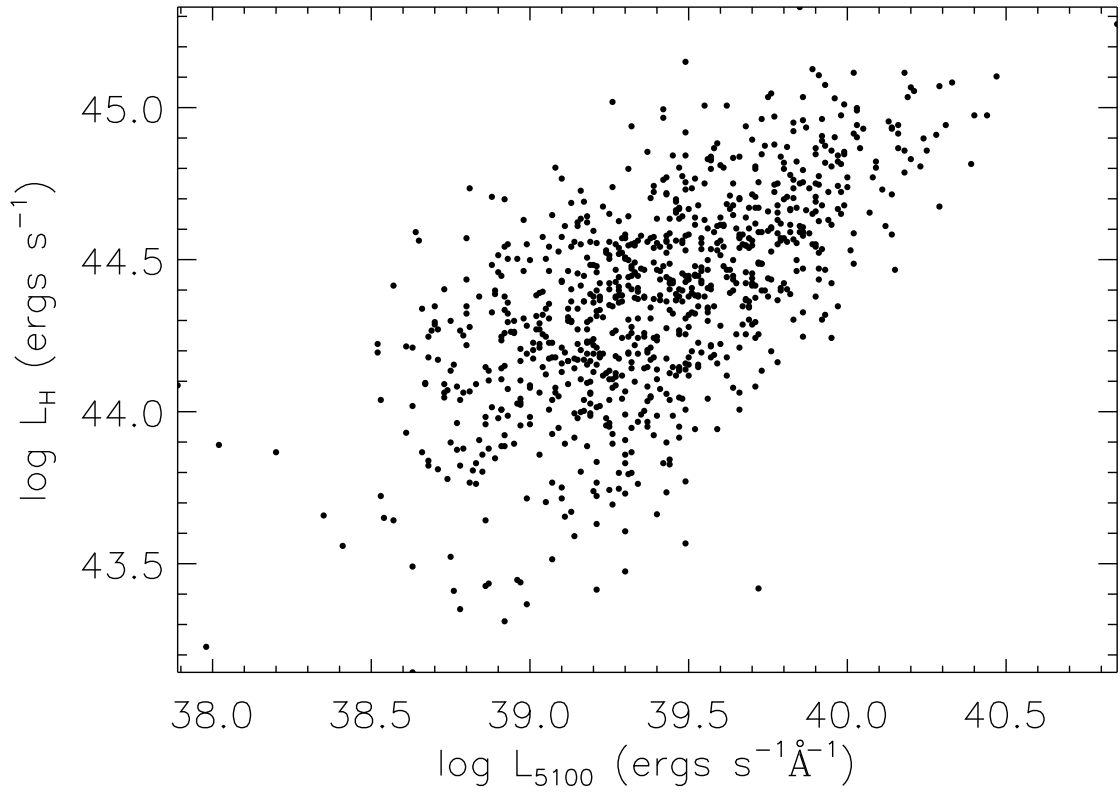}
  \caption{The distribution of the host galaxy $g$ band luminosity $L_{H}$,
  with the AGN monochromatic luminosity at 5100{\AA}.  The apparent
  correlation may be due in part to sample selection effects. \label{lhlagn}}
\end{figure}

The partial correlation tests do not determine whether the $\Delta M_{\rm
BH}$ vs.\ $L_{\rm bol}/L_{\rm Edd}$ anti-correlation is due to sample
selection effects.  Figure~\ref{vlagn} shows that there is a correlation
between AGN monochromatic luminosity and host galaxy velocity dispersion,
with a correlation coefficient of $0.569$.  This correlation can explain
the difference between the intercepts in Fig.\,\ref{vmledd}. For a
given BH mass, objects with larger values of $\sigma_*$ will generally
have larger values of $L_{5100}$, and therefore, larger Eddington
ratios.  Therefore, objects to the right of the regression line
(having larger $\sigma_*$ values) will generally have larger Eddington
ratios, and vice versa.  However, the correlation between $L_{5100}$
and $\sigma_*$ may be a consequence of the Faber-Jackson relation
(\S\,\ref{faberjackson}, Fig.\,\ref{vl}) and a correlation between
$L_{5100}$ and host luminosity $L_{H}$, shown in Fig.\,\ref{lhlagn}
(cf.~VB06).  Assuming the Faber-Jackson relation we see is not due to
selection effects, the reality of the $M_{\rm BH}-\sigma_*$ dependence
on the Eddington ratio depends upon whether or not the $L_{5100}-L_{H}$
correlation is intrinsic, or due to selection effects.  As discussed
by VB06, at least part of the correlation is due to the restriction
that both the host and AGN components contribute significantly to
the composite spectrum; this corresponds to the $0.2 < F_{H} < 0.8$
criteria described in \S\,\ref{observations}.  There are clearly many
AGNs in the universe that do not satisfy those criteria.  The partial
correlation coefficients of AGN monochromatic luminosity and host
galaxy velocity dispersion are shown in Table~\ref{table4}. Although
the correlation is weaker controlling for $L_{H}$ and/or redshift,
it is still significant. Hence, the correlation of AGN monochromatic
luminosity and host galaxy velocity dispersion is intrinsic, but is
enhanced due to a sample selection effect.  To summarize, the apparent
$M_{\rm BH}-\sigma_*$ dependence on $L_{\rm bol}/L_{\rm Edd}$ remains
after accounting for several possible interdependencies, although it
is likely that the relative luminosity criteria used to construct the
spectroscopic sample artificially strengthens the apparent dependence.

Selection effects similar to ones imposed on the spectroscopic sample
can be avoided, at least partially, in high-resolution imaging studies,
such as those carried out with the $HST$ \citep[e.g.][]{floyd04,sanchez04,
dunlop03, pagani03, hamilton02, schade00}.  In a recent compilation by
\citet{sanchez04}, for example, a strong correlation was found between
nuclear and host luminosities.  The explanation is that the data are
bracketed by lines of constant Eddington luminosity, ranging from $L_{\rm
bol}/L_{\rm Edd} \approx 0.01$ to 1.  The cutoff at an Eddington ratio
of near unity is particularly sharp, possibly indicating that very
few AGNs emit above their Eddington limit.  The low Eddington ratio
limit is more likely to be a selection effect, in that low-luminosity
nuclei are difficult to detect in high-luminosity hosts. The apparent
correlation, whether intrinsic or a selection effect, would cause at
least qualitatively the same dependence of the $M_{\rm BH}-\sigma_*$
relation on the Eddington ratio seen in Fig.\,\ref{vmledd}.

\begin{figure}
  \plotone{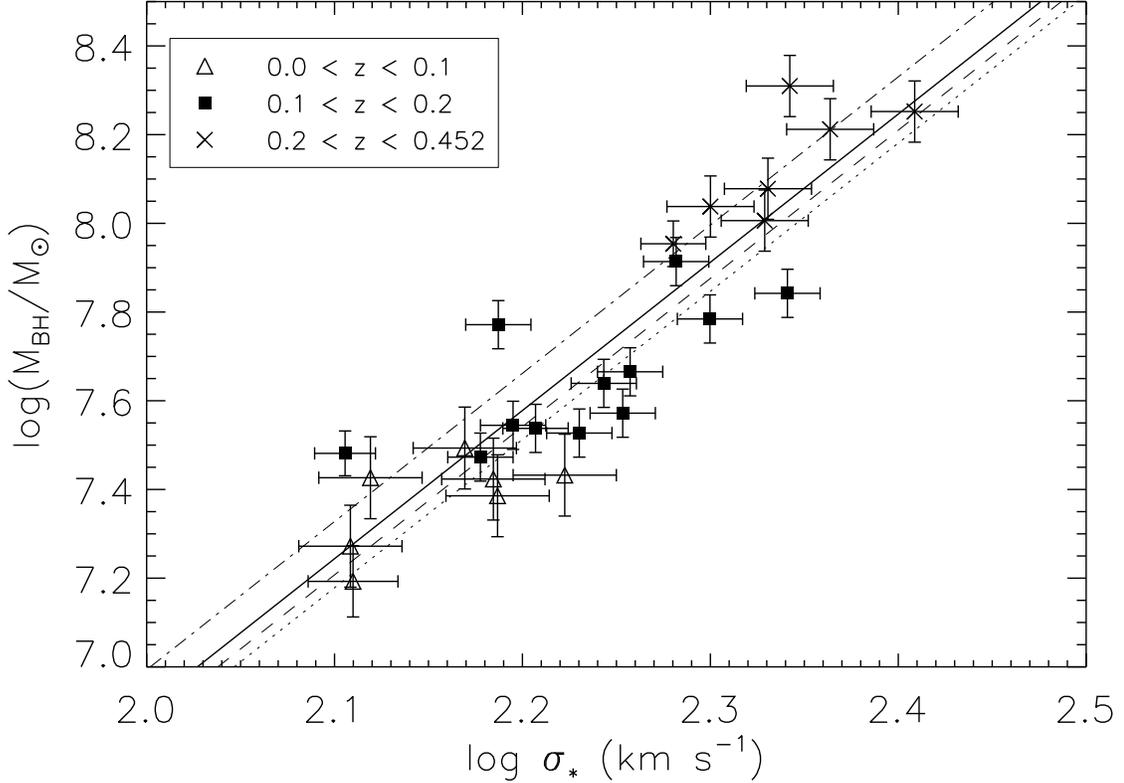}
  \caption{The $M_{\rm BH}-\sigma_*$ relationship in different redshift
  ranges. Objects with redshifts from 0 to 0.1 are represented with open
  triangles, which show the weighted average values in bins containing
  measurements from at least 25 objects.  Filled squares show the mean
  values for objects with redshifts from 0.1 to 0.2; each point has
  contributions from at least 40 objects.  Crosses show the mean values
  for objects with redshifts from 0.2 to 0.452; each point represents
  data from at least 30 objects.  The error bars are the standard
  deviation of the mean in each bin. The solid line shows the best fit
  relation from equation\,(\ref{mveq1}).  With the slope fixed at the
  value in equation\,(\ref{mveq1}), the dotted, dashed and dot-dashed
  lines show the best fits for the three bins with redshifts from low
  to high respectively. \label{vmz}}
\end{figure}

Figure\,\ref{vmz} shows the $M_{\rm BH}-\sigma_*$ relationship for data
in several different redshift bins.  Our sample was divided into three
subgroups by redshift: $0 < z < 0.1$, $0.1 < z < 0.2$, and $z > 0.2$,
shown in the figure by open triangles, filled squares, and crosses
respectively.  The data were separated into bins of host $M_{g}$, as
with the previous analysis; each point represents data from at least
25 objects.  The points clearly occupy different regions along the best
fit $M_{\rm BH}-\sigma_*$ line.  That is due to the redshift-luminosity
selection effect correlation in the sample, seen in Fig.\,\ref{zl}:
objects at higher redshifts have higher luminosities on average, so
AGNs with larger host velocity dispersions and more massive black
holes are preferentially selected at higher redshifts.  The solid
line shows the best fit to the $M_{\rm BH}-\sigma_*$ relation from
equation\,(\ref{mveq1}).  Figure \ref{vmz} shows that the intercept of
the relationship doesn't change greatly with redshift.  The points in the
highest redshift bin are on average slightly above the best fit line,
while those at lower redshifts are slightly below the line on average,
suggesting a possible redshift evolution of the $M_{\rm BH}-\sigma_*$
relation.  With the slope fixed to the value in equation\, (\ref{mveq1}),
the intercepts of the three groups from low to high redshifts are
7.85$\pm$0.02, 7.88$\pm$0.02 and 8.00$\pm$0.03 respectively.  At face
value, the results suggest a weak but significant dependence of the
$M_{\rm BH}-\sigma_*$ relation on redshift, such that active galaxies
on average have larger black hole masses for a given velocity dispersion
at high redshift as compared to low redshift.

As with the dependence of the $M_{\rm BH}-\sigma_*$ relation on Eddington
ratio, a partial correlation analysis was performed to determine
whether the apparent redshift dependence of $\Delta M_{\rm BH}$ could be
accounted for by correlations with other parameters.  The correlation
coefficient of $\Delta M_{\rm BH}$ with redshift, not accounting for
other parameters, is 0.005 (see last row of Table~\ref{table3}), which
is relatively small but may or may not be significant.  The results of
the partial correlation analysis, with $L_{5100}$, FWHM, $\sigma_{*}$,
$L_{H}$, and $L_{\rm bol}/L_{\rm Edd}$ as the control variables,
are shown in Table~\ref{table5}.  For each of the control variables,
the remaining correlation between $\Delta M_{\rm BH}$ and redshift is
relatively small.  The significance of the remaining correlations are
all quite high (probabilities $< 0.11\%$), except when $L_{H}$ is the
control variable, in which case the probability of a chance correlation
is about $11.8\%$.  However, when controlling for all five parameters,
the partial correlation coefficient is $0.059$ with a random probability
of being greater of $92.4\%$.  Thus, there is no compelling evidence for
redshift evolution in the $M_{\rm BH}-\sigma_*$ relation for the sample
presented here.

The lack of evidence for redshift evolution is in contrast to the results
of several studies \citep{Treu04,Woo06,Peng06a,Peng06b,Shields06},
that have found apparent positive correlations between the $M_{\rm
BH}-\sigma_*$ relation and redshift.  \citet{Woo06} found that their
sample of Seyfert galaxies at $z\approx 0.36$ had larger black hole
masses for a given $\sigma_{*}$, relative to the values expected from
the local relation.  \citet{Shields06} measured CO emission lines to
infer $\sigma_{*}$ for a set of nine high-redshift AGNs, and found
that the host masses are undersized by more than an order of magnitude,
given their BH masses.  \citet{Peng06a,Peng06b} found that $z\gtrsim 2$
hosts in their sample are less luminous (and therefore less massive)
than expected for a given $M_{\rm BH}$.  In contrast, \citet{Shields03}
and \citet{Salviander06}, using O\,[{\sc iii}] line widths as
surrogates for $\sigma_{*}$, found no evidence for redshift evolution
up to $z\sim3$. There is also evidence from a study of high-redshift
sub-millimeter galaxies \citep{Borys05} that at least some galaxies at
high redshift have BH masses substantially {\em smaller} than expected
for their host masses. The methods and equations used to estimate $M_{\rm
BH}$ and $\sigma_*$ were somewhat different in each study, and different
from our own, and each sample was selected in different ways, so direct
comparisons are not necessarily valid.  Our study is the first to examine
the possibility of redshift evolution for a single, large, homogeneously
collected and analyzed sample spanning both low and relatively high
redshifts; and our study is the first to determine whether an apparent
dependence of the redshift evolution could be the result of correlations
among other parameters.  It will be important to extend the studies of
the $M_{\rm BH}-\sigma_*$ relation to higher redshifts, over a wide range
of BH masses and velocity dispersions, using a variety of techniques.

\section{Discussion and Summary \label{discussion}}
The very large and homogeneous data set analyzed here has allowed
parameterization of several relations for broad-line active
galaxies. Host galaxy luminosity and velocity dispersion are described
by the Faber-Jackson relation, central black hole mass is correlated
with host bulge luminosity and velocity dispersion, the amplitude of
the $M_{\rm BH}-\sigma_*$ relation depends inversely on the Eddington
ratio, and there is no significant change in the $M_{\rm BH}-\sigma_*$
relation with redshift up to $z\approx 0.4$.

The dependence of the $M_{\rm BH}-\sigma_*$ relation on both redshift
and Eddington ratio have been matters of dispute.  In our sample,
we find no significant correlation with redshift after accounting for
possible correlations among many other parameters.  As discussed in
\S\,\ref{dependence}, there are studies that support positive, negative,
and no correlation with redshift.  The $M_{\rm BH}-\sigma_*$ relation
appears to depend on the Eddington ratio, even after accounting for
other parameters.  Apparently, the difference we have found between the
high and low $L_{\rm bol}/L_{\rm Edd}$ samples is sufficient to account
for the intrinsic scatter in the $M_{\rm BH}-\sigma_*$ relation, usually
measured to be about $0.4$\,dex in black hole mass for a given velocity
dispersion \citep[e.g.][\S\,\ref{results} of this paper]{Greene06a}.
We checked this result by dividing the \citet{Greene06a} sample by its
median $L_{\rm bol}/L_{\rm Edd}$; this sample shows the same qualitative
trend that objects with higher Eddington ratios have smaller black hole
masses for a given velocity dispersion (although the uncertainty is too
great to say whether it can account for the scatter).  As discussed
in \S\,\ref{dependence}, the dependence of the $M_{\rm BH}-\sigma_*$
relation on $L_{\rm bol}/L_{\rm Edd}$ is probably at least partly due
to selection effects --- which may also affect the \citet{Greene06a}
sample --- but a significant intrinsic component cannot be ruled out.

There is other observational support for an anti-correlation between the
$M_{\rm BH}-\sigma_*$ relation and Eddington ratio from observations
of narrow-line Seyfert 1 galaxies (NLS1s).  These active galaxies
typically have relatively small measured BH masses and high accretion
rates, which results in large Eddington ratios.  Several studies
\citep[e.g.][]{wandel02, grupe04, bian04, Nelson04} have shown that
NLS1s tend to lie below the $M_{\rm BH}-\sigma_*$ relation for inactive
galaxies.  If NLS1s are an early stage of AGN evolution, then they will
presumably move up the $M_{\rm BH}-\sigma_*$ plane over time (BH masses
growing), becoming broad-line AGNs and eventually inactive galaxies
harboring supermassive black holes.  If the objects evolve in this way
toward the $z=0$ $M_{\rm BH}-\sigma_*$ relation, the black holes would
have to be growing more rapidly than the bulges are growing.  To explain
the tight relationship between the final BH and spheroid masses, the gas
mass that builds the black hole may somehow be set at a given formation
stage of the stellar spheroid. It would then be the fate of the black hole
to grow to a particular mass, determined very early in its development
by the mass of the spheroid \citep[e.g.][]{miralda05,merritt04}.
Feedback from the central engine surrounding the BH may regulate the
BH growth process \citep[e.g.][]{robertson06, begelman05, dimatteo05,
king03}, but this feedback may have little to do with the development
of the spheroid. It is also possible that NLS1s are not a representative
class of AGN, and their evolutionary tracks in the $M_{\rm BH}-\sigma_*$
plane are different than those of other types.  Observations of AGNs
with luminosities comparable to NLS1s, but at higher redshifts than can
be obtained with the SDSS, would help clarify the issue.

Differences in the $M_{BH}-\sigma_*$ relation may also be due
to disk orientation effects, which could produce a wide range of
incorrect estimates for black hole masses.  Given that all BH masses
were determined by using the same value of the $f$ factor (equation
\ref{masseq}), the objects with high $L_{\rm bol}/L_{\rm Edd}$ may be
those in which BH masses are underestimated because their true $f$ values
are larger. \citet{MD01,MD02} argue that a flattened-disk-like geometry in
the BLR is favored over randomly-oriented orbits, and a flattened geometry
viewed over a range of orientation angles can easily result in virial BH
masses underestimated by a factor of three.  However, \citet{Collin06}
showed that the inclination effects are actually minimal when using
the line dispersions as opposed to the FWHM of the line profile.
Although the factor of 3 difference in the BH masses (0.5\,dex) can
explain the apparent dependence of the $M_{\rm BH}-\sigma_*$ relation
on $L_{\rm bol}/L_{\rm Edd}$, we cannot settle this issue here because
of our poor understanding of the BLR geometry.

Our BH masses were estimated by equation (\ref{Mbh}), where FWHM was
used to characterize the line width \citep{Vestergaard06}). However,
some studies showed that the line dispersion is a less biased
parameter in general than FWHM for black hole mass estimation
\citep[e.g.][]{Peterson04, Collin06}. The line dispersion will also
minimize the effect that may be influenced by relative source
inclination \citep{Collin06}.

The SDSS spectroscopic data set has provided a large sample of
homogeneously selected AGN from which both nuclear and host properties
can be measured.  The moderate resolution and $S/N$ of the SDSS spectra
limit the ranges of accessible luminosities, black hole masses, and
Eddington ratios. However, in principle, the eigenspectrum decomposition
technique can be applied to any spectrum with adequately strong nuclear
and host components, so it is possible with other data sets to extend
the results of this study to a larger dynamic range.

\acknowledgments

We thank Todd Boroson for providing the iron template, and Smita
Mathur, Dirk Grupe, Christy Tremonti, Ching-Wa Yip, Mariangela
Bernardi, Ohad Shemmer, David Schlegel, and Jenny Greene for useful
discussion. This research was partially supported by NSF grants
AST03-07582 and AST06-07634.  P.B.H. is supported by NSERC.

Funding for the SDSS and SDSS-II has been provided by the Alfred P. Sloan
Foundation, the Participating Institutions, the National Science
Foundation, the U.S. Department of Energy, the National Aeronautics
and Space Administration, the Japanese Monbukagakusho, the Max Planck
Society, and the Higher Education Funding Council for England.  The SDSS
Web Site is http://www.sdss.org/. The SDSS is managed by the Astrophysical
Research Consortium for the Participating Institutions. The Participating
Institutions are the American Museum of Natural History, Astrophysical
Institute Potsdam, University of Basel, Cambridge University, Case
Western Reserve University, University of Chicago, Drexel University,
Fermilab, the Institute for Advanced Study, the Japan Participation Group,
Johns Hopkins University, the Joint Institute for Nuclear Astrophysics,
the Kavli Institute for Particle Astrophysics and Cosmology, the Korean
Scientist Group, the Chinese Academy of Sciences (LAMOST), Los Alamos
National Laboratory, the Max Planck-Institute for Astronomy (MPIA),
the Max-Planck-Institute for Astrophysics (MPA), New Mexico State
University, Ohio State University, University of Pittsburgh, University
of Portsmouth, Princeton University, the United States Naval Observatory,
and the University of Washington.


%
\clearpage

%
\clearpage

%
\clearpage

%
\clearpage

%
\clearpage

%
\clearpage

%
\clearpage

%
\clearpage
%
\clearpage

%
\clearpage

%
\clearpage

%
\clearpage

%
\clearpage

%
\clearpage

%
\clearpage

%
\clearpage

%
\clearpage

%
\clearpage
\clearpage

%
\clearpage

%
\clearpage

%
\clearpage

%
\clearpage

%
\clearpage

%
\clearpage

%
\clearpage

%
\clearpage

%
\clearpage

%
\clearpage

%
\clearpage

\clearpage
\begin{center}
\begin{deluxetable}{ccrccccccccccccc}
\rotate
\tabletypesize{\scriptsize}
\tablewidth{0pt}
\setlength{\tabcolsep}{0.025in}
\tablecaption{Sample Properties}
\tablehead{\colhead{Name} & \colhead{z} & \colhead{$m_i$} & \colhead{$F_H$} &  \colhead{$\phi$} &
\colhead{$S/N$} & \colhead{$S/N$(Host)} & \colhead{M${}_g$(Host)} & \colhead{$\log L_{5100}$(1)} &
\colhead{$\log L_{5100}$(2)} & \colhead{$\sigma_*$} & \colhead{$\sigma_c$} &
\colhead{FWHM${}_{{\rm H}\beta}$} & \colhead{FWHM${}_{{\rm H}\alpha}$} & \colhead{$\log M_{\rm BH}$} &
\colhead{$L_{\rm bol}/L_{\rm Edd}$}\\
\colhead{(1)} & \colhead{(2)} & \colhead{(3)} & \colhead{(4)} & \colhead{(5)} &
\colhead{(6)} & \colhead{(7)} & \colhead{(8)} & \colhead{(9)} & \colhead{(10)} &
\colhead{(11)}  & \colhead{(12)} & \colhead{(13)} & \colhead{(14)} & \colhead{(15)} & \colhead{(16)} }
\startdata
000611.54$+$145357.2 & 0.1186 & 15.79 & 0.41 & 2.30 & 25.9 & 15.7 &  -22.47 & 39.52 & 39.90 & 181.1$\pm$13.4
& 188.9 & 3.39$\pm$0.11 &  3.13$\pm$0.05 &  6.96$\pm$ 0.07 &  0.1568 \\
000729.98$-$005428.0 & 0.1454 & 17.55 & 0.53 & 3.54 & 15.3 & 9.6 & -20.71 & 39.08 & 39.65 & 121.2$\pm$17.8
& 127.5 & 3.30$\pm$0.32 &  2.92$\pm$0.07 &  6.63$\pm$ 0.19 &  0.1217 \\
000805.62$+$145023.3 & 0.0455 & 14.93 & 0.65 & 3.70 & 44.9 & 29.2 & -20.78 & 38.67 & 39.37 & 173.1$\pm$ 7.9
& 173.4 & 6.09$\pm$0.25 &  7.02$\pm$0.07 &  6.88$\pm$ 0.09 &  0.0266 \\
000813.22$-$005753.3 & 0.1393 & 16.45 & 0.46 & 3.64 & 27.5 & 13.3 & -21.75 & 39.61 & 39.97 & 221.8$\pm$17.6
& 232.9 & 5.21$\pm$0.24 &  3.24$\pm$0.05 &  7.39$\pm$ 0.10 &  0.0717 \\
\enddata
\tablecomments{Col.(1):Name (SDSS J). Col.(2): Redshift. Col.(3): $i$ band magnitude.
Col.(4): Host galaxy fraction. Col.(5): Galaxy classification angle
(degrees). Col.(4): $i$ band spectroscopic $S/N$. Col.(7): Host galaxy $S/N$.
Col.(8): Host galaxy absolute $g$ band magnitude. Col.(9): $L_{\rm 5100}$
(ergs s${}^{-1}$\AA${}^{-1}$, host subtracted).
Col.(10): $L_{\rm 5100}$ (ergs s${}^{-1}$\AA${}^{-1}$, host not subtracted).
Col.(11): Host galaxy velocity
dispersion (km s${}^{-1}$), uncorrected for finite fiber diameter.
Col.(12): Host galaxy velocity dispersion (km s${}^{-1}$), corrected for finite fiber
diameter by equation (\ref{corr}). Col.(13): FWHM${}_{H\beta}$ (10${}^{3}$km s${}^{-1}$).
Col.(14): FWHM${}_{H\alpha}$ (10${}^{3}$km s${}^{-1}$). See
\S\,\ref{emWidth} for the error estimates of FWHM${}_{H\alpha}$ and FWHM${}_{H\beta}$.
Col.(15): $M_{\rm BH}$ ($M_{\odot}$), calculated from equation
(\ref{Mbh}) by FWHM${}_{H\beta}$.  The symbol '*' indicated where FWHM${}_{H\beta}$
was obtained from FWHM${}_{H\alpha}$ by equation (\ref{eqfwhmab}). The
$M_{\rm BH}$ uncertainties are formal uncertainties, actual uncertainties are
probably dominated by systematics in BLR geometry. Col.(16): $L_{\rm
bol}/L_{\rm Edd}$. From Col.(13) to Col.(16), some quantities are set to zeros.
These default values are given for entries with bad spectra or
measurements (see \S\,\ref{hahb}).  Table~1 is available in the electronic
edition. A portion is shown here for guidance regarding its form and content.}
\label{table1}
\end{deluxetable}
\end{center}

\newpage
\begin{deluxetable}{ccccccc}
\tablecaption{Mean properties in host galaxy luminosity bins}
\tablewidth{0pt}
\tablehead{\colhead{$M_g({\rm H})$} &
\colhead{$M_{g,faint}({\rm H})$} &
\colhead{$M_{g,bright}({\rm H})$} &
\colhead{$\log M_{\rm BH}$} & \colhead{$\Delta \log M_{\rm BH}$} &
\colhead{$\log \sigma_*$} & \colhead{$\Delta \log \sigma_*$} \\
\colhead{(1)} & \colhead{(2)} & \colhead{(3)} & \colhead{(4)} & \colhead{(5)} &
\colhead{(6)} & \colhead{(7)} }
\startdata
-23.08 & -22.82 & -23.88 & 8.31 & 0.51 & 2.38 & 0.10 \\
-22.66 & -22.55 & -22.81 & 8.10 & 0.36 & 2.36 & 0.11 \\
-22.41 & -22.31 & -22.55 & 7.92 & 0.38 & 2.33 & 0.13 \\
-22.22 & -22.14 & -22.31 & 7.96 & 0.38 & 2.30 & 0.11 \\
-22.06 & -22.00 & -22.13 & 7.91 & 0.36 & 2.32 & 0.11 \\
-21.95 & -21.91 & -21.99 & 7.90 & 0.44 & 2.27 & 0.12 \\
-21.84 & -21.76 & -21.91 & 7.74 & 0.42 & 2.25 & 0.09 \\
-21.70 & -21.64 & -21.76 & 7.57 & 0.38 & 2.22 & 0.10 \\
-21.59 & -21.54 & -21.64 & 7.63 & 0.34 & 2.25 & 0.12 \\
-21.49 & -21.45 & -21.54 & 7.62 & 0.40 & 2.24 & 0.09 \\
-21.37 & -21.30 & -21.44 & 7.60 & 0.48 & 2.23 & 0.12 \\
-21.23 & -21.16 & -21.30 & 7.70 & 0.33 & 2.18 & 0.10 \\
-21.08 & -21.02 & -21.16 & 7.53 & 0.28 & 2.20 & 0.11 \\
-20.95 & -20.89 & -21.01 & 7.44 & 0.31 & 2.17 & 0.12 \\
-20.79 & -20.71 & -20.89 & 7.50 & 0.35 & 2.16 & 0.12 \\
-20.59 & -20.46 & -20.70 & 7.38 & 0.45 & 2.15 & 0.10 \\
-20.29 & -20.08 & -20.46 & 7.48 & 0.33 & 2.11 & 0.11 \\
-19.60 & -18.41 & -20.07 & 7.20 & 0.35 & 2.11 & 0.14 \\
\enddata
\tablecomments{903 objects with reliable parameter measurements
were dividend into 18 bins; each bin contains 50 objects except
the last bin which contains
53 objects. Col.(1): Mean host galaxy $g$ band absolute magnitude
in the bin.  Col.(2): Faintest host galaxy $g$ band absolute magnitude in the bin.
Col.(3): Brightest host galaxy $g$ band absolute magnitude in the
bin. Col.(4): The weighted mean of the $\log M_{\rm BH} (M_{\odot})$ in the bin.
Col.(5): The dispersion of $\log M_{\rm BH} (M_{\odot})$ in the
bin. Col.(6): The weighted mean of velocity dispersion $\log \sigma_*$ in the bin. Col.(7):
The dispersion of $\log \sigma_*$ in the bin.}
\label{table2}
\end{deluxetable}

\newpage
\begin{deluxetable}{crrrc}
\tablecaption{Partial correlation analysis for the dependence of
  $\Delta M_{\rm BH}$ on Eddington ratio}
\tablewidth{0pt}
\tablehead{
\colhead{$x$} &
\colhead{$r_{L_{bol}/L_{Edd},x}$} &
\colhead{$r_{\Delta M,x}$}&
\colhead{$r_{\Delta M,L_{bol}/L_{Edd};x}$} &
\colhead{$P$} \\
\colhead{(1)} &
\colhead{(2)} &
\colhead{(3)} &
\colhead{(4)} &
\colhead{(5)} }
\startdata
$L_{5100}$  &  0.377 & -0.081 & -0.591 & $<0.01\%$ \\
FWHM        & -0.882 &  0.577 & -0.176 & $<0.01\%$ \\
$\sigma_{*}$ &  0.023 & -0.563 & -0.682 & $<0.01\%$ \\
$L_{H}$    &  0.030 & -0.044 & -0.576 &$<0.01\%$ \\
z          &  0.137 &  0.005 & -0.583 &$<0.01\%$ \\
\enddata
\tablecomments{Partial correlation coefficients for $\Delta M_{\rm
BH}$ with Eddington ratio. Col.(1): The control parameters in the
partial correlation test.  Col.(2): The correlation between
Eddington ratio and the control parameter. Col.(3): The correlation
between $\Delta M_{BH}$ and the control parameter. Col.(4): The
partial correlation coefficient for the inverse correlation of
$\Delta M_{BH}$ with Eddington ratio when the influence of the
control parameter is accounted for. Col.(5): The significance of the
partial correlation.}
\label{table3}
\end{deluxetable}

\newpage
\begin{deluxetable}{crrrc}
\tablecaption{Partial correlation analysis for the dependence of $L_{5100}$ on $\sigma_{*}$}
\tablewidth{0pt}
\tablehead{\colhead{$x$} & \colhead{$r_{L_{5100},x}$} &
\colhead{$r_{\sigma_{*},x}$} & \colhead{$r_{L_{5100},\sigma_{*};x}$}
&
\colhead{$P$} \\
\colhead{(1)} & \colhead{(2)} & \colhead{(3)} & \colhead{(4)} & \colhead{(5)} }
\startdata
$L_{H}$    &  0.634 & 0.533 & 0.354 & $<0.01\%$ \\
z          &  0.802 &  0.524 & 0.292 & $<0.01\%$ \\
All\tablenotemark{a} & -- & -- & 0.262 &$<0.01\%$ \\
\enddata
\tablecomments{Partial correlation coefficients for $\Delta M_{\rm
BH}$ with Eddington ratio. Col.(1): The control parameters in the
partial correlation test. Col.(2): The correlation between
$L_{5100}$ and the control parameter. Col.(3): The correlation
between $\sigma_{*}$ and the control parameter. Col.(4): The partial
correlation coefficient for the control parameter. Col.(5): The
significance of the partial correlation.}
\tablenotetext{a}{Second-order partial correlation test, accounting
for both control parameters.} \label{table4}
\end{deluxetable}

\newpage
\begin{deluxetable}{crrrc}
\tablecaption{Partial correlation analysis for the dependence of $\Delta M_{\rm BH}$ on redshift}
\tablewidth{0pt}
\tablehead{\colhead{$x$} &
\colhead{$r_{z,x}$} &
\colhead{$r_{\Delta M,x}$} &
\colhead{$r_{\Delta M,z;x}$} &
\colhead{$P$} \\
\colhead{(1)} & \colhead{(2)} & \colhead{(3)} & \colhead{(4)} & \colhead{(5)} }
\startdata
$L_{5100}$  &  0.802 & -0.081 & 0.118 & $<0.04\%$ \\
FWHM       &   0.257 &  0.577 & -0.181 & $<0.01\%$ \\
$\sigma_{*}$  &  0.524 & -0.563 & 0.428 & $<0.01\%$ \\
$L_{H}$    &  0.707 & -0.044 & 0.052 & $11.8\%$ \\
$L_{bol}/L_{Edd}$     &  0.142 & -0.576 & 0.108 & $0.11\%$ \\
All\tablenotemark{a} & -- & -- & 0.059 & $7.64\%$ \\
\enddata
\tablecomments{Partial correlation coefficients for $\Delta M_{\rm
BH}$ with redshift. Col.(1): The control parameters in the partial
correlation test. Col.(2): The correlation between redshift and the
control parameter. Col.(3): The correlation between $\Delta M_{BH}$
and the control parameter. Col.(4): The partial correlation
coefficient for the control parameter. Col.(5): The significance of
the partial correlation.}
\tablenotetext{a}{Fifth-order partial
correlation test, accounting for all 5 control parameters .}
\label{table5}
\end{deluxetable}


\clearpage
\begin{thebibliography}{}

\bibitem[Abazajian et al.(2005)]{A05} Abazajian, K., et al. 2005, \aj, 129, 1755

\bibitem[Akritas \& Bershady(1996)]{AB96} Akritas, M.G., \& Bershady, M.A. 1996, \apj, 470, 706

\bibitem[Anderson et al.(2003)]{anderson03} Anderson, S.~F., et al.\  2003, \aj, 126, 2209

\bibitem[Barth et al.(2002)]{Barth02} Barth, A., Ho, L., \& Sargent, W.L.W. 2002, \apj, 566, L13

\bibitem[Barth et al.(2004)]{Barth04} Barth, A.J., Ho, L.C., Rutledge, R.E., \& Sargent, W.L.W. 2004 \apj, 607, 90

\bibitem[Barth et al.(2005)]{Barth05} Barth, A.J., Greene, J.E., \& Ho. L.C. 2005, \apj, 619, L151

\bibitem[Becker, White, \& Helfand(1995)]{becker95} Becker, R.~H., White, R.~L., \& Helfand, D.~J.\ 1995, ApJ, 450, 559

\bibitem[Begelman \& Nath(2005)]{begelman05} Begelman, M.~C., \& Nath, B.~B.\ 2005, \mnras, 361, 1387

\bibitem[Bender (1990)]{B90} Bender, R. 1990, \aap \ , 229, 441

\bibitem[Bentz et al.(2006)]{Bentz06} Bentz, M.C.  et al. 2006, \apj, 644, 133

\bibitem[Bernardi et al.(2003a)]{B03a} Bernardi, M., et al. 2003a, \aj, 125, 1817

\bibitem[Bernardi et al.(2003b)]{B03b} Bernardi, M., et al. 2003b, \aj, 125, 1849

\bibitem[Bian \& Zhao(2004)]{bian04} Bian, W., \& Zhao, Y.\ 2004, \mnras, 347, 607

\bibitem[Blandford \& McKee(1982)]{BM82} Blandford, R.D., \& McKee, C.F. 1982, \apj, 255, 419

\bibitem[Blanton et al.(2003)]{blanton03} Blanton, M.~R., Lin, H., Lupton, R.~H., Maley, F.~M., Young, N., Zehavi, I., \& Loveday, J.\ 2003, \aj, 125, 2276

\bibitem[Bonning et al.(2005)]{Bonning05} Bonning, E.~W., Shields, G.~A., Salviander, S., \& McLure, R.~J.\ 2005, \apj, 626, 89

\bibitem[Boroson \& Green(1992)]{BG92} Boroson, T.A., \& Green., R.F. 1992, \apjs, 80, 109

\bibitem[Boroson (2003)]{Boroson03} Boroson, T.A. 2003, \apj, 585, 647

\bibitem[Boroson(2005)]{Boroson05} Boroson, T.\ 2005, \aj, 130, 381

\bibitem[Borys et al.(2005)]{Borys05} Borys, C., Smail, I., Chapman, S.~C., Blain, A.~W., Alexander, D.~M., \& Ivison, R.~J.\ 2005, \apj, 635, 853

\bibitem[Botte et al.(2005)]{Botte05} Botte, V., Ciroi, S., di Mille, F., Rafanelli, P., \& Romano, A.\ 2005, \mnras, 356, 789

\bibitem[Brinchmann et al.(2004)]{BJ04} Brinchmann, J., et al. 2004, astro-ph/0406220

\bibitem[Collin et al.(2006)]{Collin06} Collin, S., Kawaguchi, T., Peterson, B.~M., \& Vestergaard, M.\ 2006, \aap, 456, 75

\bibitem[Connolly \& Szalay(1999)]{CS99} Connolly, A. J., \& Szalay, A. S. 1999, \aj, 117, 2052

\bibitem[Di Matteo et al.(2005)]{dimatteo05} Di Matteo, T., Springel, V., \& Hernquist, L.\ 2005, \nat, 433, 604

\bibitem[Dunlop et al.(2003)]{dunlop03} Dunlop, J.~S., McLure, R.~J., Kukula, M.~J., Baum, S.~A., O'Dea, C.~P., \& Hughes, D.~H.\ 2003,   \mnras, 340, 1095

\bibitem[Faber \& Jackson(1976)]{FJ76} Faber, S.M., \& Jackson, R.E. 1976, \apj, 204, 668

\bibitem[Ferrarese \& Merritt(2000)]{FM00} Ferrarese, L., \& Merritt, D. 2000, \apjl, 539, L9

\bibitem[Ferrarese et al.(2001)]{Ferrarese01} Ferrarese, L., Pogge, R. W., Peterson, B. M., Merritt, D., Wandel, A., \& Joseph, C. L. 2001, ApJ, 555, L79

\bibitem[Fitzpatrick (1999)]{fitzpatrick99} Fitzpatrick, E.L. 1999, \pasp, 111, 63

\bibitem[Floyd et al.(2004)]{floyd04} Floyd, D.~J.~E., Kukula, M.~J., Dunlop, J.~S., McLure, R.~J., Miller, L., Percival, W.~J., Baum, S.~A., \& O'Dea,
C.~P.\ 2004, \mnras, 355, 196

\bibitem[Franx et al.(1989)]{F89} Franx, M., Illingworth, G.D., \& Heckman, T. 1989, \apj, 344, 613

\bibitem[Fukugita et al.(1996)]{F96} Fukugita, M., Ichikawa, T., Gunn, J.E., Doi, M., Shimasaku, K., \& Schneider, D.P. 1996, \aj, 111, 1748

\bibitem[Gebhardt et al.(2000a)]{G00a} Gebhardt, K., et al. 2000, \apjl, 539, L13

\bibitem[Greene \& Ho(2005)]{Greene05} Greene, J.E., \& Ho L.C. 2005, \apj, 630, 122

\bibitem[Greene \& Ho(2006a)]{Greene06a} Greene, J.E., \& Ho L.C., 2006a, \apjl, 641, L21

\bibitem[Greene \& Ho(2006b)]{Greene06b} Greene, J.E., \& Ho L.C. 2006b, \apj, 641, 117

\bibitem[Grupe \& Mathur(2004)]{grupe04} Grupe, D., \& Mathur, S. 2004, \apjl, 606, L41

\bibitem[Gunn et al.(1998)]{gunn98} Gunn, J.E., et al. 1998, \aj, 116  3040

\bibitem[Gunn et al.(2006)]{gunn06} Gunn, J.E., et al. 2006, \aj, 131, 2332

\bibitem[Hamilton et al.(2002)]{hamilton02} Hamilton, T.~S., Casertano, S., \& Turnshek, D.~A.\ 2002, \apj, 576, 61

\bibitem[Heckman et al.(2004)]{H04} Heckman T., et al. 2004, \apj, 613, 109

\bibitem[Hogg et al.(2001)]{Hogg01} Hogg, D. W., Schlegel, D. J., Finkbeiner, D. P., \& Gunn, J. E. 2001, \aj, 122, 2129

\bibitem[Ivezi{\'c} et al.(2002)]{I02} Ivezi{\'c}, {\v Z}., et al. 2002, \aj, 124, 2364

\bibitem[J{\o}rgensen et al.(1995)]{J95} J{\~o}rgensen I., Franx, M., \& Kj{\ae}rgaard, P. 1995, \mnras, 276, 1341

\bibitem[Kaspi et al.(2000)]{K00} Kaspi, S., Smith P.S., Netzer, H., Maoz, D., Jannuzi, B.T., \& Giveon, U. 2000, \apj, 533, 631

\bibitem[Kaspi et al.(2005)]{K05} Kaspi, S., Maoz, D., Netzer, H., Peterson, B.M., Vestergaard, M., \& Jannuzi, B.T. 2005, \apj, 629, 61

\bibitem[King(2003)]{king03} King, A. 2003, \apj, 596, L27

\bibitem[Kollatschny (2003)]{Kollatschny03} Kollatschny, W. 2003, \aap, 407, 461

\bibitem[Kormendy \& Richstone (1995)]{KR95} Kormendy, J., \& Richstone, D. 1995, \araa, 33, 581

\bibitem[Kuhlbroadt et al.(2004)]{K04} Kuhlbroadt, B., Wisotzki, L., \& Jahnke, K. 2004, \mnras, 349, 1027

\bibitem[Laor (1998)]{L98} Laor, A. 1998, \apj, 505, L83

\bibitem[Lynden-bell (1969)]{L69} Lynden-Bell, D. 1969, Nature, 223, 690

\bibitem[Magorrian et al.(1998)]{M98} Magorrian, J., et al. 1998, \aj, 115, 2285

\bibitem[Mclure et al.(2000)]{M00} McLure, R.J., Dunlop, J.S., \& Kukula, M.J. 2000, \mnras, 318, 693

\bibitem[Mclure \& Dunlop(2001)]{MD01} McLure, R.J., \& Dunlop, J.S. 2001, \mnras, 327, 199

\bibitem[Mclure \& Dunlop(2002)]{MD02} McLure, R.J., \& Dunlop, J.S. 2002, \mnras, 331, 795

\bibitem[Mclure \& Jarvis (2002)] {Mclure02} Mclure, R.J., \& Jarvia, M.J. 2002, \mnras, 337, 109

\bibitem[Merritt \& Poon(2004)]{merritt04} Merritt, D., \& Poon, M.~Y.\ 2004, \apj, 606, 788

\bibitem[Miralda-Escud{\'e} \& Kollmeier(2005)]{miralda05} Miralda-Escud{\'e}, J., \& Kollmeier, J.~A.\ 2005, \apj, 619, 30

\bibitem[Moultaka et al.(2004)]{MJ04} Moultaka, J., Ilovaisky, S.A., Prugniel, P. \& Soubiran, C. 2004, \pasp, 116, 693

\bibitem[Murray \& Chiang(1997)]{murray97} Murray, N., \& Chiang, J.\ 1997, \apj, 474, 91

\bibitem[Nelson \& Whittle(1996)]{Nelson96} Nelson, C.~H., \& Whittle, M.\ 1996, \apj, 465, 96

\bibitem[Nelson (2000)]{Nelson00} Nelson, C.H. 2000, \apj, 544, L91

\bibitem[Nelson et al.(2004)]{Nelson04} Nelson, C.H., Green, R.F., Bower, G., Gebhardt, K., \& Weistrop, D. 2004, \apj, 615, 652

\bibitem[Netzer \& Peterson (1997)]{NP97} Netzer, H., \& Peterson, B. M. 1997, in Astronomical Time Series, ed. D.Maoz, A. Sternberg, \& E. M. Leibowitz
(Dordrecht: Kluwer), 85

\bibitem[Netzer (2003)]{Netzer03} Netzer, H. 2003, \apj, 583, L5

\bibitem[Onken \& Peterson (2002)]{Onken02} Onken, C.A., \& Peterson, B.M. 2002, \apj, 572, 746

\bibitem[Onken et al.(2004)]{Onken04} Onken, C.A., Ferrarese, L., Merritt, D., Peterson, B.M., Pogge, R.W., Vestergaard, M., \& Wandel, A. 2004, \apj, 615,
645

\bibitem[Pagani et al.(2003)]{pagani03} Pagani, C., Falomo, R., \& Treves, A.\ 2003, \apj, 596, 830

\bibitem[Peng et al.(2002)] {Peng02} Peng, C.Y., Ho, L.C., Impey, C.D., \& Rix, H. 2002, \aj, 124, 266

\bibitem[Peng et al.(2006a)]{Peng06a} Peng, C.~Y., Impey, C.~D., Ho, L.~C., Barton, E.~J., \& Rix, H.-W.\ 2006a, \apj, 640, 114

\bibitem[Peng et al.(2006b)]{Peng06b} Peng, C.~Y., Impey, C.~D., Rix, H.-W., Kochanek, C.~S., Keeton, C.~R., Falco, E.~E., Leh{\'a}r, J., \& McLeod, B.~A.\
2006b, \apj, 649, 616

\bibitem[Peterson (1993)]{Peterson93} Peterson, B.M. 1993, \pasp, 105, 247

\bibitem[Peterson \& Wandel (1999)]{Peterson99} Peterson, B.M., \& Wandel, A. 1999, \apj, 521, L95

\bibitem[Peterson \& Wandel (2000)]{Peterson00} Peterson, B.M., \& Wandel, A. 2000, \apj, 540, L13

\bibitem[Peterson et al.(2004)]{Peterson04} Peterson, B.M., et al. 2004, \apj, 613, 682

\bibitem[Peterson et al.(2005)]{Peterson05} Peterson, B.M., et al. 2005, \apj, 632, 799

\bibitem[Pier et al.(2003)]{pier03} Pier, J. R., Munn, J. A., Hindsley, R. B., Hennessy, G. S., Kent, S. M., Lupton, R. H., \& Ivezi{\'c}, {\v Z}. 2003,
\aj, 125, 1559

\bibitem[Press et al.(1992)]{press92} Press, W. H., Teukolsky, S. A., Vetterling, W. T., \& Flannery, B. P. 1992, Numerical Recipes in C (Second ed.;
cambridge: Cambridge Univ. Press), 660

\bibitem[Rees (1984)]{R84} Rees, M. J. 1984, \araa, 22, 471

\bibitem[Richards et~al.(2002)]{richards02} Richards, G.~T.~et al.\ 2002, \aj, 123, 2945

\bibitem[Richards et~al.(2006)]{richards06} Richards, G.~T.~et al.\ 2006, astro-ph/0601558

\bibitem[Rix \& White(1992)]{RW92} Rix, H.W., \& White, S.D.M. 1992, \mnras, 254, 389

\bibitem[Robertson et al.(2006)]{robertson06} Robertson, B., Hernquist, L., Cox, T.~J., Di Matteo, T., Hopkins, P.~F., Martini, P., \& Springel, V.\ 2006,
\apj, 641, 90

\bibitem[Salviander et al.(2006)]{Salviander06} Salviander, S., Shields, G.~A., Gebhardt, K., \& Bonning, E.~W.\ 2006, New Astronomy Review, 50, 803

\bibitem[S{\'a}nchez et al.(2004)]{sanchez04} S{\'a}nchez, S.~F., et al.\ 2004, \apj, 614, 586

\bibitem[Sargent et al.(1977)]{S77} Sargent, W.L.W., Schechter, P.L., Boksenberg, A., \& Shortridge, K. 1977, \apj, 212, 326

\bibitem[Schade et al.(2000)]{schade00} Schade, D.~J., Boyle, B.~J., \& Letawsky, M.\ 2000, \mnras, 315, 498

\bibitem[Schlegel et al.(1998)]{schlegel98} Schlegel, D.J., Finkbeiner, D.P., \& Davis, M. 1998, \apj, 500, 525

\bibitem[Schneider et al.(2005)]{Schneider05} Schneider, D.P. et al. 2005, \aj, 130, 367

\bibitem[Shang et al.(2005)]{Shang05} Shang, Z. et al. 2005, \apj, 619, 41

\bibitem[Shields et al.(2003)]{Shields03} Shields, G.A., Gebhardt, K., Salviander, S., et al. 2003, \apj, 583, 124

\bibitem[Shields et al.(2006)]{Shields06} Shields, G.~A., Menezes, K.~L., Massart, C.~A., \& Vanden Bout, P.\ 2006, \apj, 641, 683

\bibitem[Smith et al.(2002)]{Smith02}Smith, J. A., et al. 2002, AJ, 123, 2121

\bibitem[Spergel et al.(2006)]{SD06} Spergel, D.N. et al. 2006, astro-ph/0603449

\bibitem[Stoughton et al.(2002)]{stoughton02} Stoughton, C.~et al.\ 2002, \aj, 123, 485

\bibitem[Strauss et al.(2002)]{strauss02} Strauss, M.~A.~et al.\ 2002, \aj, 124, 1810

\bibitem[Tonry \& Davis (1979)]{TD79} Tonry, J., \& Davis, M. 1979, \aj, 84, 1511

\bibitem[Tremaine et al.(2002)]{T02} Tremaine, S., et al. 2002, \apj, 574, 740

\bibitem[Treu et al.(2004)]{Treu04} Treu, T., Malkan, M.~A., \& Blandford, R.~D.\ 2004, \apjl, 615, L97

\bibitem[Vanden Berk et al.(2005)]{vandenberk05} Vanden Berk, D.~E., et al.\ 2005, \aj, 129, 2047

\bibitem[Vanden Berk et al.(2006)]{V06} Vanden Berk, D.~E., Shen, J., Yip, C.-W., Schneider, D.~P., et al. 2006, \aj, 131, 84

\bibitem[Vestergaard (2002)]{Vestergaard02} Vestergaard, M. 2002, \apj, 571, 733

\bibitem[Vestergaard (2004)]{Vestergaard04} Vestergaard, M. 2004, \apj, 601, 676

\bibitem[Vestergaard \& Peterson (2006)]{Vestergaard06} Vestergaard, M., \& Peterson, B.M. 2006, \apj, 641, 689

\bibitem[Wall \& Jenkins (2003)]{wall} Wall,J.V. \& Jenkins C.R. 2003, Practical Statistics for Astronomers, P66, Cambridge University Press

\bibitem[Wandel et al.(1999)]{Wandel99} Wandel, A., Peterson, B.M., \& Malkan, M.A. 1999 \apj, 526, 579

\bibitem[Wandel(2002)]{wandel02} Wandel, A. 2002, \apj, 565, 762

\bibitem[Warner et al. (2003)]{Warner03} Warner, C., Hamann, F., \& Dietrich, M. 2003, \apj, 596, 72

\bibitem[Wegner et al.(1999)]{Wegner99} Wegner, G., Colless, M., Saglia, R. P., McMahan, R. K., Davies, R. L., Burstein, D., \& Baggley, G. 1999, \mnras,
305, 259

\bibitem[Woo et al.(2006)]{Woo06} Woo, J.-H., Treu, T., Malkan, M.~A., \& Blandford, R.~D.\ 2006, \apj, 645, 900

\bibitem[Wu et al. (2004)]{Wu04} Wu, X.B., Wang, R., Kong, M.Z., Liu, F.K.,   \& Han, J.L. 2004, \aap

\bibitem[V\'eron-Cetty et al.(2004)]{veron04} V\'eron-Cetty, M.-P., Joly, M., \& V\'eron, P. 2004, \aap, 417, 515

\bibitem[Yip et al.(2004a)]{YiP04a} Yip, C.W., et al. 2004a, \aj, 128, 585

\bibitem[Yip et al.(2004b)]{YiP04b} Yip, C.W., et al. 2004b, \aj, 128, 2603

\bibitem[York et al.(2000)]{York00} York, D.G., et al. 2000, \aj, 120, 1579

\end{thebibliography}
\end{document}